%% file: cancel.tex
\documentclass[11pt]{article}

\usepackage{latexsym}
\usepackage{amssymb}
\def\thefootnote{\fnsymbol{footnote}}

\input{MaryK.tex}

\begin{document}

\begin{titlepage}
\begin{center}

\hfill UCB-PTH-14/37 \\
\hfill Nikhef-2014-044\\
\hfill October 2014\\[.1in]

{\large {\bf THE ANOMALY STRUCTURE OF REGULARIZED
    SUPERGRAVITY}}\footnote{This work was supported in part by the
  Director, Office of Science, Office of High Energy and Nuclear
  Physics, Division of High Energy Physics, of the U.S. Department of
  Energy under Contracts DE-AC02-05CH11231 and DE-AC03-76SF00098, in
  part by the National Science Foundation under grants PHY-0457315,
  PHY-1002399 and PHY-0098840, in part by the Australian Research
  Council (grant No. DP1096372), in part by a UWA Research Development
  Award, by the ERC Advanced Grant no. 246974, {\it ``Supersymmetry: a
    window to non-perturbative physics''} and by the European
  Commission Marie Curie International Incoming Fellowship grant
  no. PIIF-GA-2012-627976.}  \\[.1in]

Daniel Butter$^1$ and Mary K. Gaillard$^2$\\[.1in]

{\em $^1$Nikhef Theory Group,
Science Park 105,
1098 XG Amsterdam, The Netherlands}\\

{\em $^2$Department of Physics and Theoretical Physics Group,
 Lawrence Berkeley Laboratory, \\
 University of California, Berkeley, California 94720}\\[.3in] 

\end{center}

\begin{abstract}

  On-shell Pauli-Villars regularization of the one-loop divergences of
  supergravity theories is used to study the anomaly structure of
  supergravity and the cancellation of field theory anomalies under a
  $U(1)$ gauge transformation and under the T-duality group of modular
  transformations in effective supergravity theories with three
  K\"ahler moduli $T^i$ obtained from orbifold compactification of the
  weakly coupled heterotic string.  This procedure requires
  constraints on the chiral matter representations of the gauge group
  that are consistent with known results from orbifold
  compactifications.  Pauli-Villars regulator fields allow for the
  cancellation of all quadratic and logarithmic divergences, as well
  as most linear divergences. If all linear divergences were canceled,
  the theory would be anomaly free, with noninvariance of the action
  arising only from Pauli-Villars masses.  However there are linear
  divergences associated with nonrenormalizable gravitino/gaugino
  interactions that cannot be canceled by PV fields.  The resulting
  chiral anomaly forms a supermultiplet with the corresponding
  conformal anomaly, provided the ultraviolet cut-off has the appropriate
  field dependence, in which case total derivative terms, such as
  Gauss-Bonnet, do not drop out from the effective action.
%
%
  The anomalies can be partially canceled by the four-dimensional
  version of the Green-Schwarz mechanism, but additional counterterms,
  and/or a more elaborate set of Pauli-Villars fields and couplings,
  are needed to cancel the full anomaly, including D-term
  contributions to the conformal anomaly that are nonlinear in the
  parameters of the anomalous transformations.

\end{abstract}
\end{titlepage}

\newpage

\renewcommand{\thepage}{\roman{page}}
\setcounter{page}{2}
\mbox{ }

\vskip 1in

\begin{center}
{\bf Disclaimer}
\end{center}

\vskip .2in

\begin{scriptsize}
\begin{quotation}
  This document was prepared as an account of work sponsored by the
  United States Government. While this document is believed to contain
  correct information, neither the United States Government nor any
  agency thereof, nor The Regents of the University of California, nor
  any of their employees, makes any warranty, express or implied, or
  assumes any legal liability or responsibility for the accuracy,
  completeness, or usefulness of any information, apparatus, product,
  or process disclosed, or represents that its use would not infringe
  privately owned rights.  Reference herein to any specific commercial
  products process, or service by its trade name, trademark,
  manufacturer, or otherwise, does not necessarily constitute or imply
  its endorsement, recommendation, or favoring by the United States
  Government or any agency thereof, or The Regents of the University
  of California.  The views and opinions of authors expressed herein
  do not necessarily state or reflect those of the United States
  Government or any agency thereof, or The Regents of the University
  of California.
\end{quotation}
\end{scriptsize}

\vskip 2in

\begin{center}
\begin{small}
{\it Lawrence Berkeley Laboratory is an equal opportunity employer.}
\end{small}
\end{center}

\newpage
\renewcommand{\theequation}{\arabic{section}.\arabic{equation}}
\renewcommand{\thepage}{\arabic{page}}
\setcounter{page}{1}
\def\thefootnote{\arabic{footnote}}
\setcounter{footnote}{0}

\input{./Introduction.tex}
\input{./Section2.tex}
\input{./Section3.tex}
\input{./Section4.tex}

\input{./Section5.tex}
\input{./Section6.tex}

\vskip .3in
\noindent{\bf Acknowledgments.} We are happy to acknowledge the
contributions of Andreas Birkedal, Choonseo Park, Matthijs Ransdorp
and So-Jong Rey to various early stages of this work.  We thank Dan
Freedman, Joel Giedt, Brent Nelson, Claudio Scrucca, Marco Serone, Tom
Taylor and Lewis Tunstall for discussions and helpful input.  MKG
would like to thank the Kavli Institute of Physics at U.C. Santa
Barbara for hospitality during part of the completion of this work.
This work was supported in part by the Director, Office of Science,
Office of High Energy and Nuclear Physics, Division of High Energy
Physics, of the U.S. Department of Energy under Contract
DE-AC02-05CH11231, in part by the National Science Foundation under
grants PHY-0457315 and PHY-1002399.

\newpage
\appendix

\def\ksubsection{\Alph{subsection}}
\def\theequation{\ksubsection.\arabic{equation}} 

     
\catcode`\@=11

\def\thesubsection{\Alph{subsection}}
\def\thesubsubsection{\Alph{subsection}.\arabic{subsubsection}}
\noindent{\large \bf Appendix}

\input{./AppendixA.tex}
\input{./AppendixB.tex}
\input{./AppendixC.tex}

\input{./AppendixD.tex}
\input{./AppendixE.tex}

\input{./AppendixF.tex}

\end{document}

%% file: MaryK.tex
\DeclareMathAlphabet   {\mathsc}{OT1}{cmr}{m}{sc}

\newcommand{\myref}[1]{(\ref{#1})}
\newcommand{\mysec}[1]{Section~\ref{#1}}
\newcommand{\myapp}[1]{Appendix~\ref{#1}}

\newcommand{\pr}{{\it Phys.\ Rev. }}
\newcommand{\pl}{{\it Phys.\ Lett. }}

\def\ppd{{\partial\over\partial p}\cdot}
\def\dpp{\cdot{\partial\over\partial p}}

\def\Del{\Delta}
\def\half{{1\over2}}
\def\bea{\begin{eqnarray}}
\def\eea{\end{eqnarray}}
\def\beq{\begin{equation}}
\def\eeq{\end{equation}}
\def\bfm{{\bf m}}
\def\pr{{1\over -p^2}}
\def\bL{\bar{\Lambda}}
\def\ux{$U(1)_X$}
\def\uk{$U(1)_K$}
\def\uo{$U(1)$}
\def\dx{\delta_X}
\def\vx{V_X}

\def\cx{c_X}

\def\aax{a_X}
\def\superint{\int d^{4}\theta}
\def\M{\bar{M}}
\def\N{\bar{N}}
\newcommand{\WaWa}{ W_a^{\alpha} W^a_{\alpha}}
\newcommand{\WbWb}{ W^a_{\dot{\alpha}}W_a^{\dot{\alpha}}}

\newcommand{\DbDb}{{\cal D}_{\dot{\alpha}}{\cal D}^{\dot{\alpha}}}
\newcommand{\Da}{{\cal D}_{\alpha}}
\newcommand{\Db}{{\cal D}^{\dot{\beta}}}
\newcommand{\Lb}{L^{\dot{\beta}}}
\newcommand{\Ld}{L_{\dot{\beta}}}
\newcommand{\Dc}{{\cal D}^{\alpha}}
\newcommand{\Dd}{{\cal D}_{\dot{\beta}}}
\newcommand{\Wa}{W_{\alpha}}
\newcommand{\Wb}{W^{\dot{\beta}}}
\newcommand{\Wc}{W^{\alpha}}
\newcommand{\Wd}{W_{\dot{\beta}}}
\newcommand{\Ya}{Y_{\alpha}}

\newcommand{\Yc}{Y^{\alpha}}
\newcommand{\Yd}{Y_{\dot{\beta}}}
\def\bv{\bar{\varphi}}
\def\bsig{\bar{\sigma}}
\def\brh{\bar{\rho}}
\newcommand{\phia}{\varphi_{\alpha}}
\newcommand{\phib}{\bv^{\dot{\beta}}}
\newcommand{\phic}{\varphi^{\alpha}}
\newcommand{\phid}{\bv_{\dot{\beta}}}
\def\chid{\chi_{\dot{\beta}}}
\def\sigd{\sigma_{\dot{\beta}}}
\newcommand{\Xa}{X_{\alpha}}
\newcommand{\ka}{k_{\alpha}}

\newcommand{\Xc}{X^{\alpha}}
\newcommand{\La}{L_{\alpha}}
\newcommand{\Lc}{L^{\alpha}}
\newcommand{\fa}{f_{\alpha}}
\newcommand{\fc}{f^{\alpha}}
\newcommand{\ga}{g_{\alpha}}
\newcommand{\gc}{g^{\alpha}}
\newcommand{\Xb}{X^{\dot\beta}}
\newcommand{\Xd}{X_{\dot\beta}}

\def\dotb{\dot{\beta}}

\def\re{{\rm Re}}
\def\im{{\rm Im}}

\def\bM{\bar{M}}

\def\tg{\tilde{g}}

\def\G{{\cal G}}

\def\cW{{\cal W}}

\def\cO{{\cal O}}

\def\tr{{\tilde r}}
\def\tG{{\widetilde G}}
\def\tX{{\widetilde X}}

\def\tF{{\tilde F}}
\def\tZ{{\widetilde Z}}
\def\tY{{\widetilde Y}}
\def\dZ{{\dot Z}}
\def\dY{{\dot Y}}

\def\tL{{\tilde L}}

\def\tN{{\tilde N}}
\def\tL{{\tilde L}}
\def\hD{{\hat D}}
\def\hG{{\hat G}}
\def\hZ{{\widehat Z}}
\def\hY{{\widehat Y}}

\def\hF{{\hat F}}

\def\D{{\cal D}}
\def\hN{{\hat N}}

\def\bD{\bar{\D}}

\def\pp{\partial}

\def\notd{\not{\hspace{-.03in}\pp}}
\def\notp{\not{\hspace{-.03in}p}}

\def\notj{\not{\hspace{-.03in}J}}
\def\notD{\not{\hspace{-.05in}D}}

\def\notG{\not{\hspace{-.04in}\G}}

\def\notZ{\not{\hspace{-.05in}Z}}

\def\lll{{\Lambda^2\over 32\pi^2}}
\def\lln{{\ln\Lambda^2\over 32\pi^2}}
\def\ibar{\bar{\imath}}

\def\[{\left [}
\def\]{\right ]}
\def\({\left (}
\def\){\right )}
\def\lbr{\left\{}
\def\rbr{\right\}}
\def\r{\right|}
\def\l{\left.}

\def\J{\bar{J}}
\def\H{\bar{H}}
\def\h{\bar{h}}
\def\T{\bar{T}}

\def\z{\bar{z}}
\def\q{\bar{q}}
\def\del{\delta}
\def\R{{\cal{R}}}
\def\cZ{{\cal{Z}}}

\def\heta{{\hat\eta}}
\def\teta{{\tilde\eta}}

\def\S{{\bar{S}}}

\def\STr{{\rm STr}}
\def\Tr{{\rm Tr}}
\def\cA{{\cal A}}
\def\K{{\cal K}}

\newcommand{\Gaa}{\Gamma_{\alpha}}

\def\Ti^{T^{(i)}}

\def\f{\bar{f}}
\def\bR{\bar{R}}

\def\L{{\cal L}}
\def\R{{\cal{R}}}

\def\hM{\widehat{M}}

\def\hD{\hat{D}}
\def\hG{\hat{G}}
\def\hV{\hat{V}}
\def\n{\bar{n}}
\def\m{\bar{m}}
\def\s{\bar{s}}
\def\cM{{\cal{M}}}
\def\cN{{\cal{N}}}
\def\cP{{\cal{P}}}
\def\cO{{\cal{O}}}
\def\bth{\bar{\theta}}

\def\bc{\bar{\chi}}

\def\tbM{{\widetilde{\M}}}
\def\tM{{\widetilde M}}

\def\tm{{\tilde m}}

\def\A{\bar{A}}

\def\y{{\bar{y}}}

\def\Y{{\bar{Y}}}
\def\Z{{\bar{Z}}}
\def\tbZ{{\widetilde\Z}}
\def\tbY{{\widetilde\Y}}
\def\hbZ{{\widehat\Z}}
\def\hbY{{\widehat\Y}}
\def\dbZ{{\dot\Z}}
\def\dbY{{\dot\Y}}
\def\bb{\bar{b}}

\def\bph{\bar{\phi}}
\def\bPh{\bar{\Phi}}
\def\hph{\hat{\varphi}}
\def\hf{\hat{\phi}}
\def\hbp{\hat{\bv}}

\def\tph{\tilde{\varphi}}
\def\tbp{\tilde{\bv}}
\def\tphi{\tilde{\phi}}

\def\bF{\bar{F}}
\def\Q{\bar{Q}}

\def\bj{\bar{\jmath}}
\def\a0{\alpha_0}
\def\chiproj{(\bD^2 - 8R)}
\def\bchiproj{(\D^2 - 8 \bar R)}

\def\eee{\nonumber \\ &=&}
\def\ddd{\nonumber \\ &&}
\def\mmm{\nonumber \\ &&\qquad}
\def\nnn{\nonumber \\ }
\def\hc{ + {\rm h.c.}}

\def\bU{U^{-1}}

\def\DX{\l\Dc X_\alpha\r}
\def\Df{\l\Dc f_\alpha\r}

%% file: Introduction.tex
\section{Introduction}
\hspace{0.8cm}\setcounter{equation}{0} It has been
shown~\cite{pv1}--\cite{pvdil} that on-shell Pauli-Villars (PV)
regularization of one-loop quadratic and logarithmic ultraviolet
divergences in general $N=1$ supergravity~\cite{crem,bggm} is
possible, subject to constraints on the matter representations of the
gauge group that are consistent with the spectra found, for example in
orbifold compactifications.  Supergravity derived from weakly coupled
string theory typically has classical symmetries that are broken at
the quantum level by conformal and chiral anomalies which arise,
respectively, from logarithmic and linear divergences in the light
field loops.  If they can be canceled by PV loops in the regulated
theory, the remaining noninvariance under the classical symmetries
arises from the noninvariance of the PV mass terms, provided all other
PV couplings are invariant.  In this paper we investigate the anomaly
structure and anomaly cancellation in a class of $Z_N$ orbifolds with
just three ``diagonal'' K\"ahler moduli $T^I = T^{II}$ $(I=1,2,3)$.

The anomalous symmetries that we consider are the target space duality
transformations, hereafter referred to as modular transformations, and
a gauge transformation under an Abelian gauge group, hereafter
referred to as \ux.  These symmetries are perturbatively
unbroken~\cite{mod} in the underlying string theory and therefore must
be canceled by some combination of loop contributions from heavy
string and Kaluza-Klein modes (``string threshold corrections'') and
of counterterms, including a four-dimensional version of the
Green-Schwarz (GS) mechanism~\cite{gs}.  The anomaly canceling
contributions to the Yang-Mills (YM) Lagrangian have been determined
for large classes of orbifolds (including the $Z_N$ orbifolds
considered here) by matching field theory and string loop
calculations~\cite{ant}--\cite{linear}.  More recently, a string
theory analysis~\cite{ssanom} of a limited class of $Z_N$ heterotic
orbifolds found cancellation of all anomalies through a universal GS
mechanism.  Here we approach the same problem from the point of view
of the effective four dimensional supergravity theory~\cite{bglett}.

In the following section we define our notation and display the
ultraviolet divergent part of the low energy effective Lagrangian
obtained from light particle loops~\cite{us}--\cite{uslett} in the
form of superfield operators that will be convenient for the
subsequent analysis. In Section \ref{pvreg} we will use the results
of~\cite{pvcan} and~\cite{pvdil}, hereafter referred to as I and II,
respectively, to construct invariant couplings of PV supermultiplets
needed to cancel the light loop divergences.  Mechanisms for anomaly
cancellation and constraints on PV masses will be discussed in Section
\ref{ancan}, and the explicit form of the anomalies will be displayed
in Section \ref{modan}.  Our results are summarized in Section
\ref{sum}.

Some calculational details are presented in a series of appendices.
As discussed below, requiring the cancellation of quadratic and
logarithmic UV divergences that were identified in
\cite{us}--\cite{uslett} does not uniquely fix the couplings of the PV
sector. In \myapp{lindiv} we derive additional constraints that assure
the cancellation of almost all linear divergences. While the
constraints from the requirement of cancellation of (on-shell)
quadratic divergences automatically assures the cancellation of the
chiral K\"ahler anomaly arising from the fermion spin connection,
there is a residual linear divergence associated with the affine
connection of the gravitino.  In addition there is an off-diagonal
gravitino/gaugino connection that has no counterpart in the PV sector.
The corresponding contributions to the chiral anomaly must have 
supersymmetric counterparts; these can be obtained by introducing a
field-dependent UV cut-off
\beq \Lambda = \mu_0e^{K/4},\label{uvcut}\eeq
where $\mu_0$ is a constant parameter that may be set to infinity at
the end of the one-loop calculation, and $K$ is the K\"ahler potential
of the light field theory.  This has only the effect that total
derivatives with nonvanishing coefficients of $\ln\Lambda$ do not drop
out of the S-matrix elements of the regulated theory.  Once these
procedures have been implemented we recover the standard form of the
anomaly coefficients of the Yang-Mills and curvature field strengths
as well as agreement with string theory results.  However the
anomalous coefficients of operators that themselves depend on the modular
weights of the light fields depend on details of the PV regularization
procedure, and are not uniquely fixed by the cancellation of ultraviolet
divergences.

In \myapp{chiralan} we calculate the chiral anomaly for a general
supergravity theory.  As has been recently emphasized~\cite{fk}, in
supergravity the fermion connections and corresponding field strengths
contain many more operators than the
Yang-Mills~\cite{ant}--\cite{linear} and space-time
curvature~\cite{ant2}-\cite{gg} terms that have been studied previously
in the context of anomaly cancellation. These additional operators
include~\cite{co,us,us2} the K\"ahler \uk\, connection for all
fermions, the reparameterization connection for chiral fermions, an
axion coupling in the gaugino connection and a (matrix-valued)
connection~\cite{us2} linear in the Yang-Mills field strength in the
gaugino-gravitino sector.\footnote{In supergravity without a GS term
  to cancel the \ux\ anomaly, there is an additional
  connection~\cite{bdkv} which must also be included~\cite{fk,efk}.}
In addition, besides the spin connection common to all fermions, the
gravitino connection includes a term proportional to the affine
connection, and the gauginos have an additional connection that
involves the dilaton and its axionic superpartner.  The anomaly is ill
defined in an unregulated theory.  The authors of Ref.~\cite{fk} study
supergravity theories in which a subgroup of the invariance group of
the K\"ahler metric (group of modular transformations) is gauged.  They
use consistency conditions analogous to those used to obtain the
``consistent anomaly''~\cite{bard}--\cite{wbbz} for Yang-Mills
theories. They also note that some of the operators in their
expression can be removed by counter terms~\cite{fk}.  Here we are
working with a regulated theory in which the ambiguity is removed,
except for those terms for which a linear divergence remains.  The
squared field strength $\G_{\mu\nu}\G^{\mu\nu}$ corresponding to the
full fermion connection appears in the coefficient of $\ln\Lambda$.
It combines with other contributions from both light and PV loops to
cancel the UV divergence, up to a total derivative.  Part of the
conformal anomaly, including the standard Yang-Mills term and part of
the curvature term, is determined by the field-dependence of the PV
masses in such a way that it combines with the part of the chiral
anomaly arising from PV masses to form a superfield.  The remainder of
the conformal anomaly arises from the residual total derivative
logarithmic divergence mentioned above, and combines with the part of
the chiral anomaly associated with the residual linear divergence to
form a superfield, provided the cut-off has the correct
field-dependence.  Specifically, the choice \myref{uvcut} assures a
supersymmetric result for the full anomaly coefficient of the
curvature field strength term.  When combined with the PV contribution
we recover an anomaly coefficient that is consistent with string loop
calculations~\cite{ant2}.  Similarly, the chiral anomaly associated with the
off-diagonal gaugino-gravitino connection, which depends on the
Yang-Mills field strength, forms a supermultiplet with a contribution
to the conformal anomaly from a total derivative in the coefficient of
$\ln\Lambda$, when the field-dependence of the cut-off \myref{uvcut}
is included.  This contribution to the anomaly is canceled by a PV loop
contribution for a particular choice of the relevant PV mass ratio.

In the absence of linear divergences, the chiral anomaly arises solely
from the noninvariance of PV masses.  In \myapp{pvchiral} we sketch
how this may be demonstrated by comparing the direct chiral anomaly
calculation with an indirect method which assumes that no linear
divergences are present.  We also show that under appropriate
assumptions the ``consistent anomaly''~\cite{bard}--\cite{wbbz} is
recovered.  As described in~\cite{pv1}--\cite{pvdil}, and reviewed in
\mysec{pvreg} below, all the quadratic and logarithmic UV divergences
of supergravity can be regulated with PV fields in chiral
supermultiplets and Abelian vector supermultiplets.  The Abelian
vector fields acquire \ux\, and modular invariant masses through the
superhiggs mechanism and do not contribute to the anomaly, except for
a contribution, mentioned above, that cancels the gravitino-gaugino
mixed loop contribution.  As a consequence, the part of the PV sector
relevant to the bulk of the anomaly does not contain any connections
associated with the nonrenormalizable couplings of the
gravitino-gaugino sector, and the only field strength bilinears that
appear in their contribution to the anomaly coefficient are those
associated with the spin connection, the YM gauge connection, the \uk\
connection and the scalar reparameterization connections associated
with the K\"ahler metric for PV chiral superfields.  As shown in
\myapp{fullan}, the latter can be chosen such that PV fields with
noninvariant masses have very simple reparameterization connections.

In the case of an anomalous Yang-Mills theory with constant PV masses,
the anomaly is uniquely determined, as illustrated in \myapp{pvchiral}
by the standard result \myref{chiral}. However in PV
regulated supergravity, some PV masses are necessarily field
dependent, and there is considerable leeway in the choice of these
masses, which can only be fixed by a detailed knowledge of
string/Planck scale physics. As explained in \mysec{strat}, the QFT
anomaly cannot be completely removed, even in the absence of linear
divergences. Terms linear in gauge charges and modular weights are
fixed by the requirement that quadratic divergences cancel; this
uniquely fixes the PV contribution to the coefficient of
$r_{\mu\nu\rho\sigma}\tr^{\mu\nu\rho\sigma}$ as given in
\myref{delsr}, but terms cubic in these parameters are not fixed by
the requirement of cancellation of quadratic (or logarithmic) UV
divergences.  However the expansion of the parity odd part of the
fermion determinant [see \myref{tminus}], given explicitly in~\cite{us2},
has, in addition to a linear divergence proportional to $\Tr A$ that
generates an $r_{\mu\nu\rho\sigma}\tr^{\mu\nu\rho\sigma}$ term in the
anomaly, a linear divergence proportional to $\Tr A^3$, where $A$ is
the {\rm axial} current in the fermion connection.  This contains the
$U(1)_K$ connection as well as the (symmetric) YM gauge connections
and the (axial part of the) scalar reparameterization connections
for chiral fermions.\footnote{As in any regularization
  procedure, the definition of $\gamma_5$ is {\it a priori} ambiguous;
  as discussed in~\cite{pv1}--\cite{pvdil} the separation into
  ``vector'' and ``axial'' currents is dictated by the finiteness
  requirement and nonrenormalization of the
  $\theta$-parameter~\cite{russ}.}  Requiring that this linear
divergence vanish imposes cubic constraints, but as mentioned above,
does not completely determine the anomaly.

In \myapp{fullan} we consider simple parameterizations of the PV
sector that are motivated by physical considerations, and are
consistent with perturbative modular invariance of the K\"ahler
potential and cancellation of quadratic and logarithmic UV
divergences, as well as cancellation of all linear divergences from
chiral multiplet loops.  Under these assumptions we calculate the
bosonic part of the variation of the Lagrangian under a variation of
the noninvariant PV masses.  We then identify the corresponding
anomaly superfields and compare them with operators that could
potentially cancel anomalies in a generalized GS term; these include a
generalization to K\"ahler superspace of the (F-term) operator found
in Ref.~\cite{gg} for the case of pure supergravity, as well as new
D-term operators~\cite{bglett}, that can be inferred~\cite{dan} by
first working in conformal supergravity and then gauge-fixing to
K\"ahler superspace supergravity which we use in this paper.

In order to evaluate the anomaly coefficients for specific orbifold
models, in \myapp{orbs} we construct simple examples of PV sectors
that can be used to regularize the matter sector in such a way that
those PV fields that contribute to the renormalization of the K\"ahler
potential have modular and \ux\ invariant masses.  Implementation of
the GS anomaly cancellation mechanism for the F-term imposes
constraints on the modular weights and gauge charges.  Cancellation of
UV divergences assures that some of these constraints are satisfied,
but terms nonlinear in the charges under the anomalous symmetries are
not completely determined by finiteness conditions and depend on the
specifics of the PV spectrum and couplings.  The simple procedure
adopted in \myapp{fullan} and \myapp{orbs} is not sufficient to assure
the factorization needed for implementation of anomaly cancellation by
a universal GS term.\footnote{The difficulty of obtaining
  factorization in a direct field theory approach was noted in
  \cite{ssanom}.}  In addition, some D-term anomaly operators are
nonlinear in the parameters of the anomalous transformations and
cannot obviously be canceled by a straightforward generalization of
the standard four dimensional GS term; it is possible that the
additional counterterms that may be needed could correspond to 4-d
remnants of the conformal analogue of the 10-d GS~\cite{gs} term that
cancels the 10-d chiral anomaly.

The construction of the GS term is discussed in \myapp{gsterm}, and
our notations and conventions are summarized in \myapp{notation}.

%% file: Section2.tex
\section{One-Loop On-Shell Ultraviolet Divergences}\label{uvdiv}
\hspace{0.8cm}\setcounter{equation}{0}

In this paper we consider supergravity theories with classical
K\"ahler potential $K$, superpotential $W$ and gauge kinetic function
$f$ given by \bea K(Z,\Z) &=& - \ln(S +\S) + G(T + \T,\Phi,\bPh) = k +
G, \qquad W(Z) = W(T,\Phi), \nonumber \\f_{ab}(Z) &=& \delta_{ab}S,
\qquad \l S\r = \delta_{ab}(x+i y),
\label{sdil}\eea
which are the classical functions found in string compactifications with
affine level one.\footnote{The results can be generalized to higher
affine level with $f_{ab} = \delta_{ab}k_af,\;k_a=$ constant, 
by making the substitutions $F^a_{\mu\nu}\to k_a^{1\over2}F^a_{\mu\nu}, 
\;A^a_\mu\to k_a^{1\over2}A^a_\mu, \; T^a\to k_a^{-{1\over2}}T^a.$}
$S$ is the dilaton superfield in the chiral formulation used here,
$T^i$ are the moduli chiral superfields, and $\Phi^a$ are superfields
for gauge-charged matter.

The ultraviolet divergent part of the one-loop corrected bosonic
supergravity Lagrangian was calculated in~\cite{us}-\cite{uslett}. As
shown in II, the result for the logarithmically divergent part can be
interpreted as the bosonic part of a superfield expression.  After a
Weyl redefinition to put the Einstein term in canonical form, one
obtains \beq \L_{eff} = \L\(g,K_R\) + \sqrt{g}\lln\(L_D + L_F\) +
\L_Q,
\label{l0}\eeq
where $\L(g,K)$ is the standard Lagrangian~\cite{crem,bggm} for $N=1$
supergravity coupled to matter with space-time metric $g_{\mu\nu}$,
K\"ahler potential $K$ and superpotential $W$.  If $N_G$ is the
dimension of the gauge group, the renormalized K\"ahler potential
$K_R$ is given by\footnote{We denote by $z^i = \l Z^i\r$ the lowest
(scalar) components of the light chiral superfields $Z^i$.}
\bea K_R &=& K + \lln\[e^{-K}A_{ij}\A^{ij} -2\hV + (N_G - 10)M^2 -
4\K^a_a -16\D\], \nonumber \\ \K^a_b &=& {1\over
x}(T^az)^i(T_b\z)^{\m}K_{i\m}, \qquad A = e^KW = \A^{\dag}, \qquad
A_{ij} = D_i D_jA,\label{kr}\eea
where $V = \hV + \D$ is the classical scalar potential with 
\bea \hV &=& e^{-K}A_i\A^i - 3M^2,\qquad A_i = D_iA,\qquad
\D = (2x)^{-1}\D^a\D_a,\nnn \D_a &=&
K_i(T_az)^i,\qquad M^2 = e^{-K}A\A, \eea
where $M^2$ is the field-dependent squared gravitino mass, and $D_i$
is the scalar field reparameterization covariant derivative:
\beq A_i = \pp_i A, \qquad A_{i j} = \pp_i\pp_j A - \Gamma^k_{i
j}\pp_k A, \qquad \ldots\label{DA}\eeq
Scalar indices are lowered and raised with the
K\"ahler metric $K_{i\m}$ and its inverse $K^{i\m}$.  In the K\"ahler
$U(1)$ superspace formulation~\cite{bggm} of supergravity, which we
use throughout, a general ``F-term'' Lagrangian takes the form
\beq L_F = L(\Phi) = {1\over2}\superint{E\over R}\Phi + {\rm
h.c.},\label{fterm} \eeq 
where $\Phi$ is a chiral superfield of K\"ahler $U(1)$ weight $w(\Phi)
= 2$, and a general ``D-term'' Lagrangian has the form \beq L_D =
L(\phi) = \superint E\phi = - {1\over16}\superint{E\over R}\(\bD^2 -
8R\)\phi + {\rm h.c.},
\label{dterm}\eeq
with $\phi$ a real superfield of K\"ahler weight $w(\phi)=0.$ These
contributions to (\ref{l0}) are given by\footnote{The signs of the
$\Phi^a_{YM}$ terms were inadvertently flipped in transcribing the
results of I to II.  In addition, the sign of $W_{a b}$
as defined in I is incorrect, as is the sign of (A12) (see \myapp{gsterm}).  
Here we have normalized some of the
operators differently from those in II: $\l\Phi_{\rm Y M}\r_{\rm here} = 
4\l\Phi_{\rm YM}\r_{\rm I I},\;\l\Phi_{W}\r_{\rm here} = 
6\l\Phi_{W}\r_{\rm I I},\;\l\Phi_{X}\r_{\rm here} = 
- 2\l\Phi_{\alpha}\r_{\rm I I}.$}
\bea \Phi &=& \Phi_1 + \Phi_2 + {3\over2}C_a\Phi_{YM}^a + {1\over36}
\Phi_X\(N - 9N_G - 79\) + {1\over6}\Phi_W\(41 + N - 3N_G\) +
N_G\Phi'_g, \nonumber \\ \Phi_X &=& X^\beta X_\beta,
\qquad \Phi_W = W^{\alpha\beta\gamma}W_{\alpha\beta\gamma},
\qquad \Phi^a_{YM} = W_a^\alpha W^a_\alpha, \nnn 
\Phi_1 &=& - \half\(C^M_a\Phi_{YM}^a +
\Gamma_j^{i\alpha}\[\Gamma^j_{i\alpha} + 2(T_a)^j_iW^a_\alpha\]\),\nonumber \\
\Phi_2 &=& {1\over3}X^\alpha\[\Gamma_\alpha +
2(T_a)^i_iW^a_\alpha\], \qquad \Gamma_\alpha =
\Gamma^i_{i\alpha}. \label{fops}\eea
for the F-terms with $N$ the number of chiral supermultiplets
$Z^i = S,T,\Phi$, and 
\bea \phi &=& \phi_3 - 4\phi_{\cW} - 4\phi_{\cW k} + 8\hf_0 + \phi'_0
+ N_G\phi'_g + {1\over2}\phi_\chi\(41 + N + 9N_G\),\nonumber \\ \phi_3
&=& {1\over2}R^{\alpha k\;\;l}_{\;\;\;\;\alpha} R^{\dotb}_{\;\; k\dotb
  l} + \(R^{\alpha k\;\;l}_{\;\;\;\;\alpha}e^{-K/2}A_{kl} + {\rm
  h.c.}\), \qquad
\phi'_0 = 3i\Wc_a W^{\dot\beta a}\D_{\alpha\dot\beta}(S-\S)\nonumber \\
\phi_{\chi} &=& {1\over3}\(G_{\dot\beta\alpha}G^{\alpha\dot\beta} -
4R\bR\), \qquad \hf_0 =
K^{\alpha\dot{\beta}}K_{\alpha\dot{\beta}} - 4R\bR, \nonumber \\
\phi_{\cW^a_b} &=& {x^2\over4}\Wc_a\Wa^b\Wd^a\Wb_b, \quad \phi_{\cW k}
= {1\over8x}\Wc_a\Da S\Wd^a\Db\S,\label{dops}\eea
for the D-terms, where we used the on-shell relation 
\beq G_{\alpha\dot\beta} = K_{\alpha\dot\beta} - {x\over2}\Wa\Wd\eeq
to simplify the expression for $\phi_\chi$ given in (2.15) of II, and
rewrote the ``F-term'' contribution $\Phi'_0$ of that paper as the
``D-term'' contribution $\phi'_0$.  Here $x$ is understood as the
superfield $\half(S + \S)$, and $G_{\alpha\dot\beta}$ and $R =
\bR^\dag$ are auxiliary fields~\cite{bggm} of the supergravity
supermultiplet.  The chiral superfield $\Wa^a$ is the Yang-Mills
superfield strength, with $a$ a gauge index, and $L(\Phi_{YM})$ gives
the standard gauge charge renormalization.  $C_a$ and $C_a^M$ are the
quadratic Casimirs in the adjoint and matter representations,
respectively, of the gauge subgroup $\G_a$:
\beq \Tr(T_aT_b)_{\rm adj} = \delta_{ab}C_a, \qquad
\Tr(T_aT_b)_{\rm matter} = \delta_{ab}C_a^M,\eeq 
with $T_a$ a generator of $\G_a$ and $T_b$ any generator.
$W_{\alpha\beta\gamma}$ is the Weyl chiral superfield~\cite{bggm}, and
$L(\Phi_W)$ contains the Gauss-Bonnet (GB) term and the pseudoscalar
operator that is bilinear in the Riemann tensor; its explicit
expression is given by \myref{d3} (with $H(Z)=1$).  The other chiral
superfields in (\ref{fops}) are constructed from superfields of
the general form
\beq T_\alpha = - {1\over8}\(\DbDb - 8R\)\hat{T}_\alpha, \qquad
\hat{T}_\alpha = T_i\D_\alpha Z^i,
\label{tft} \eeq
where $T_i(Z,\Z)$ is any (tensor-valued) zero-weight
function of the chiral and
anti-chiral superfields $Z^i$ and $\Z^{\ibar}$, respectively.  
In particular, the chiral superfield 
\beq K_\alpha = X_\alpha = - {1\over8}\(\DbDb - 8R\)\D_\alpha K,
\label{defx} \eeq
was introduced in~\cite{bggm}; the lowest 
component of its spinorial derivative $-{1\over2}\D^\alpha X_\alpha |$ 
is the kinetic term for matter fields in the classical Lagrangian.
For the D-terms we further introduced the zero-weight real superfields
\bea T^{\alpha\dot{\beta}}_{\alpha\dot{\beta}} &=& {1\over16}\Dc Z^i\Da
Z^j \Dd\Z^{\m}\Db\Z^{\n}T_{ij\m\n},\nnn
T_{\alpha\dot{\beta}} &=& {1\over4}\Da Z^i
\Dd\Z^{\m}T_{i\m},\qquad T^\alpha_\alpha = {1\over2}\Dc
Z^i\Da Z^jT_{ij} + {\rm h.c.}\label{tdt}\eea
Thus $\phi_3$ is constructed from the Riemann tensor
\beq R^i_{j k\m} = K^{i\n}R_{\n j k\m} = D_{\m}\Gamma^i_{j k}\eeq
associated with the K\"ahler metric.  As discussed\footnote{In the
expression (2.2) of II, $L'_g = L(\Phi'_g) + L(\phi'_g).$} in II, the
superfields $\Phi'_g,\phi'_g$ are equivalent--up to terms that vanish
on shell--to linear combinations of the the generic operators
introduced above and D-terms that involve supergravity
superfields:  $G_{\alpha\dot{\beta}}K_{s\s}\D^\alpha S\D^{\dot{\beta}}\S,\ldots$.
As shown in Appendix C of II, these terms must be exactly canceled by
PV Abelian gauge multiplets that couple to the dilaton.  In the
string-derived models considered here and in II, the masses of these
PV fields are invariant under T-duality and the anomalous $U(1)$, and
therefore do not contribute to the field theory anomalies.

The various terms in the bosonic part of (\ref{l0}) are given in
component form in I and II, where total derivatives (such as the GB
term) were dropped.  These terms cannot be dropped if, after
regularization, the constant cut-off is replaced by a field-dependent
PV mass that plays the role of effective cut-off (or if the cut-off
itself is field-dependent).  All terms relevant to the anomaly
structure and anomaly cancellation will be included in Sections
\ref{modan}.

The quadratically divergent part $\L_Q$ of (\ref{l0}) 
\bea \L_Q &=& -\sqrt{g}\lll\[{1\over2}(3+N_G- N)\l\Dc\Xa\r + 
\(\hV + M^2\)(7 + 3N_G - N)\right.
\nonumber \\ & & \l+ N_G\l\Dc\ka\r + \l\Dc\Gaa\r + {2\over x}\D_a\Tr T^a\]
+ {\rm fermion\; terms}\label{lquad}\eea
can be interpreted as a further renormalization of the K\"ahler
potential $K_Q$ only after regularization~\cite{pv1,pvcan}; $K_Q$ is
governed by the PV squared masses.

%% file: Section3.tex
\section{Invariant PV Regularization}\label{pvreg}
\hspace{0.8cm}\setcounter{equation}{0}
In this section we construct gauge and modular invariant PV couplings
to light fields that are needed to cancel the ultraviolet divergences
from light field loops in the theory defined by \myref{sdil}. 

\subsection{Chiral multiplet loops}\label{chiloops}
To regulate the loops arising from dimension-four operators involving
only chiral supermultiplets $Z^i$, we need to introduce at least one
set of PV chiral superfields $\dZ^I$ with negative metric and the same
gauge charges, K\"ahler metric, and quadratic superpotential couplings
as the light chiral multiplets.  This assures cancellation of the
ultraviolet divergences associated with $\Phi_1$ and $\Phi_2$ in
(\ref{fops}), and $\l\D^\alpha\Gamma_\alpha\r$ in (\ref{lquad}).
Since the dilaton does not have a classical superpotential, and its
K\"ahler metric is easily reproduced, we concentrate in this
subsection on the chiral supermultiplets $Z^p=T^i,\Phi^a$ and their PV
counterparts $\dZ^P = \dZ^I,\dZ^A$: \bea K_0^{\dZ} &=&
\sum_{P,Q=p,q}\lbr K_{p\q}\dZ^P\dbZ^{\Q} + {1\over2}\[\(\pp_p\pp_qK -
K_pK_q\)\dZ^P\dZ^Q + {\rm h.c.}\]\rbr, \nonumber \\ W_{02}^{\dZ} &=&
{1\over 2}\sum_{P,Q=p,q}\dZ^P\dZ^Q W_{pq},
\label{pvz0}\eea
where the subscript 2 on $W_{02}$ signals the presence of a further
contribution: we will denote by $W_1$ the part of the PV superpotential
that gives PV fields order Planck scale masses and that will be
constructed explicitly in Section \ref{pvmass}.  From (\ref{pvz0}) it
is straightforward to determine that~\cite{pvcan}
\bea R^{\dZ}_{I\m J\n} &=& R_{i\m j\n} + K_{i\m}K_{j\n} + K_{i\m}K_{j\n},\qquad
A^{\dZ}_{IJ} = A_{ij}, \qquad \A_{\dZ}^{IJ} = \A^{ij}, \nonumber \\
R^{\dZ}_{I\m J\n}\(R^{\dZ}\)^{I\;\; J}_{\;k\;\;\ell} &=& 
R_{i\m j\n}R^{i\;\; j}_{\;k\;\;\ell} + 4R_{k\m\ell\n}  
+ 2\(K_{k\m}K_{\ell\n} + K_{\ell\m}K_{k\n}\), \nonumber \\
A^{\dZ}_{IJ}\A_{\dZ}^{IJ} &=& A_{ij}\A^{ij}, \qquad
R^{\dZ}_{I\m J\n}\A_{\dZ}^{IJ} = R_{i\m j\n}\A^{ij} + 2\A_{\m\n},\label{zmes} \eea
assuring~\cite{pvcan} 
cancellation of the divergences arising from the first term
in brackets in (\ref{kr}) and from $\phi_3$ in (\ref{dops}).
The couplings in (\ref{pvz0}) are gauge invariant, provided $\dZ^P$
transforms the same way as $\dZ^p$ under gauge transformations.
Now consider a chiral field redefinition that effects a K\"ahler
transformation 
\bea Z^p &\to& Z'^p, \qquad K\to K' = K(Z',\Z') = 
K + F(Z) + \bF(Z'), \nonumber \\ W &\to& W' = W(Z') = e^{-F}W,
\qquad dZ'^p = M^p_q dZ^q, \qquad M^p_q = {\pp Z'^p\over\pp Z^q}.
\label{genkahl}\eea
The part of (\ref{pvz0}) involving the PV K\"ahler metric $K_{P\Q}
= K_{p\q}$ is invariant under (\ref{genkahl}) provided
\beq \dZ^P\to \dZ'^P = M^p_q \dZ^Q,\label{zkahl}\eeq
but the other terms are not.  Nevertheless the matrix elements
(\ref{zmes}) are covariant, depending only on the covariant scalar
Riemann tensor and on the covariant holomorphic 
scalar derivatives of the K\"ahler covariant operator 
\beq A = e^KW, \qquad A' = e^{\bF}A,\qquad A'_p = e^{\bF}A_p, \qquad
etc. \label{delA}\eeq
This suggests that (\ref{pvz0}) can be made invariant without modifying
(\ref{zmes}).  For example, if we introduce uncharged fields
$\dZ^N$ and add to (\ref{pvz0}) expressions of the form 
\bea \Delta K^{\dZ} &=& \sum_{N}\[\rho_N
\sum_{P=p}K_p\dZ^P\dZ^{N} + {1\over2}\rho'_N
\(\dZ^N\)^2 + {\rm h.c.}\],\nonumber \\ 
\Delta W^{\dZ}_{2} &=& \sum_{N} \[\rho_N \sum_{P=p}
W_p\dZ^P\dZ^N -  {1\over2}\rho'_N\(\dZ^N\)^2W\],
\label{pvz1}\eea
where $\rho,\rho'$ are constants, 
the one-loop corrections to $K_R$ and $\phi_3$ are unchanged:
\beq  A_{PN} = R_{\n PN\m} = 
A_{NM} = R_{\n NM\m} = 0. 
\label{zeros2}\eeq
This trick was used in II to construct an explicitly
invariant (covariant) expression for $K^{\dZ}\; (W^{\dZ}_2)$
in no-scale models with the special properties: $G_{pq}
\propto G_pG_q,$ $G^{p\q}G_pG_{\q} = 3$,
 $\phi^pW_p = 3W$. The general covariance of 
(\ref{zmes}) suggests that this should be possible in general
modular invariant models, {\it e.g.}, including the twisted sector
in orbifold compactification.
In general we have under (\ref{genkahl})
\bea K'_p &=& N^q_p\(K_q + F_q\), \qquad K'_{p\m}
=N^k_pN^{\n}_{\m}K_{k\n},\qquad N = M^{-1},\nonumber \\ K'_{p q} &=&
N^n_pN^m_q\big[K_{n m} + F_{mn} - N^l_k\(K_l + F_l\)\pp_nM^k_m\big]
,\label{delk}\eea
and
\bea W'_p &=& e^{-F}N^q_p\(W_q - F_q W\), \nonumber \\ W'_{pq} &=&
e^{-F} N^k_pN^m_q\big[W_{km} -F_kW_m - F_mW_k - \(F_{km} - F_kF_m\)W
\nonumber \\ & &\qquad - N^l_n\(W_l -
F_l\)\pp_kM^n_m\big].\label{delw}\eea
Restoring invariance/covariance of these expressions in the general
case would be cumbersome at best.  For this reason, we restrict our
attention to effective supergravity theories with three moduli $T^i$,
that have some promising features for a viable phenomenology (see for
example~\cite{iban}).  These theories are classically invariant under
the modular transformation:
\bea T^i &\to& T'^i = {aT^i - ib\over icT^i + d}, 
\qquad ad - bc = 1,\nonumber \\
\Phi^a &\to& \Phi'^a = e^{-q_i^a F^i}\Phi^a = e^{-F^a}\Phi^a, 
\qquad F^i = \ln\(icT^i + d\),
\label{modt} \eea
which effects the K\"ahler transformation (\ref{genkahl}) with
\beq F(Z) = \sum_i F^i(T^i). \label{ft}\eeq
The parameters $q_a^i$ are the modular weights of $\Phi^a$.
In the absence of a twisted sector potential ($W_i=0$), the
classical invariance is $SL(2,\R):\;a,b,c,d,\in \cal R.$ In the
presence of a twisted sector potential, this is broken to
$SL(2,\cZ):\;a,b,c,d,\in \cZ.$  We have 
\bea M^a_b &=& e^{-F^a}\delta^a_b, \qquad M^i_j = \delta^i_je^{-2F^i},
\qquad M^i_b = 0, \qquad M^a_j = - F^a_je^{-F^a}\Phi^a,
\nonumber \\
 N^a_b &=& e^{F^a}\delta^a_b, \qquad N^i_j = \delta^i_je^{2F^i},
\qquad N^i_b = 0, \qquad N^a_j = F^a_je^{2F^j}\Phi^a,
\nonumber \\ F_i &\equiv& F_{t^i} = F^i_{t^i}, 
\qquad F_{ij} = - \delta_{ij}F^2_i. \label{fmod}\eea
Writing the superpotential $W$ as a sum of monomials:
\beq W = \sum_\alpha W^\alpha, \qquad
W^\alpha = c_\alpha\prod_{n = 1}^{N_\alpha}
\Phi^a\prod_i[\eta(iT^i)]^{2\(\sum_nq^q_i - 1\)},\eeq
where $\eta(iT)$ is the Dedekind function:
\beq \eta(iT'^i) = e^{{1\over2}F^i}\eta(iT^i),\label{ded}\eeq
the PV K\"ahler potential and superpotential can be made
modular invariant and covariant, respectively, if we 
introduce three PV chiral superfields $\dZ^N$ that 
transform under (\ref{modt}) as
\beq \dZ'^N = \dZ^N - \dot a F^n_i\dZ^I,\label{xkahl}\eeq
and modify (\ref{pvz0}) follows:
\bea K^{\dZ} &=& K_0^{\dZ} - \dot a^{-1}\[2\sum_{A,N=a,n}\(1 - q^a_n\)\dZ^N\dZ^AK_a +
2\sum_{I,M=i,m}\(1 -\delta_{im}\)K_i\dZ^I\dZ^M\right. \nonumber \\ & &
\qquad  \l + \dot a^{-1}\sum_{N,M=n,m}\[\delta_{nm} \(1 + q_n\) - 1 -
q_{nm}\]\dZ^N\dZ^M\hc\],\nonumber \\ W^{\dZ}_{2} &=& W_{02}^{\dZ} -
2\dot a^{-1}\sum_{\alpha;A,N=a,n}\(1 - q^a_n\)\dZ^N\dZ^A W^\alpha_a -
2\dot a^{-1}\sum_{\alpha;I,M=i,m}\(1 - \delta_{im}\)W^\alpha_i\dZ^I\dZ^M
\nonumber \\ & & \qquad + \dot a^{-2}\sum_{\alpha;N,M=n,m}\[\delta_{nm}\(1 +
q^\alpha_n\) - 1 - q^\alpha_{nm}\] W^\alpha \dZ^N \dZ^M,
\nonumber \\ q_i &=& \sum_aq^a_i\phi^aK_a, \qquad
q_{ij} = \sum_aq^a_iq^a_j\phi^aK_a, \nonumber \\
q^\alpha_i &=& \sum_aq^a_i\phi^a\pp_a\ln W^\alpha, \qquad
q^\alpha_{ij} = \sum_aq^a_iq^a_j\phi^a\pp_a\ln W^\alpha,\label{pvz}\eea
which reduces to the result found in II for the untwisted sector:
$\phi^a\to\phi^{ia},\;q^{ia}_j = \delta^i_j.$
Note that $q^\alpha_i,q^\alpha_{ij}$ are constants, whereas
$q_i,q_{ij}$ are not.  Nevertheless we still get
\beq A_{PN} = R_{\n PN\m} = A_{NM} = R_{\n N M\m} = 0.
\label{zeros3}\eeq
so the one-loop corrections to $K_R$ and $\phi_3$ are unchanged,
provided $K^{P\Q}$ does not change.  We
must also introduce a modular invariant K\"ahler metric for $\dZ^N$;
leaving $K^{P\Q}$ unchanged requires that it be of the form
\beq K(\dZ^N) = \sum_{n=N,m=M}\xi_{n\m}\(\dZ^N +
\sum_{p=P}\chi^n_p\dZ^P\) \(\dbZ^{\M} +
\sum_{q=Q}\bc^{\m}_{\q}\dbZ^{\Q}\),\label{xkahl2}\eeq
where the metric $\xi_{n\m}(Z^i,\Z^{\bj})$ is modular invariant and
\beq \chi'^n_p(Z',\Z') = N^q_p\(\chi^n_q(Z,\Z) + \dot a F_q\).
\label{chitransf}\eeq
As shown in \myapp{orbs}, for a general orbifold metric, this leads to
unwanted contributions to $\L_Q$ and $\Phi^{PV}_{1,2}$ unless
$\xi_{n\m} = \del_{n m}$ and $\chi^n_p$ is holomorphic, {\it e.g.}
$\chi^n_p = 2\dot a\pp_p\ln\eta(iT^n)$. Alternatively, most of the
unwanted terms can be canceled by including several copies of the
above: $\dZ\to \dZ_\alpha$, $\alpha = 1,\ldots, 2n_{\dZ}+1$ with
signatures $\eta^{\dZ}_\alpha$ and parameters $\dot a_\alpha$ such
that 
\beq \sum_\alpha\eta_\alpha^{\dZ} = -1,\qquad
\sum_\alpha\eta_\alpha^{\dZ}\dot a^2_\alpha =
\sum_\alpha\eta_\alpha^{\dZ}\dot a^4_\alpha = 0.\label{zcond}\eeq
Only one of these with negative metric need have the
superpotential and additional K\"ahler potential couplings in
\myref{pvz} required to regulate the K\"ahler potential corrections
and $\phi_3$. Then the only remaining unwanted contribution is from
the curvature terms associated with the metric $\xi_{n\m}$, which
vanish if this metric is flat, or can be canceled by the introduction
of additional gauge-singlet chiral PV multiplets with positive signature
and just the metric $\xi_{n\m}$.  The correct procedure cannot be
determined without further knowledge of the string loop corrections,
and/or higher order terms in the classical twisted sector metric.  We
will consider only the case $\xi_{n\m} = \del_{n m}$, with just a few
simple choices for $\chi^n_p$.

\subsection{Gauge couplings}\label{gauge}
To regulate light gauge field loops, we must introduce at least three
Pauli-Villars chiral supermultiplets $\varphi^a,\tph^a,\hph^a$, with
positive signature, that transform according to the adjoint
representation of the gauge group.  Their contributions cancel the
term proportional to $C_a$ in (\ref{fops}).  To cancel the divergences
associated with the term proportional to $\K^a_a$ in (\ref{kr}),
we need at least one chiral supermultiplet $\hY_P$ that transforms
according to the representation of the gauge group conjugate to that
of $Z^P$, with positive signature, and superpotential 
coupling~\cite{pvcan}
\beq W_2^\phi = 2\varphi^a\hY_P(T_aZ)^p,\label{w2phi} \eeq
which is modular covariant if under (\ref{modt}) $\hY_P$ and $\varphi^a$
transforms as
\beq \hY'_P = N_p^q\hY_Q, \qquad \varphi'^a =
e^{-F}\varphi^a.\label{ytransf}\eeq
Then the K\"ahler potentials
\beq K_0^{\hY} = \sum_{P,Q=p,q}K^{p\q}\hY_P\hbY_{\Q}, \qquad
K^\varphi = e^G\sum_a|\varphi^a|^2 \label{ykahl}\eeq
are gauge and modular invariant, and
\beq A^{\hY\varphi}_{Pa} = 2e^K(T_az)^p, \qquad 
 \A_{\hY\varphi}^{Pa} = 2e^{-G}(T^a\z)^{\m}K_{p\m},
\qquad 2\sum_{Pa}A^{\hY\varphi}_{Pa}\A_{\hY\varphi}^{Pa} = 4\K^a_a
\label{yvarphi0}.\eeq
However $\hY_P$ loops contribute terms proportional to $\Phi_1,\Phi_2$
and $C_a^M\Phi_{YM}$ in (\ref{fops}), and $\l\D^\alpha\Gamma_\alpha\r$
in (\ref{lquad}).  Therefore we must introduce a set $\hY_P^\alpha$ of
such fields with signatures $\eta^{\hY}_\alpha$ such that
$\sum_\alpha\eta^{\hY}_\alpha = 0$, with at least one of these, say
$\hY_P^1,\;\eta^{\hY}_1 = +1$, participating in the coupling
(\ref{w2phi}).

In order to introduce PV masses that are gauge invariant under the
nonanomalous gauge group, we have to introduce equal numbers
of PV fields that transform like $Z^p$ and its conjugate
representation.  This requires that we also introduce additional
charged PV fields $U_\beta^A$ and $U^\beta_A$ that transform according to
the representation $R^a_A$ and its conjugate, respectively, under the
nonanomalous gauge group factor $\G_a$, and/or fields $V^A_\beta$ that
transform according to a (pseudo)real representation that is
traceless and anomaly-free.  Their gauge couplings must satisfy
\beq \sum_{A\in U,V} \eta^A_\beta C^a_A = C^a_M,\label{cascond}\eeq
where 
\beq \Tr_R\(T^aT^b\) = \delta_{ab} C^a_R, \label{casconst}\eeq 
which may imply a constraint on the matter representations of the
gauge group in the light spectrum, as discussed in I.  This constraint
can be trivially satisfied for all $U(1)$'s. It is satisfied in any
supersymmetric extension of the Standard Model (SM). In addition to
two Higgs doublets, its minimal supersymmetric extension, the MSSM,
has $2N_f$ fundamental representations (reps) ${\bf n}$ of each factor
$\G_n=SU(n),\; n=2,3$, where $N_f$ is the number of quark
flavors. Their Casimirs can be mimicked by $N_f$ real PV reps $({\bf
n} + {\bf\n})$. Further extensions necessarily involve real reps of
the SM gauge group, so that the additional states can get SM gauge
invariant masses.  The condition (\ref{cascond}) is also satisfied in
the hidden sectors~\cite{joel} that can accompany the SM-like $Z_N$
orbifolds found in~\cite{iban}.  These hidden sectors also come in
even numbers of reps + antireps, except for one hidden sector that
contain 3 {\bf 16}'s of $SO(10)$ that contribute $C^{SO(10)}_M = 6$,
which can be mimicked by a real PV rep with 6 {\bf 10}'s, and another
hidden sector with 3 $({\bf 5} + {\bf 10})$'s and 6 ${\bf\bar 5}$'s of
$SU(5)$ with $C^{SU(5)}_M = 9$, that can be mimicked by 9 real PV reps
$({\bf 5} + {\bf\bar5})$.  Since the underlying theory is finite when
all degrees of freedom are included, one would expect (\ref{cascond})
to have a solution for general superstring compactifications.

\subsection{Nonrenormalizable couplings}\label{nonren}

In order to cancel the term proportional to $A_p\A^p$ in (\ref{kr}) we
need a PV superpotential coupling of the form~\cite{pvcan} \beq W_2
\ni W_2^{\hY} = \sum_\alpha\[\hat a_\alpha W_p\tZ^P_\alpha
\hY^\alpha_0 + W\tZ^P_\alpha \hY^\alpha_P\],\label{w2y}\eeq
where $\tZ^P,\hY_P$ transform like $Z^p,\Z^{\bar p}$, respectively,
under the gauge group, and ($G_{p\q}\equiv K_{p\q}$, $Z^q\ne S$)
\bea K_\alpha^{\tZ} &=& \sum_{P,Q=p,q}
G_{p\q}{\tZ}_\alpha^P{\tbZ}_\alpha^{\Q}, \nonumber \\ K_\alpha^{\hY}
&=& \sum_{P,J=p,j} G^{p\q}\hY^\alpha_P\hbY^\alpha_{\Q} -
\hat a_\alpha\sum_{P=p}\(\hY^\alpha_P\hbY^\alpha_0G^p + {\rm h.c.}\) +
|\hY^\alpha_0|^2\[1 + \hat a_\alpha^2G^pG_p\], \nonumber \\
\eta_\alpha^{\hY} &=& \eta_\alpha^{\tZ},\qquad \hN =
\sum_\alpha\eta^{\hY}_\alpha = 0,\qquad G^p= G^{p\q}G_{\q}, \qquad
G^{p\q}G_{\q k} = \delta^p_k.\label{kahly}\eea
The superpotential (\ref{w2y}) is covariant and the K\"ahler
potential (\ref{kahly}) are invariant under modular transformations
provided that under (\ref{genkahl})
\beq \hY'^\alpha_P = N_p^q\(\hY^\alpha_Q + a_\alpha F_q\hY_0\), \qquad
\hY'_0 = \hY_0,\qquad \tZ'^Q_\alpha = M_p^q\tZ^P_\alpha.\label{pvt0}\eeq
The same form of the K\"ahler potential is in fact needed for the PV
fields $\hY_P$ that couple to $\varphi^a$ in (\ref{w2phi}) in order to
cancel~\cite{pvcan,pvdil} the term proportional to $\D$ in (\ref{kr}).
We therefore generalize these to a subset of the $\hY_\alpha$'s with the
same signatures as a subset of $\varphi^a_\alpha$, with the full set
satisfying $\sum_\alpha\eta_\alpha^\varphi =1$, and take for the full
superpotential $W_2$
\bea W_2 &=& W_2^{\dZ} + W_2^{\hY} + 2\sum_\alpha g_\alpha
\varphi^a_{1 + \alpha}\hY^\alpha_P(T_aZ)^p + c_\alpha \phi^S_\alpha
\phi^\alpha_SW,\label{w2}\eea
where $\phi^S,\phi_S$ are gauge singlets needed to cancel dilaton loop
contributions, with a K\"ahler potential
\beq K^S = \sum_{\alpha}\[e^{-2k}|\phi_S^\alpha|^2 +
2|\phi_0^\alpha|^2 - e^{-k}\(\bph_{\S}^\alpha\phi_0^\alpha + {\rm
h.c.}\) + e^{2k}|\phi^S_\alpha|^2\], \label{kahls}\eeq
that involves an additional chiral superfield $\phi_0$, analogous to
$\hY_0$; this field is also needed [see \myref{kin1}] to regulate
dilaton couplings.  The superpotential (\ref{w2}) is modular covariant
and the K\"ahler potential (\ref{kahls}) is modular invariant if the
superfields $\phi^r = \phi^S,\phi_S,\phi_0$ are modular invariant.
Note that because $F(Z)$ is holomorphic and gauge invariant it
satisfies
\beq (T^aZ)^pF_p = 0,\eeq 
so (\ref{w2phi}) remains modular covariant with the modification
(\ref{kahly}) of the K\"ahler potential for $\hY_I$.  To cancel the
remaining divergences arising from nonrenormalizable couplings we
introduce gauge singlets $\phi^\gamma$, as well as $U(1)$ vector
supermultiplets $V_\gamma = V^0_\gamma, V^s_\gamma$, with signatures
$\eta^0_\gamma,\eta^s_\gamma,$ respectively, that form massive vector
supermultiplets with chiral multiplets $\Phi^{0,s}_\gamma =
e^{\theta^{0,s}_\gamma}$ of the same signature and $U(1)_\beta$
charges $q_\gamma\delta_{\gamma\beta}$.  We also need additional
chiral multiplets $\tph^\alpha,\hph^\alpha$ in the adjoint
representation of the low energy gauge group; one of these sets,
$\hph^\alpha$, participates in the matrix-valued gauge kinetic
function that couples the PV superfield strengths $W^0_\gamma,
W^s_\gamma$ and the light gauge fields $W^a$ to one another in the
Yang-Mills kinetic term:
\bea f^{ab} &=& \delta^{ab}\(S + \sum_\alpha h^S_\alpha\phi^S_\alpha
\phi_0^\alpha\), \qquad f_s^{a\gamma} = 0, \nonumber \\
f^0_{\gamma\beta} &=& \delta_{\gamma\beta},\qquad f^s_{\gamma\beta} =
\delta_{\gamma\beta}S, \qquad f_0^{a\gamma} = \sum_\beta
e^{\gamma\beta}\hph^a_\beta.\label{kin1} \eea
These couplings are modular invariant provided $\theta_\gamma$ and
$\hph^a_\alpha$ are modular invariant.  The matrices
$d_{\alpha\beta},e_{\alpha\beta}$, are nonvanishing only when they
couple fields of the same signature.  Including the fields $U,V$, as
well as additional chiral supermultiplets $\phi_\gamma$ needed to
regulate gravity loops~\cite{pv1}, the full K\"ahler potential takes
the form
\bea K_{PV} &=& \sum_\gamma\[\half f_{\phi_\gamma}\phi^\gamma\bph_\gamma +
{1\over2} \nu_\gamma(\theta_\gamma + \bth_\gamma)^2 +
\sum_A\(f_{U_\gamma^A}|U_\gamma^A|^2 + f_{U^\gamma_A}|U^\gamma_A|^2 
+ f_{V^A_\gamma}|V^A_\gamma|^2\)\]
\nonumber \\ & & + \sum_{\alpha,a}\(e^G\varphi_\alpha^a\bv_a^\alpha +
e^k\hph_\alpha^a\hbp_a^\alpha + \tph_\alpha^a\tbp_a^\alpha \) +
\sum_{\alpha}\(K_\alpha^{\tZ} + K_\alpha^{\hY}\) + K^{\dZ} + K^S.
\label{kahl}\eea
As shown in I and II, the ultraviolet divergences from light loops are
canceled if the PV coupling constants satisfy\footnote{There are
extraneous factors of $e^K$ and $W$ in the expression for $\bar
A^{IJ}_{Z_\alpha,Y_\alpha}$ in (2.36) of I.}
\bea \sum_\alpha\eta_\alpha^{\hY}\hat a_\alpha^2 &=& -2, \qquad
\sum_\alpha\eta_\alpha^{\hY}\hat a_\alpha^4= +2,\nonumber \\
\sum_\alpha\eta^{\phi^S}_\alpha c^2_\alpha &=& 5, \qquad
\sum_\alpha\eta_\alpha^{\hY} g_\alpha^2\hat a_\alpha^2 = -1, \qquad
\sum_\alpha\eta_\alpha^{\hY} g_\alpha^2 = 1, \nonumber \\
{1\over2}\sum_{\alpha,\beta}\eta^{\hph}_\alpha e_{\alpha\beta}^2 &=& -
4 = {3\over4}\sum_{\alpha\beta\gamma\delta}\eta^{\hph}_\gamma
e_\alpha^\beta e_\beta^\gamma e_\gamma^\delta e^\alpha_\delta, \qquad
\sum_\alpha\eta^{\phi^S}_\alpha h^S_\alpha c_\alpha = 1,\nonumber \\
\sum_\alpha\eta^{\phi^S}_\alpha(h^S_\alpha)^2 &=& 2, \qquad
\sum_\gamma\eta^s_\gamma = - N_G, \label{cond0} \eea
and the PV signatures satisfy
\bea \sum_\alpha\eta^{\varphi^a}_\alpha &=&
\sum_\alpha\eta^{\hph^a}_\alpha = \sum_\alpha\eta^{\tph^a}_\alpha = 1,
\qquad \eta^{\varphi^a}_{1 + \alpha} = \eta^{\hY}_\alpha, \qquad
\eta^{\varphi^a}_1
= +1, \qquad \eta^{U^A}_\alpha = \eta^{U_A}_\alpha, \nonumber \\
\sum\eta^{\dZ}_\alpha &=& \sum_\alpha\eta^{\phi^S}_\alpha = - 1, \qquad
\sum_\alpha\eta^{\tZ}_\alpha = 0, \qquad \eta^{\tZ}_\alpha =
\eta^{\hY}_\alpha, \qquad \eta^{\phi^S}_\alpha = \eta^{\phi_S}_\alpha =
\eta^{\phi_0}_\alpha,\nonumber \\ \sum_\gamma\eta^\theta_\gamma &=&
\sum_\gamma\eta^s_\gamma + \sum_\gamma\eta^0_\gamma = - 12 - N_G =
N'_G,\qquad \sum_P\eta^C = -29 -N = N',\label{sigs}\eea
where $C$ is any PV chiral superfield and $N(N_G)$ is the number of
chiral (gauge) superfields in the light sector.  Cancellation of
quadratic divergences requires that the pre-factors $f_C$ in
\myref{kahl} satisfy:
\beq \sum_C\eta_C\pp_{i}\ln f_C = -10K_i,\qquad
\sum_C\eta_C\pp_{\ibar}\ln f_C = -10K_{\ibar},\label{cond1}\eeq
and cancellation of logarithmic divergences requires
\beq \sum_C\eta_C\pp_i\ln f_C\pp_{\bj}\ln f_C = -4K_i K_{\bj} + 2k_i k_{\bj}.
\label{condquad}\eeq
where in \myref{cond1} $\phi^C$ is any PV chiral superfield except
$\dZ,\tZ,\hY,\phi^S,\phi_{S,0}$; for example 
\beq f_\varphi = e^{K - k},\qquad f_{\hph} = e^k, \qquad f_{\tph} = 1
.\label{phiparams}\eeq
%
%
%
%

The K\"ahler potentials for $\varphi_\alpha^a$ and $\hph_\alpha^a$ are
determined by their couplings so as to cancel light field divergences,
while the choice for $\tph_\alpha^a$ assures the K\"ahler anomaly
matching condition for the gauge loop contribution to the term
quadratic in the Yang-Mills field strength, which requires~\cite{tom}
that, averaged over the adjoint PV fields, $\langle\ln
f\rangle_\varphi = K/3$.  The parameters $\nu_\gamma$ in \myref{kahl}
determine the squared PV masses of the PV vector supermultiplets and
the corresponding eaten chiral supermultiplets $\theta_\gamma$; these
masses play the role of effective (squared) cut-offs. The PV masses
$\mu_\alpha$ of the remaining PV chiral multiplets will be introduced
through superpotential terms in section \ref{pvmass}.

%% file: Section4.tex
\section{Anomaly Cancellation}\label{ancan}
\hspace{0.8cm}\setcounter{equation}{0} 

In the context of orbifold compactification of the heterotic string
the known mechanisms for cancellation of the effective field theory
anomalies are the four-dimensional GS mechanism and string loop
threshold corrections.  The latter can be parametrized as
moduli-dependent PV masses.  In this section we outline the needed
generalization of the GS mechanism, show that it restricts the form of
PV mass terms, and construct an explicit PV superpotential that
satisfies these restrictions.

\subsection{Strategies for anomaly cancellation}\label{strat}
The regularized theory would be invariant under the classical
symmetries if it were possible to introduce PV mass terms that respect
these symmetries and also cancel all the ultraviolet divergences.
That this is not possible can easily be seen by looking at the
quadratic divergences.  For example, in (\ref{lquad})
\bea \l\Dc\Gaa\r &=& 2x^{-1}\D_aD_p(T^az)^p -
2R_{p\m}\(e^{-K}\A^pA^{\m} + \D_\mu z^p\D^\mu\z^{\m}\),\nonumber \\
D_p(T^az)^p &=& \Tr T^a + \Gamma^p_{pq}(T^az)^q .\eea
If \ux\, is anomalous, Tr$T^X\ne 0$, one cannot regulate the quadratic
divergences without introducing at least one mass term for a pair of
PV chiral superfields $U,U'$ with $U(1)_X$ charges $q_X + q'_X\ne 0$.
Similarly, if the low energy theory possesses a classical invariance
under a nonlinear symmetry that effects a K\"ahler transformation
(\ref{genkahl}), a mass term $W_U = \mu UU'$ with constant $\mu$ is
covariant if $K_{U\bar U}K_{U'\bar U'} = e^K$. In this case
\beq \l\Dc\Gaa^U\r + \l\Dc\Gaa^{U'}\r = 2\l\Dc\Xa\r + 2x^{-1}\D_X \(q_X
+ q'_X\).\label{uumass}\eeq
The first term on the RHS of (\ref{uumass}) cancels the $U,U'$
counterpart of the term proportional to $N\l\Dc\Xa\r$ in
(\ref{lquad}).  Thus, as discussed in I, quadratic divergences
associated with scalar curvature cannot be regulated by PV fields with
invariant masses.  For the orbifold-derived supergravity models
considered here, only a discrete group of modular transformations is
expected to survive at the quantum level.  As will be seen below it is
always possible to construct PV masses that are invariant under this
discrete group by including holomorphic functions of the moduli that
have well-defined modular weights: $w(T') =
e^{\sum_iF^i\omega_i}w(T)$.  The appearance of these factors in the
effective one-loop Lagrangian would be interpreted as string-loop
threshold effects.  However it is known from string-loop
calculations~\cite{ant,dixon} in orbifold compactifications that these
effects do not fully cancel the anomaly.  Indeed there are no
threshold corrections to the Yang-Mills Lagrangian in $Z_3$ and $Z_7$
orbifolds, where the anomaly is completely canceled by the four
dimensional analogue~\cite{ovrut,anomalies} of the Green-Schwarz term.
Moreover, the \ux\, anomalies can be canceled only by such a mechanism
\cite{dine}.

The GS mechanism is most easily displayed in the linear multiplet
formalism~\cite{linear} for the dilaton, where the GS term takes the
form
\beq \L_{GS} = \superint EL\[bg(Z,\Z) - \half\dx V_X\]
\equiv \superint ELV_{GS},\label{lgs}\eeq
where $b$ and $\dx$ are constants, $\vx$ is the \ux\, vector 
superfield, and $g$ is a real superfield which, under a modular
transformation (\ref{genkahl}), transforms as
\beq g(Z',\Z') = g(Z,\Z) + F(Z) + \bF(\Z).\label{ggs}\eeq
The real superfield $L$ satisfies the modified linearity conditions
\beq
\chiproj L = - \WaWa, \qquad
\bchiproj L = - \WbWb ,\label{modlin}\eeq
where $\Wa^a$ is the gauge superfield strength.  Under (\ref{ggs}) and
a gauge transformation:
\beq V_X \to V'_X = V_X + \(\Lambda + \bL\),
\label{gaugetr}\eeq
where $\Lambda$ is a chiral superfield, (\ref{lgs}) transforms as,
using (\ref{modlin}), \bea \L_{GS}&\to&\L'_{GS} = \L_{GS} +
\Delta\L_{GS}, \nonumber \\ \Delta\L_{GS} &=& \superint EL\(bF
-{\dx\over2}\Lambda\) + {\rm h.c} = -\superint{E\over8 R}\(\bD^2 -
8R\)L\(bF -{\dx\over2}\Lambda\) + {\rm h.c}\nonumber \\ &=&
\superint{E\over8 R}\WaWa\(bF -{\dx\over2}\Lambda\) + {\rm h.c}, 
\label{deltagv}\eea
which has the same form as the quantum anomaly from renormalizable
couplings of particles charged under the gauge group $\G_a$ that also
carry modular weights and/or \ux\, charge.  The constants $b,\dx$ can
be chosen to cancel the anomalous term for any one $\G_a$. In orbifold
compactifications of the heterotic string, the couplings satisfy
constraints that allow universal \ux\, anomaly cancellation. Modular
anomaly cancellation is also universal in orbifolds with no $N=2$
twisted sector.  Otherwise there are moduli-dependent threshold
corrections that contribute to the anomaly cancellation.  As discussed
in I, these can be included in the PV regulator masses.

The GS mechanism can be generalized to cancel anomalous coefficients
of higher dimension operators by generalizing the modified linear
condition:
\beq \chiproj L = - \Phi, \qquad
\bchiproj L = - \bPh ,\label{genlin}\eeq
where $\Phi$ is a chiral superfield with chiral K\"ahler weight
$w(\Phi) = 2$, provided higher dimension operators are also
included in the tree Lagrangian:
\bea \L_{tree}(L) &=& -\superint E\[3 - 2Ls(L)\]
 \nonumber \\  &=& \superint {E\over16R}
\chiproj\[3 - 2Ls(L)\] + {\rm h.c.} \nonumber \\ &=& \superint{E\over8R}
s(L)\Phi + {\rm h.c.} + \cdots, \label{ltreel}\eea
where the ellipsis represents the kinetic terms for the various
components of $L$ and the supergravity multiplet. The K\"ahler
potential \myref{sdil} in this formalism is $K = k(L) + G(Z,\Z)$, and
the relation
\beq k'(L) + 2Ls'(L) = 0\label{einst} \eeq
between the real functions $k(L)$ and $s(L)$ assures
a canonical Einstein term.  It follows from (\ref{genlin})
that $\Phi$ is the chiral projection of a real field $\Omega$
with K\"ahler weight $w_K(\Omega) = 0$ and Weyl weight 
$w_W(\Omega) = w_W(L) = - w_W(E) = 2$, and that it satisfies
a generalized Bianchi identity:
\beq \(\bD^2 - 24R\)\Phi - \(\D^2 - 24\bR\)\bPh
= {\rm total \; derivative}.\label{bianchi}\eeq
When $\Phi = \WaWa$ as in (\ref{modlin}), the first term on the RHS
of (\ref{ltreel}) reduces to the Yang-Mills kinetic term, and $\Omega
= \Omega_{Y M}$, the Chern-Simons (CS) superfield.  To obtain the
Lagrangian for the chiral multiplet formulation we make a superfield
duality transformation by writing~\cite{linear}
\beq \L = -\superint E\[3 - 2Ls(L) + \(L + \Omega\)\(S + \S\)\],\eeq
where $L$ is
unconstrained and $S$ is chiral.  Writing $S$ as the chiral projection
of an unconstrained complex field $\Sigma$: $S = \chiproj\Sigma$, the
equations of motions for $\Sigma$ and $\Sigma^{\dag}$ give the
generalized linearity constraint (\ref{genlin}).  If instead we solve
the equations of motion for $L$, we obtain~\cite{linear} $L$ as a
function of $S + \S$ such that
\beq s(L) = (S + \S)/2, \label{dual}\eeq
and we recover the standard chiral superfield formulation~\cite{crem}
of supergravity, with a canonical Einstein term provided (\ref{einst})
holds, except that the standard Yang-Mills term is generalized to
\beq \L_{Y M}\to \superint{E\over8R}S\Phi + {\rm
h.c.},\label{sgs}\eeq 
provided that
\beq {\delta\over\delta L}\(E\Omega\)
= 0.\label{opcond} \eeq
When the GS term \myref{lgs} is included, \myref{dual} is modified
to
\beq s(L) = (S + \S - V_{GS})/2. \label{gsdual}\eeq
Since $L$ is invariant under modular transformations \myref{ggs} and
gauge transformations \myref{gaugetr}, the chiral superfield $S$ is
not invariant:
\beq S \to S' = S + bF(Z) - \dx\Lambda/2.\label{strsansf}\eeq
Then from either \myref{lgs} in the linear formulation or 
\myref{sgs} in the chiral formulation, we get a variation in the
Lagrangian
\beq \Delta\L = \superint {E\over8R}\Phi\(bF -
\dx\Lambda/2\)\hc.\label{delgv2}\eeq
In \mysec{pvreg} we specified the PV Lagrangian in terms of modular
and \ux\, invariant K\"ahler potential $K_{PV}$ and gauge kinetic
functions \myref{kin1}.  As a consequence the massive Abelian PV
vector fields necessarily have modular invariant masses that do not
contribute to the anomaly.  We further imposed \ux\, invariance and
modular covariance on the part $W_2$ of the PV superpotential that
cancels divergences from light chiral supermultiplet loops.  The
chiral and conformal anomalies of the low energy quantum field theory
arise from linear and logarithmic divergences, respectively. If we
assume that they are canceled when the theory is properly regulated,
the only noninvariance in the regulated theory appears in the part
$W_1$ of the superpotential that gives a large supersymmetric mass
matrix $m$ to PV chiral supermultiplets.  In this case the one loop
action can be written as
\beq \L_1 = \L_{inv} + \L_{ni}, \qquad \L_{ni} = {i\over2}\STr\ln\[D^2
+ H(m)\]_{ni} + T_{ni}(m),\label{even}\eeq
where $\L_{inv}$ is modular invariant, $T$ is the helicity-odd fermion
contribution, and $\L_{ni}$ arises only from chiral supermultiplet
loop contributions.  As shown in I, under a transformation on the PV
fields that leaves the tree Lagrangian, the PV K\"ahler potential and
the PV gauge couplings invariant, with $W_2$ covariant:
\bea \Phi' &=& g\Phi, \qquad m'(\Phi) = m(\Phi') \nonumber\\ 
\L'_1 &=& \L_{inv} + \L_{ni}(\tm), \qquad \tm = 
g^{-1}m'g,\label{lprime}\eea
because all the operators in the determinants are covariant except the
matrix $m_{ni}$ for the chiral multiplets whose PV
masses arise from the noncovariant part of $W_1$.  The residual quantum
anomaly can be canceled by the GS term provided 

\bea \L_{ni}(m) &=& \superint E\STr\[\Omega\ln(m\m)\], \qquad \tm_{ni} =
e^{\sum_iQ_i F^i + Q_X\Lambda_X}m_{ni}, \label{lni}\eea
with, for example
\bea \STr\(\Omega_{\rm Y M}Q_i\) &=& - b\Omega_{\rm Y M},\quad
\quad  \STr\(\Omega_{\rm Y M}Q_X\) = \dx\Omega_{\rm Y M}/2,\eea
where $\Omega_{\rm Y M}$ is the (matrix valued) Yang-Mills Chern-Simons form:
\beq \chiproj\Omega_{\rm Y M} = \WaWa.\label{OYM}\eeq
As mentioned in the introduction, the linear divergences are not completely
canceled in the PV regulated theory.  Cancellation of {\it on-shell}
quadratic divergences imposed the two constraints given in (2.20) of I,
which can be recast in the form 
\beq N + N' - N_G - N'_G = 3 + 2\alpha, \qquad N + N' - 3N_G - 3N'_G = 7,
\qquad \alpha = \sum_C\eta_C\ln f_C/K.\label{2quadconds} \eeq
The first of these assures the cancellation of the linear divergence
associated with the spin connection; as a consequence the associated
anomaly arises only from PV masses, and, in the absence of
threshold corrections, one gets a contribution [see \myref{newdefs}]
\bea {16\pi^2\over\sqrt{g}}\Del\L^\chi_{\rm spin} &=&
- {1\over48}r\cdot\tr\sum_n\(N' - N'_G  - 2\alpha
 + 2\sum_p q^p_n\)\im F^n\eee  {1\over48}r\cdot\tr\sum_n\(N
  - N_G - 3 - 2\sum_p q^p_n\)\im F^n.\label{spinan}\eea
  However the linear divergence associated with the affine connection
  in gravitino loops is not canceled, and we get an additional
  contribution to the chiral anomaly:
\beq {16\pi^2\over\sqrt{g}}\Del\L^\chi_{\rm affine} = 
{24\over48}r\cdot\tr\im F^n.\label{affan}\eeq
The conformal anomaly counterpart of \myref{spinan}, namely
\bea {16\pi^2\over\sqrt{g}}\Del\L^c_{\rm spin} &=& 
\sum_n\(N - N_G - 3 - 2\sum_p q^p_n\)L_{G B}\re F^n,\label{cspinan}\\ 
L_{G B} &=& {1\over48}\(r_{\mu\nu\rho\sigma}r^{\mu\nu\rho\sigma}
- 4r_{\mu\nu}r^{\mu\nu} + r^2\),\label{defLGB}\eea
automatically combines with \myref{spinan} to form the supersymmetric
combination contained in the operator $\Phi_W$ in \myref{fops}. This
is because Pauli-Villars regularization manifestly preserves
supersymmetry, in contrast to a straight momentum cut-off procedure,
which does not.  

The second constraint in \myref{2quadconds} determines the PV
mass-independent coefficient of the Gauss-Bonnet term in the
regularized one-loop effective Lagrangian:
\beq {16\pi^2\over\sqrt{g}}\L^c_{G B} = L_{GB}(N + N' - 3N_G - 3N'_G +
41) \ln\Lambda = 48L_{G B}\ln\Lambda.\label{LGB}\eeq
Since $L_{G B}$ is a total derivative, this drops out of the effective action
if $\Lambda$ is constant. If instead, $\Lambda = \mu_0e^{\alpha_\Lambda K}$,
there is a finite, anomalous contribution to the effective action  with
\beq {16\pi^2\over\sqrt{g}}\Del\L^c_{G B} = 96\alpha_\Lambda L_{GB}\re
F.\label{LGBanon}\eeq
Supersymmetry  requires that we take $\alpha_\Lambda = {1\over4}$, in which case
the residual contribution to the anomaly in K\"ahler superspace is 
given by
\beq \Del\L^{\rm res}_{G B} =  {1\over8\pi^2}\superint E F^n(T)
\Omega_{G B}\hc, 
\qquad\Omega_{G B} = - 8\Omega_W - {4\over3}\Omega_X -
3\phi_\chi,\label{resan}\eeq
which is of the required form, \myref{opcond}, with $\Omega_W$ 
the Chern-Simons form for the curvature superfield
strength, and $\Omega_X$ the Chern-Simons
superfield for the chiral superfield $\Phi_X$ introduced in \myref{fops}:
\beq \chiproj\Omega_W = W^{\alpha\beta\gamma}W_{\alpha\beta\gamma} = 
\Phi_W,\qquad \chiproj\Omega_X = \Phi_X,\label{OWX}\eeq   
and $\phi_\chi$ is defined in \myref{dops}.  Including the PV
contribution, the total anomaly involving the curvature strength
becomes
\beq \Del\L_{G B} = {1\over8\pi^2}{1\over24}
\sum_n\(N - N_G + 21 - 2\sum_p q^p_n\)\superint E F^n(T)\Omega_{G B}\hc,
\label{LW}\eeq
The off-diagonal gravitino-gaugino connection leads to a contribution
to the chiral anomaly (see \myapp{chigauge})
\bea {16\pi^2\over\sqrt{g}}\Del\L^\chi_{\rm off} &=&
- 2\im F\D^\mu\(xF_a^{\rho\nu}\D_\rho\tF^a_{\mu\nu}\),
\label{offan}\eea
which combines with the conformal anomaly associated with total
divergences from mixed gauge-gravity loop contributions:
\bea {16\pi^2\over\sqrt{g}}\Del\L^c_{\rm mixed} &=&
- 8\alpha_\Lambda\re F\D^\mu\(xF_a^{\rho\nu}\D_\rho F^a_{\mu\nu}
- xF^a_{\mu\nu}\D_\rho F_a^{\rho\nu} - 2\D_a\D^\nu F^a_{\nu\mu}\) 
+ \cdots,\label{mixan}\eea
to give an overall contribution 
\beq \L_{\rm M x} = {1\over8\pi^2}\superint E F\Omega_{\rm M x}\hc,
\label{LMx}\eeq
where 
\bea \Omega_{\rm M x} &=& - {1\over4}\[\D^{\alpha},\D^{\dot\beta}\]\[(S +
\S)\Wa\Wd\] - {1\over16}\(\D^2\[(S + \S)\Wc\Wa\]\hc\) + \cdots,
\label{OMx}\eea
and the ellipses in \myref{mixan} and \myref{OMx} represent terms with
derivatives of the dilaton superfield\footnote{Determination of the
  explicit expression for these terms requires restoring all the total
  derivatives dropped in~\cite{us2}; we have done this only for the
  term appearing explicitly in \myref{mixan}.} $S$ (as well as
fermionic terms).  As discribed in \myapp{chigauge}, this contribution
to the anomaly is cancelled by mixed PV gauge-chiral matter loops for
a specific choice of the $V_0$-$\theta_0/\varphi^a$-$\hph^a$
squared PV mass ratio. We will adopt this choice.

In addition there are total derivatives with logarithmically divergent
coefficients that generate a D-term conformal anomaly with no chiral
counterpart.  Specifially there is a contribution from chiral loops
and mixed chiral-gravity loops
\beq {16\pi^2\over\sqrt{g}}\Del\L^c_{\rm\chi G} = - 8\alpha_\Lambda\re
F K_{i\m}\D^\mu\[\D_\mu\z^{\m}\(\A^i A e^{-K} + D_\nu\D^\nu z^i\) -
D_\nu\D_\mu\z^{\m}\D^\nu z^i\hc\].\label{chiganom}\eeq
Given that it is not trivial to identify all the total derivatives that
were dropped in~\cite{us} and~\cite{us2}, we cannot guarantee that
there are no other such D-term anomalies, so for present purposes we
will subsume these into an operator that we will call $\Omega'_D$.

In order to preserve the correct form of the anomaly, we require
that PV chiral supermultiplet mass terms have well-defined modular
weights.  Those supermultiplets that regulate dilaton loops, and/or
that contribute to \myref{kin1}, must have invariant masses.
Operators that do not satisfy (\ref{opcond}) include those that
contribute to the renormalization of the K\"ahler potential.  These
get contributions from all fields with couplings in $W_2$.  We
therefore require that these PV fields also have modular and \ux\,
invariant masses; a subset of these also regulate $\phi_3$ in
(\ref{dops}).  The requirement that these PV fields have invariant
masses (covariant PV superpotential terms) has implications for soft
supersymmetry breaking scalar masses.

Other operators $\Omega_n$ that satisfy (\ref{opcond}) include those,
like $\Omega_{Y M}$, $\Omega_W$ and $\Omega_X$, whose chiral
projections $\Phi_n$ are bilinear in generalized chiral superfield
strengths. In the present formalism $\Omega_X$, defined in \myref{OWX}
and \myref{fops}, is generalized to the operator:
\beq \chiproj\Omega_{X^m} = X^\alpha_m X_\alpha^m, \qquad
X_\alpha^m = {3\over8}\chiproj\Da\ln\cM^2 + \Xa,\label{defxm}\eeq
where the real superfield $\cM^2(Z,\Z,\vx) = \cM\overline{\cM}$ is the
squared mass of a pair of PV fields.
$\Omega_X$ and $\Omega_{X^m}$ will be explicitly constructed in
\myapp{csfields} following the construction of $\Omega_{Y M}$
in~\cite{gg}.   

In addition to the above ``F-term'' operators, which have no bosonic terms in
their lowest components, there are ``D-term'' operators:
\bea \Omega_D = \cM^2\bchiproj\cM^{-2}R^m\hc,
\qquad-\Omega_G = G_m^\mu G^m_\mu = \half G_m^{\alpha\dot\beta}G^m_{\alpha\dot\beta},
\qquad \Omega_R = R^m\bR^m,\label{omR}\eea
where 
\bea R^m = -{1\over8}\cM^{-2}\chiproj\cM^2 = -{1\over8}\cM^{-2}\bD^2\cM^2
+ R,
\qquad G^m_{\alpha\dot\beta} = \half\cM[\Da,\Dd]\cM^{-1}
+ G_{\alpha\dot\beta},\label{Gm}\eea
are generalizations of the supergravity auxiliary fields
$R,\;G_{\alpha\dot\beta}$.  Although the operators \myref{omR} satisfy
the constraint \myref{opcond}, their chiral projections cannot be
entirely included in $\Phi$, {i.e}., the right hand side of the
generalized linearity condition \myref{genlin}, because they contain
terms that are not invariant under the anomalous transformations (or
do not transform into a linear multiplet).  Therefore full anomaly
cancellation appears to require that we add additional counterterms.

To determine which additional terms of the form \myref{sgs} are
actually present in the linearity condition \myref{genlin} requires
evaluating, in superstring derived supergravity, the higher dimension
terms from the underlying string theory.  However some insight might
be gained by considering the zero-slope limit that gives a
supergravity theory in 10 dimensions.  The components of the linear
multiplet defined in (\ref{modlin}) include a three-form
$h_{\lambda\mu\nu}$ that is a linear combination of the curl of a
two-form $b_{\mu\nu}$ and the Yang-Mills CS form $\omega^{\rm Y
  M}_{\lambda\mu\nu}$. Similarly, the three-form $H_{LMN}$ of 10-d
supergravity includes the curl of a two-form $B_{MN}$ and the
Yang-Mills CS form $\omega^{\rm Y M}_{LMN}$. However in order to
cancel all the anomalies of 10-d supergravity, the 3-form $H$ must be
modified to include a term proportional to the Lorentz CS form as
well.  This must also then be the case in the 4-d effective theory,
and supersymmetry then implies that in (\ref{genlin}) and
\beq \Phi = c_{\rm Y M}\Phi_{\rm Y M} + c_{\rm G B}\Phi_{\rm G B} +
\cdots, \qquad \Omega = c_{\rm Y M}\Omega_{\rm Y M} + c_{\rm G B}
\Omega_{\rm G B} + \cdots.\label{omega}\eeq
Then (\ref{ltreel}) contains a term proportional to the GB term which
must also have a 10-d counterpart.  When compactified to four
dimensions the 10-d Riemann tensor includes, in addition to the 4-d
Riemann tensor, derivatives of the 10-d dilaton $\Phi$ and the
breathing mode(s) ({\it e.g.}, $\sigma = \ln\det g_{mn}$, $m,n =
5\ldots 9$). However, in the class of models considered here, with
only three diagonal moduli,
\beq g^i_{m n} = \del_{m n}e^{\sigma^i} = \del_{m n}\re(t^i s^{1\over3}),\eeq
we find for the Chern-Simons 3-forms
\beq (\omega^{\rm Y M}_{\mu\nu\rho})_{\rm 10d} = (\omega^{\rm Y
  M}_{\mu\nu\rho})_{\rm 4d}, \qquad (\omega^{\rm
  Lor}_{\mu\nu\rho})_{\rm 10d} =(\omega^{\rm Lor}_{\mu\nu\rho})_{\rm
  4d},
\label{3form}\eeq
and there are no nonvanishing elements of $\omega_{M N R}$ involving
scalar derivatives.  The 10-d expression for the Green-Schwarz counter
term is~\cite{gs}
\bea L_c^{(10)} &=& \int B\wedge X_8 - \({2\over3} +
\alpha\)\int\(\omega^{\rm Y M} - \omega^{\rm Lor}\)\wedge X_7 \eee \int\[H +
\({1\over3} - \alpha\) \(\omega^{\rm Y M} - \omega^{\rm Lor}\)\]\wedge
X_7,\label{10dGS}\eea
where $\alpha$ is an arbitrary parameter, the 8-form curl $X_8$ of the
7-form $X_7$, $X_8 = d X_7$, is constructed from the Yang-Mills and
curvature field strengths, and the 3-form $H$ is
\beq H = dB + \omega^{\rm Y M} -  \omega^{\rm Lor}.\eeq
In compactifications of the heterotic string the 10-d field strengths
\beq \langle  R_{m n}\rangle = \langle F_{m n}\rangle \eeq
are nonvanishing in the 4-d vacuum, but 
\beq \left\langle\(\omega^{\rm Y M} - \omega^{\rm Lor}\)\right\rangle
= 0,\eeq
and no new couplings appear to be generated by direct truncation of
\myref{10dGS} to four dimensions.  However this expression, constructed
to cancel the 10-d chiral anomaly, should have a 10-d conformal
anomaly counterpart, and it is possible that its 4-d remnant could be
at the origin of some of the D-term contributions found here.

\subsection{PV masses}\label{pvmass}

Given a set\footnote{These may be, e.g., any of $\dZ,\hZ,\tZ$; see
\myapp{notation}.} of PV chiral superfields $Z^P$ with K\"ahler
metric $K_{P\Q}$, a mass term with a well-defined modular weight can
be constructed by coupling them to a set $Y_P$ with with inverse
K\"ahler metric multiplied by a light field-dependent function that is
modular invariant up to factors $e^{q_ig^i}$:
\beq K^{Y} = f_{Y}(Z^p,S,\vx)Y_P\Y_{\Q}K^{P\Q},\eeq
where $V_X$ is the vector supermultiplet\footnote{In Yang-Mills
superspace~\cite{bggm} only the gauge covariant superfield strengths
$\Wa^a$ are introduced and the chiral superfields $\Phi^b$ are
covariantly chiral with a nonholomorphic gauge transformation
superfield operator.  In the presence of an anomalous gauge symmetry
$\G_X$, it is necessary to introduce the corresponding gauge potential
superfield $V_X$.  For \ux\, we use ``partial'' \ux\, superspace; see
\myapp{anmass}.} of the anomalous \ux.  The K\"ahler potentials are
modular invariant if under (\ref{modt}) for an appropriate choice of
the matrix $M$
\beq Z'^P = M^P_QZ^Q,\qquad Y'_Q = e^{-q^{Y}_iF^i}N^P_QY_P, \qquad N =
M^{-1},\qquad f' = e^{q^{Y}_i\(F^i + \bF^i\)}f, \eeq
and the superpotential 
\beq W_1^Z = \mu_Z(T^j)Z^PY_P, \quad W'^Z_1 = e^{-Q^Z_iF^i}W_1^Z,
\qquad Q^Z_i = - \omega^Z_i + q^{Y}_i, \quad \mu'_Z = \mu_Z
e^{\sum_i\omega^Z_iF^i}\label{supZ}\eeq
has modular weights $Q^Z_i$; for example, $\omega_i = n_i$ if $\mu(T^i)
= \mu\prod_i[\eta(T^i)]^{2n_i}$.  A modular covariant mass term $W_1^X$
has $Q^X_i=1$.  Note also that the squared mass matrix
\beq (m^2_Z)^P_Q = K_Z^{P\bR}W_{1\bR}^{\S}K^{\Y}_{\S M}W_{1Q}^M =
f^{-1}\delta^P_Q|\mu_Z(T)|^2\label{diagm2}\eeq
is diagonal and commutes with the operators that appear in the
one-loop effective action \myref{l0}. The moduli-dependence of the PV
masses we have introduced in \myref{supZ} can be interpreted as a
parameterization of the threshold corrections~\cite{dixon,uth}
corrections from higher KK and string mode loops. This would be
consistent with an interpretation of the PV masses as fully
parameterizing Planck/string scale physics which provides the
cancellations that restore the finiteness of the underlying theory, as
well as restoring the perturbative symmetries of string theory, up to
the terms that require including a field dependence in the (infinite)
UV cut-off.  The requirement of modular covariant mass terms entails
the introduction of factors of Dedekind eta functions $\eta(iT^n)$;
these functions, and therefore PV masses~\cite{tomt} with
$\omega_i>0$, are exponentially suppressed in the (strong coupling)
limit of large $\re t$. However we do not expect our perturbative
treatment to be applicable in this strong coupling limit.

If the fields $Z^P$ have the same K\"ahler metric as the light fields
$Z^p$: $K_{P\Q} = K_{p\q}$, the $Z^P$ and $Y_P$ loop contributions to
terms linear in $\Gamma_\alpha$ in $\Phi_{1,2}$ and $\L_Q$ cancel,
while they give a double contribution to terms quadratic in
$\Gamma_\alpha$ in $\Phi_1$.  Therefore we need to either 1) introduce
other PV fields (with negative signature) with a simpler metric that
can mimic the contribution of the light fields or 2) couple one set of
(negative signature) fields $Z^P$ to fields $\Phi_P$ with the opposite
gauge charge but trivial metric, {\it e.g.} $K^{\Phi} = \sum_P
e^{g^P}|\Phi_P|^2.$ Option 1) is possible for sigma models whose
K\"ahler potentials have the special property
\beq G(Z^p) = \sum_ng^n(Z_n^p), \qquad g^n_{pq} = c_n
g^n_pg^n_q,\label{fact}\eeq
as shown in II for no-scale models that characterize the untwisted
sector of orbifold compactifications.  This allows full regularization
of the theory in a simple way if the twisted sector fields are set to
zero in the background.  This was justified in II on the grounds that
the K\"ahler potential for orbifolds is not known beyond leading
(quadratic) order in the twisted sector superfields $\Phi_T^a$.  As a
consequence the $O(|\Phi_T|^2)$ loop corrections cannot be determined.
However since in realistic models~\cite{iban} the SM spectrum includes
twisted sector fields, one would like to include them in the effective
one-loop Lagrangian using a general modular invariant parametrization
of the K\"ahler potential.  Therefore we adopt option 2), which
entails a squared-mass matrix that is not of the form \myref{diagm2}
and generally does not commute with other operators in the one-loop
effective action.  As shown in Appendices B--D, cancellation of
on-shell UV divergences can be achieved with a PV sector such that the
only masses that are not invariant under the anomalous symmetries are
of the form \myref{diagm2}, and indeed we argued above that the $\dZ$
mass matrix should be modular and \ux\, invariant.  In fact we will
adopt the simplest possible approach, with noninvariant masses
generally present only for PV fields $\Phi^C$ with a K\"ahler metric
of the form $K_{C\bar D} = f_C\del_{C D}.$

To construct invariant masses for $\dZ^\sigma = \dZ^P,\dZ^N$, we
introduce fields $\dY_\sigma = \dY_P,\dY_N,$ with K\"ahler potential
\beq K^{\dY} = \sum_Ae^{G^A}|\dY_A|^2 + \sum_N e^{G^N}\[g_n^{i\bj}\(\dY_I
- \chi^n_i\dY_N\)\(\dbY_{\J} - \bc^{\n}_{\bj}\dbY_{\N}\) +
|\dY_N|^2\],\label{kdy}\eeq
where $g = \sum_n g^n$ is either the K\"ahler potential for just the
moduli with $\phi^a$ all gauge-charged matter, or it is the
K\"ahler potential for the untwisted sector with $\phi^a$ the twisted
sector,\footnote{In this case we have to slightly modify \myref{kdy}
and \myref{wzi2}; see \myapp{orb3}.}, and 
\beq g_n^{i\bj}g^n_{\bj k} = \del^i_k,\qquad 
G^{A,N} = \sum_m q^{A,N}_m g^m + \ell^{A,N},\label{GAN}\eeq 
where $\ell^{A,N}$ is a modular and gauge invariant function of the
chiral superfields.  The $\dY_A$ are PV counterparts of the light
fields $Z^a$ not included in the $g_n$; they transform according to
the gauge group representation conjugate to that of the $Z^a$ and have
modular weights $q^A_n$. The K\"ahler potential (\ref{kdy}) is modular
invariant provided under (\ref{modt})
\bea \chi'^n_i(Z',\Z') &=& n^j_i\(\chi^n_j(Z,\Z) + \dot a F_q\), \qquad
\dY'_A = e^{-F^A}\dY_A, \qquad\dY'_I = e^{-F^N}n_i^j\(\dY_J +
\dot a F^n_j\dY_N\), \nnn \dY'_N &=& e^{-F^{N}}\dY_N, \qquad F^{N,A} =
\sum_m q^{N,A}_m F^m.\label{trdy} \eea
where $n^i_j$ is the submatrix of $N^p_q$, defined in
(\ref{fmod}), that acts only on the fields $Z^i$ in $\sum_n g^n$.  
Then the terms in the superpotential
\bea W_1^{\dZ} &=& \sum_a W_a + \sum_n W_n,\qquad
W_n
= \dot\mu_n(T^i)\(\sum_{P\in n}\dZ^P\dY_P + \dZ^N\dY_N\), \qquad
\dot\mu_n(T'^i) = e^{F^N-F}\dot\mu_n(T^i),  \nnn W_a &=& \dot\mu_a
(T^i)\(\dZ^A\dY_A - \dot a^{-1}q^a_n Z^a\dZ^N\dY_A\), 
\qquad \dot\mu_a(T'^i) = e^{F^a + F^A-F}\dot\mu_a(T^i),\label{wzi2}\eea
are modular covariant.  The UV-divergent contributions of the $\dY$ to
the loop corrections can be canceled by the fields $U$ and $V$ along
with additional fields $\Phi$ with very simple transformation
properties, K\"ahler potential and superpotential $W_1(U,V,\Phi),$
that are given explicitly in \myapp{orbs}.

To give masses to the fields introduced in \mysec{nonren}, we
introduce additional fields $\hZ$ and $\tY$ that transform under the
nonanomalous gauge group like $Z^p$ and its conjugate, respectively,
as well as gauge singlets $\hf_r$, with K\"ahler potentials
\bea K_\alpha^{\hZ} &=& f_{\hZ_\alpha}
\[\sum_{P,Q=p,q}\hZ_\alpha^{P}\hbZ_\alpha^{\Q}\(G_{p\q} 
+ a^2_\alpha G_pG_{\q}\) + a_\alpha\(\hZ_\alpha^P\hbZ_\alpha^0G_p +
{\rm h.c.}\) + |\hZ_\alpha^0|^2 \], \nonumber \\ 
K_\alpha^{\tY} &=& f_{\tY_\alpha}
\sum_{P,Q=p,q}G^{p\q}\tY^\alpha_P\tbY^\alpha_{\Q}, \nnn 
K_\alpha^{\hf} &=& e^K\lbr e^{\beta_0^\alpha k}\[2e^{2k}|\hf^S_\alpha|^2 +
|\hf^0_\alpha|^2 + e^{k}\(\hat{\bph}^{\S}_\alpha\hf^0_\alpha + {\rm
h.c.}\)\] + e^{\beta^S_\alpha k}|\hf_S^\alpha|^2\rbr,\label{hatyz}\eea
that are modular and \ux\, invariant with the appropriate
transformation properties.\footnote{The K\"ahler potential for
$\hf^{S,0}$ and $\phi_{S,0}$ are the same as for $\hZ^\sigma$ and
$\hY_\sigma$, respectively, with the substitution $G(Z^p,\Z^{\bar p})
\to k(S,\S)$ and $a^{\hf}_\alpha = -1,\;f_{\hf_\alpha}=\exp(K +
\beta^\alpha_0k)$.}  We thus modify the signature constraints
\myref{sigs} to read
\bea \eta^{\hZ}_\alpha &=& \eta^{\hY}_\alpha = \heta_\alpha, \qquad
\eta^{\tY}_\alpha = \eta^{\tZ}_\alpha = \teta_\alpha, \qquad
\sum_\alpha\teta_\alpha + \sum_\alpha\heta_\alpha \equiv \tN + \hN =
0,\nnn \eta^{\hf^r}_\alpha &=& \eta^{\phi^r}_\alpha, \qquad \phi^r =
(\phi^S,\phi_S,\phi_0), \label{hatkahl}\eea
and the first two constraints in
(\ref{cond0}) are modified to read
\beq  \sum_\alpha\eta_\alpha^{\hY}\hat a_\alpha^2 = -1,
\qquad  \sum_\alpha\eta_\alpha^{\hY}\hat a_\alpha^4=
+1.\label{newcond}\eeq
because the contribution from $\hZ_\alpha^I$ to $\Phi_1^{PV}$ doubles
that from ${\hY}^\alpha_I$.  

The remaining PV masses can be generated, for example, by the
superpotential
\bea W_1 &=& W^{\dZ} + W^{\hY} + W^{\tZ} + W^\varphi + W^{\tph}+ W^S +
W^\phi + W^V + W^\Phi, \nnn W^{\hY} &=&
\sum_\alpha\mu^{\hY}_\alpha\sum_{P=
A,I,0}\hZ^P_\alpha\hY^\alpha_P, \qquad W^{\tZ} = \sum_\alpha
\mu^{\tZ}_\alpha\sum_{P=A,I}\tZ^P_\alpha\tY^\alpha_P, \nnn
W^\varphi &=&
\sum_{\alpha,a}\mu^\varphi_\alpha\varphi^a_\alpha\hph^a_\alpha, \qquad
W^{\tph} = {1\over2}\mu^{\tph}_\alpha\tph^a_\alpha\tph^a_\alpha, \qquad
W^V = \sum_{A,\gamma}\(\mu_\gamma^U U_A^\gamma U^A_\gamma +
{1\over2}\mu_\gamma^V(V_A^\gamma)^2\), \nnn W^S &=&
\sum_{\alpha,r}\mu_\alpha^S\phi_\alpha^r\hf^\alpha_r, \qquad W^\phi =
{1\over2}\sum_{\alpha,\beta}\mu^\phi_{\alpha\beta}
\phi^\alpha\phi^\beta, \qquad \eta^\phi_\alpha = \eta^\phi_\beta \;\;
{\rm if} \;\;\mu^\phi_{\alpha\beta}\ne0,\label{w1} \eea
where $W^{\dZ}$ is given by \myref{wzi2}, and $W^\Phi$ is given in
\myapp{orbs}. The K\"ahler potential for the fields $\Phi$, whose only
role is to cancel unwanted curvature terms generated by the K\"ahler
potential \myref{kdy}, is also given in \myapp{orbs}. Note that
$W^\varphi,\; W^S$ and $W^{\dZ}$ (for at least one set $\dZ,\dY$) are
\ux\, invariant and modular covariant.  The mass parameters $\mu$ can
in general depend on the moduli: $\mu = \mu(T^i)$.  When all the
above additional PV fields are included, the prefactors in their
K\"ahler potentials and their \ux\, charges are understood to be
included in the sums in (\ref{cond1}) and \myref{condquad}, and in
\myref{sigs} $N'$ includes all PV chiral superfields. We further
require
\beq \sum_\alpha\eta_\alpha^Y \ln f_\alpha^Y -
\sum_\alpha\eta_\alpha^Z\ln f_\alpha^Z = 0,
\qquad Y = \hY,\tY,\qquad Z = \hZ,\tZ.\label{condfyz}\eeq

The simplest form of the prefactors needed for the cancellation of
all UV divergences is 
\beq f = e^{\alpha K + \beta k + \sum_n q_n g^n + q\vx}.\label{basicf}\eeq
Although we are working generally in Yang-Mills superspace~\cite{bggm}
in which the gauge potential superfields $V_a$ do not appear
explicitly in the superpotential, as explained in \myapp{anmass}
fields with superpotential mass terms that are not \ux\, invariant
must transform in such a way that the mass term remains holomorphic
under a \ux\, transformation; this requires the presence of $q\vx$ in
the exponent of the prefactor for these fields.  As shown in
\myapp{orbs}, all the anomalies arising from gauge-charged matter in
the light sector can be absorbed into the mass terms of $U,V,\Phi$.
This means in particular that we need to include factors $e^{q\vx}$
only for these fields.  These fields also have prefactors of the form
$e^{\sum_n q_n g^n}$, similar to $e^{G^{A,N}}$ in \myref{kdy}, that
accommodate the charged matter and $T$-moduli contributions to the
modular anomaly.  For the other PV fields we take the simpler form
\beq f_C = e^{\alpha_C K_C + \beta_C k_C}.\label{basicf2}\eeq
A similar prefactor may be included in the K\"ahler potential for any
chiral superfield $\Phi\ne\theta$ that does not have other couplings
except in $W_1$, provided the moduli-dependence of $W_1$ assures
modular covariance as needed; this includes additional fields with the
same K\"ahler metric, gauge charges and superpotential terms in $W_1$ as
$\dZ,\hZ,\tZ,\dY,\hY,\tY$ but no other couplings.  All PV fields with
couplings that contribute to the renormalization of the superpotential
must have invariant mass terms. Here we will assume
only the minimal number of these fields needed to regulate light field
contributions to the renormalization of the K\"ahler potential.  Thus
we include $\dZ,\dY$ only with the K\"ahler potential given in
\myref{pvz}, \myref{xkahl2} and \myref{kdy}.  We take all $\hY,\tZ$ to
have the K\"ahler potentials given in \myref{kahly}, and $\hZ,\tY$ to
have, respectively, the inverse K\"ahler metrics with prefactors
\beq f_{\hZ_\alpha,\tY_\alpha} = e^{\alpha^{\hZ,\tY}_\alpha K},\eeq
such that including the $T^i$-dependence of $\mu^{\hY,\tZ}_\alpha$,
the masses are modular invariant, and the constraint \myref{condfyz},
which reduces to
\beq \sum_\alpha\hat\eta_\alpha\(\alpha^{\tY}_\alpha - 
\alpha^{\hZ}_\alpha\) = 0,\label{alphayz}\eeq
is satisfied.

\subsection{Quadratic PV mass terms}\label{quadm}

We have constructed the PV Lagrangian in such a way that PV fields
that couple to one another in $W_1$ have no other common coupling.
The the term quadratic in masses takes the form, in a basis where
the squared mass matrix is diagonal
\bea \L^{PV}_Q &=& 
{1\over32\pi^2}\[\(3\sum_\gamma\ell^\theta_\gamma -
\sum_C\ell_C\)\(\hV + M^2\) - {1\over2} \(\sum_C\ell_C -
\sum_\gamma\ell^\theta_\gamma\)\l\Dc\Xa\r \right.
\nonumber \\ & &\l +
\sum_\gamma\ell_\gamma^s\l\Dc\ka\r +
\sum_C\ell_C\l\Dc\Gamma^C_{C\alpha}\r\], \eea
where here $C$ refers to all heavy chiral supermultiplets
$\Phi^C\ne\theta$, and
\beq \ell_X = m^2_X\ln(\Lambda^2/m^2_X)\eta_X, \qquad X =
\Phi^C,\theta.\eeq 
Finiteness requires that the coefficient of $\ln\Lambda$ vanish:
\beq 0 = \sum_\gamma m^2_{0\gamma}\eta^0_\gamma = \sum_\gamma
m^2_{s\gamma}\eta^s_\gamma = \sum_C m^2_C\eta_C = \sum_C
m^2_C\eta_C\l\Dc\Gamma^C_{C\alpha}\r.\label{finreq}\eeq
If $\Phi^C$ couples to $\Phi^D$ in $W_1$, with 
\beq m^2_C = m^2_D = \mu^2_C|w_C(t)|^2e^{K - 2(q_C +
q_D)\vx}f_C(z,\z)\equiv \mu^2_C\Lambda^2_C,\label{pqmass} \eeq
then $\Da\(\Gamma^C_{C\alpha} + \Gamma^D_{D\alpha}\)$ is completely
determined in terms of $(q_C+q_D)$ and scalar derivatives of $f_C$, so
the third equality in \myref{finreq} insures the fourth, since the coefficient of each
term of fixed $\Lambda^2_C\ne$ constant must vanish separately.  We
have regulated the theory such that all masses are invariant except
for some chiral superfields $\Phi^{C_\alpha}$ with masses of the form
\myref{pqmass}.  These give contributions to the renormalized K\"ahler
potential of the form
\bea K_Q &\ni& \sum_C{\lambda_C\Lambda^2_C\over32\pi^2} 
-2{\lambda_0\Lambda^2_0\over32\pi^2}, \qquad \lambda_X = 
\sum_{\gamma =1}^{n_X}\eta^{X_\gamma}\mu_{X_\gamma}^2\ln(\mu^2_{X_\gamma}), 
\nnn m^2_{X_\gamma} &=& 
\mu^2_{X_\gamma}\Lambda^2_X, \qquad \sum_{\gamma=1}^{n_X}
\eta^{X_\gamma}\mu^2_{X_\gamma} = 0.\label{quadcond}\eea
We require $\lambda_{C} = 0$ if $\Lambda^2_{C}$ is not
 modular and \ux\,
invariant.  Note that in order to cancel the chiral anomaly and the
logarithmic divergences, we have necessarily 
\beq \sum_{\alpha=1}^{n_C}\eta^{C_\alpha} \ne0 \eeq 
for at least some
sets $\{C,D\}$.  For these the constraint $\lambda_C = 0$
requires\footnote{See Appendix C of \cite{sigma}.} $n_C \ge5$.


%% file: Section5.tex
\section{Superfield structure of the anomaly and anomaly 
cancellation}\label{modan}
\hspace{0.8cm}\setcounter{equation}{0} 

The bosonic part of the anomaly under infinitesimal modular and \ux\,
transformations that arises from the noninvariance of PV masses is
given in \myref{totanom} of \myapp{fullan}.  Referring to \myref{Om2}
and \myref{E10}, when combined with the additional contributions
described in \mysec{strat}--with the cancellation of
\myref{OMx} imposed--that expression is the bosonic part of an 
infinitesimal variation of the superfield operator
\bea \L_{\rm anom} &=& \L_0 + \L_1 + \L_r = \superint E\(L_0 + L_1+
L_r\),\label{SFtotanom}\\
L_0 &=& {1\over8\pi^2}\[\Tr\eta\ln\cM^2\Omega_0 + K\(\Omega_{G B} 
+ \Omega'_D\)\],\qquad L_r = - {1\over192\pi^2}\Tr\eta\int
d\ln\cM\Omega_r.\label{totomega}\eea
$\L_1$ is defined by its variation:
\beq \Del L_1 = {1\over8\pi^2}{1\over192}\Tr\eta\Del\ln\cM^2\Omega'_L 
= {1\over8\pi^2}{1\over192}\Tr\eta H\Omega'_L\hc,\eeq
and $\Omega'_D$ is the ``D-term'' anomaly that arises from certain
total derivatives with logarithmically divergent coefficients as
discussed in \mysec{strat}.
The superfield $\Omega_r$ is defined by \myref{E10} and \myref{E12},
$\Omega_{G B}$ is defined in \myref{resan}, the real
superfield $\cM^2$ is the PV squared mass matrix,
\beq \Del\ln\cM^2 = 2\(H + \bar H\),\eeq
with $H$ holomorphic, and
\bea \Omega_0 &=& {1\over3}{\Omega}_W + \Omega^0_{\rm YM}
- {1\over36}\Omega_X + {1\over192}\Omega_L
- {1\over48}\(6\phi_\chi + \D^2R + \bD^2\bR\) - {1\over192}\Omega'_L\eee
- {1\over24}\Omega_{G B} + \Omega^0_{\rm Y M} 
- {1\over48}\Omega_\phi,\nnn
\Omega'_L &=& \Omega_L - 16\Omega_X\label{Om}, 
\qquad \Omega_\phi = 12\phi_\chi + \D^2R + \bD^2\bR,
\label{OprimeL}\eea
with the Chern-Simons superfields in \myref{Om} defined in
\myref{OYM}, \myref{OWX}, and \myref{E7}.  The above results were also
found~\cite{dan} by performing a superspace calculation of the anomaly
in PV regulated superconformal supergravity, and then gauge-fixing to
K\"ahler \uk\, superspace~\cite{bggm}.  Note that the part of $\Omega'_L$
that is independent of the \ux\, charges and modular weights, namely
\beq \Tr\eta H\Omega'_L 
= 16\sum_C\tph^+_C\[(1 - 2\alpha^C)^2 - 1\]\Omega_X + O(q_X,q_i),
\label{OXcancel}\eeq
\noindent
drops out of (5.3) by virtue of (D.103) and (D.113) [with (D.114)].
In the present approach, the coefficient of the Chern-Simons
superfield $\Omega_{YM}^X$ comes from a combination of terms in
$\Omega_{X_m}$ and $\Omega_G$ in \myref{Om2}. Because the anomalous
\ux\, is not incorporated into the superspace geometry, the superfield
strength in the chiral projection of $\Omega^0_{\rm Y M}$, defined in
\myref{Om2}, does not include $\Wa^X$.  Instead, the PV masses depend
on the the \ux\, vector field strength $\vx$, giving a \ux-dependent
contribution from $\Omega_L$. We note here that the operator
$\Omega_0$ in \myref{totomega} has contributions that contain a
dependence on the dilaton field, such as $\Omega_X$ and $\Omega_L$;
these do not satisfy the condition \myref{opcond} of \mysec{strat}.  We
show in \myapp{csfields} that the linear/chiral multiplet duality
outlined in \mysec{strat} still holds in the presence of these
operators; although some intermediate equations are modified the
action for the dilaton coupling to $\Omega$ is the same in both the
linear and chiral multiplet formulations, and still takes the form
\myref{ltreel} or \myref{sgs}.  

There are additional \ux-dependent contributions from the
``D-terms''in $\Omega_{r}$.  Integrating \myref{E10} gives
\bea \L_r  &=& - {1\over8\pi^2}{1\over24}\superint E\Tr\eta\lbr
\ln\cM\(\half\Dc\La + 2\Dc\Xa\)(\ln\cM)^2
+ G^{\alpha\dot\beta}\Da\ln\cM\Dd\ln\cM\right.\mmm\qquad\l
- {1\over4}\(\D^2\ln\cM\Dd\ln\cM\Db\ln\cM\hc\)
\right.\mmm\qquad\l
- \ln\cM\[{1\over8}D^2\bD^2\ln\cM + \Dc(R\Da\ln\cM)\hc\]
\right.\mmm\qquad\l
- \half\Dc\ln\cM\Da\ln\cM\Dd\ln\cM\Db\ln\cM\rbr,\label{Lr}\eea
up to a linear superfield and a total derivative; the variation of $L_r$
under an anomalous transformation is given in \myref{E11}.

In this section we evaluate the anomalies and discuss the counterterms
needed to cancel them.  Possible connections of these counterterms to
the 10d GS term were discussed in \mysec{strat}.  For simplicity we
present explicit results only for the case without threshold
corrections as in $Z_3$ and $Z_7$ orbifold compactifications; the
result for any specific compactification with threshold corrections can
be reconstructed from the results of \myapp{anmass}, by matching the
parameters $\omega^C_n$ to the threshold corrections specific to the
model.

\subsection{Generalization of the 4d GS mechanism}

The traces in $\L_0$ are completely determined, up to possible
threshold corrections, in terms of the quantum numbers of the light
spectrum.  Taking the traces subject to the constraints given in
\myref{cond0}--\myref{condquad}, \myapp{lindiv} and \myapp{anmass}, we
obtain a contribution that can be written in the form (up to terms that
are invariant under modular\footnote{In the case that string threshold
corrections are present, these include terms of the form
$\sum_{n,A}b^A_n\ln\[(T^n + \T^n)|\eta(T^n)|^2\]\Omega_A$.} and
\ux\, transformations and terms that do not contribute to the
variation of the action)
\bea L_0 &=& \(b g - \half\dx\vx\)\Omega_0 + {g\over8\pi^2}
\(\Omega'_D + \Omega_\phi\).\label{LF}\eea
In contrast, the traces in $\L_1$ depend on the details of the PV
regularization procedure.  For example if we use either \myref{ellp1}
or \myref{ellp2} we get an expression of the form
\bea {1\over192}\Tr\eta H\Omega'_L &=& {1\over3}\sum_l F^l\[C_{G
  S}\Omega^X_{Y M} + A_l\Omega_{W_X X} - \sum_{m,n}A_{m n l}\Omega_{n
  m}\right.\mmm\qquad\qquad\l - \sum_n\(2B_{n l}\Omega_{W_X n} - A_{n
l}\Omega_{n X}\)\]\ddd + {1\over3}\Lambda\[3C'_{G S}\Omega^X_{Y M} +
a\Omega_{W_X X} - \sum_{m,n}a_{m n}\Omega_{n m} -
\sum_n\(2b_{n}\Omega_{W_X n} - a_{n}\Omega_{n X}\)\]\label{TrOL},\eea
with
\bea \chiproj\Omega_{W_X X} &=& \Wc_X\Xa, \qquad 
\chiproj\Omega_{n m} = \gc_n\ga^m, \nnn
\chiproj\Omega_{n X} &=& \gc_n\Xa, \qquad 
\chiproj\Omega_{W_X n} = \Wc_X\ga^n, \eea
the expressions for the parameters $A,B,a,b,c$ are given in
\myref{D110}--\myref{D116} for the PV sectors of \myapp{orb1} and
\myapp{orb2} with the choice \myref{ellp1}, and
\beq C_{G S} = 8\pi^2b,\qquad C'_{G S} = - 4\pi^2\del_X,
\label{5.5}\eeq
In writing the above we used the known string theory constraints, which
for orbifold compactifications with no threshold corrections read
\bea 8\pi^2b &=& {1\over24}\(2\sum_p q^p_n - N + N_G - 21\)\quad\forall\quad n\eee
C_a - C^M_a + 2\sum_b C^b_a q^b_n 
\quad \forall\;\; n,a, \quad C_a = \Tr(T_a)^2_{\rm adj}, \quad 
C^M_a = \Tr(T_a)^2_{\rm mat},\label{wthconds} \\
2\pi^2\dx &=& - {1\over24}\Tr T_X = - {1\over3}\Tr T_X^3 = 
- \Tr(T^2_aT_X)\quad \forall\quad a\ne X\label{uxconds}.\eea
If the expression \myref{TrOL} were to factorize:
\bea {1\over192}\Tr\eta H\Omega'_L &=& \(C_{G S}F + C'_{G S}\)\Omega'_0,\\
\Omega'_0 &=& \[\Omega^X_{Y M} 
+ A\Omega_{W_X X} - \sum_{m,n}A'_{m n}\Omega_{n m}
- \sum_n\(2B'_{n}\Omega_{W_X n} - A'_{n}\Omega_{n X}\)\]\label{dream},\eea
the full F-term contribution to the anomaly
could be canceled provided the tree-level Lagrangian includes
a contribution~\cite{bglett} (in the chiral formulation for the dilaton)
\beq \L_{S} = - \int d^4\theta(S + \S)\Omega = {1\over8}\superint
{E\over R}S\Phi\hc,\qquad\Phi = \chiproj\Omega,
\qquad \Omega = \Omega_0 + \Omega'_0,\label{stree}\eeq
with $S$ transforming under modular and \ux\, transformations as
\beq S \to S - {\dx\over2}\Lambda + bF,\label{DelS}\eeq
so that the dilaton K\"ahler potential 
\beq K_S = k(S + \S + {\dx\over2}\vx - b G)\label{ktree}\eeq
is invariant.  Note that the coefficient of $\Omega^X_{Y M}$ in
$\Omega$ is fixed, since the dilaton is known to couple universally to
the gauge field strength (for affine level $k=1$).  The Lagrangian
\myref{stree} and the K\"ahler potential \myref{ktree} can be obtained
by a duality transformation from the formulation where the dilaton is
the lowest component of a modified linear multiplet, as described in
\mysec{strat}.
The new couplings in \myref{stree} of course also contribute
ultraviolet divergent contributions in the effective QFT.  We expect
that these can be regulated by PV fields with modular and \ux\,
invariant masses, as we have shown to be the case for the dilaton
coupling to $\Phi_{YM}$, so that they will not induce any further
contributions to the anomaly.  In the case that string loop
corrections of the form
\bea \L_{\rm thresh} &=& {1\over4\pi^2}\sum_{n,A}\ln|\eta(T^n)|^2
b^A_n\Omega_A,\\
\Del \L_{\rm thresh} &=& {1\over8\pi^2}\sum_{n,A}\[F(T^n) + \bF(\T^{\n})\]
b^A_n\Omega_A,\label{th1}\eea
are present, the constraints \myref{wthconds} and \myref{uxconds} are
modified as in \myref{orbcond} and \myref{D98}, and the quantum
anomaly arising from the F-term should be canceled by a the
combination of \myref{DelS} and \myref{th1}.

If we make a different choice than \myref{ellp1}
or \myref{ellp2} for the invariant functions $\ell$ in \myref{GAN},
there will be additional terms on the right hand side of \myref{TrOL}.
However this model-dependence is removed if we evaluate it with only
the K\"ahler moduli nonvanishing among the background chiral superfields, 
such that 
\beq \ell \to 0, \qquad K \to g = \sum_n g^n,\qquad \Omega_{n X}\to
\sum_m\Omega_{n m},\qquad \Omega_{W_X X}\to
\sum_n\Omega_{W_X n}, \eeq
so that \myref{TrOL} reduces to
\bea {1\over192}\Tr\eta H\Omega'_L &=& {1\over3}\sum_l F^l\[C_{G
  S}\Omega^X_{Y M} + \sum_n\(A_l- 2B_{n l}\)\Omega_{W_X n} -
\sum_{m,n}\(A_{m n l} - A_{n l}\)\Omega_{n m}\]\ddd 
+ {1\over3}\Lambda\[3C'_{G S}\Omega^X_{Y M} + \sum_n\(a - 2b_{n}\)\Omega_{W_X n} -
\sum_{m,n}\(a_{m n} - a_{n}\)\Omega_{n m}\]
\label{TrOL2}.\eea
One can check that this does not factorize for the models listed in
\myapp{fiqs} if we use the results \myref{D110}--\myref{D116} that
apply if the prescription of \myapp{orb1} or \myapp{orb2} is used.
With this prescription, the modular weights and \ux\, charges of the PV
fields that enter the sums in \myref{D110}--\myref{D116} are exactly
those of the light chiral supermultiplets. They arise from the contributions
of PV fields $\Psi^C$ with, in the absence of threshold corrections
($\omega^C_n = 0\;\forall\; \Phi^C$) modular weights 
\beq q^{\Psi^C}_n = 1 - q^C_n.\eeq
In contrast, the more constrained prescription of \myapp{orb3} replaces
the gauge singlet PV ``moduli'' $\Phi^N,\Phi^n$ with
\beq (q^N_m, q^n_m) = (2\del^N_m,0)\eeq
by the same number of PV ``moduli'' but with 
\beq (q^N_m, q^n_m) = (\del^N_m,\del^n_m)\eeq
This leaves sums linear in the modular weights unchanged, but slightly
modifies the nonlinear sums:
\beq \Del\sum_C q^{\Psi^C}_{n_1}\ldots q^{\Psi^C}_{n_M} = (2 - 2^M)\prod_{k=1}^{M-1}
\del_{n_k n_{k+1}}.\eeq
This does not achieve factorization, but suggests that an even more
constrained PV sector, such as ``option 1'', discussed in
\mysec{pvmass} and briefly outlined in the beginning of \myapp{orbs},
might allow a significantly different set of modular weights for the
PV fields that have noninvariant masses.  Note also that in the
effective field theory for the $\mathbb{Z}_7$ case of \myapp{fiqs},
there are mixed anomalies associated with $\Tr T_a q_n\ne0$ and $\Tr
T_a q_n^2\ne0$.  No anomalies are present in the corresponding string
theory~\cite{ssanom} (which has no Wilson lines and therefore no
anomalous \uo).  Therefore in the appropriately regulated effective
field theory these anomalies should disappear.  For the case where an
anomalous \ux\, is present, we also need a modification of $\Tr
T_X^2q_n$.  In addition there may be larger symmetries, or partial
symmetries--such as the ``Heisenberg symmetry'' of the untwisted
sector K\"ahler potential that is used in \myapp{orb3}--that should be
respected in the PV sector.\footnote{In writing the transformation
  properties \myref{modt} and \myref{ded} we neglected constant
  phases~\cite{Ferrara:1989bc} and matrices~\cite{Chun:1989se} that
  mix chiral superfields with the same modular weights.  These can be
  trivially incorporated and do not affect the conclusions.}  For
example, the K\"ahler metric of the untwisted sector for
$\mathbb{Z}_3$ models, including the more realistic FIQS model with
Wilson lines and a \ux, have a larger symmetry than the
$[SL(2,\mathbb{Z})]^3$ group of modular transformations that we have
been using.  If that symmetry is preserved in the PV sector, this
might also give a different result for terms nonlinear in modular and
\ux\, charges.

Cancellation of the remaining contributions to the anomaly, namely the
the term proportional to $g$ in \myref{LF}--that has not been
considered previously in this context--and the D-term anomaly $\L_r$
in \myref{Lr} appears to require additional counterterms, although
possibly the former, and certainly the latter, will be modified by any
modification of the PV sector that can provide factorization in
\myref{TrOL2}.  Because $\ln\cM$ contains a term $(\half - \alpha)K$,
it is not impossible that the the former term can be canceled by a
contribution from the latter, in the same way that that the $\Omega_X$
term in $\Omega'_L$ is canceled in \myref{OXcancel}; this cancellation
seems to be independent of the details of the PV sector.


%% file: Section6.tex
\section{Summary of Results}\label{sum}
\hspace{0.8cm}\setcounter{equation}{0} 

We used on-shell Pauli-Villars regularization of the one-loop
ultra-violet divergences of supergravity to determine the anomaly
structure of these theories.  This regularization procedure requires
constraints on the chiral matter representations of the gauge group
that are satisfied in any supersymmetric extension of the Standard
Model, and by hidden sectors that have been found in orbifold
compactifications of the heterotic string.

We showed that the logarithmic and quadratic one-loop divergences in
the S-matrix can be canceled by a PV sector that respects the
classical symmetries of the theory except for certain superpotential
terms that generate large, noninvariant PV masses for a subset of PV
chiral superfields that have a very simple K\"ahler metric.  A PV
sector with this feature was explicitly constructed for effective
supergravity theories with three K\"ahler moduli $T^i$ obtained from
orbifold compactifications of the weakly coupled heterotic string.
These theories are classically invariant under a \ux\, gauge
transformation and under the T-duality group of modular
transformations that are anomalous at the quantum level.  

If all linear divergences were canceled in the regulated theory, it
would be anomaly free, with noninvariance of the action arising only
from that of the Pauli-Villars masses.  Here we calculated their
contribution to the bosonic part of the anomaly in the component
Lagrangian. Because the regularization procedure is manifestly
supersymmetric the full anomaly generated by PV loops necessarily
forms a superfield. Since only chiral superfields contribute, and
since they have a diagonal K\"ahler metric, we were able to draw on
the recent superfield calculation~\cite{dan} of the chiral multiplet loop
contribution to the UV divergences and anomalies in supergravity.
This provided a check of our component results, and greatly simplified
the identification of the superfield form of the anomaly.

Pauli-Villars regulator fields allow for the cancellation of all
quadratic and logarithmic divergences, but there remain residual
linear divergences associated with nonrenormalizable gravitino/gaugino
interactions.  This result follows directly from the constraints
imposed by cancellation of on-shell quadratic divergences, and is
therefore independent of the details of the PV sector.  These residual
divergences result in additional contributions to the chiral anomaly.
There is also an additional conformal anomaly that appears if the UV
cut-off $\Lambda$ is field dependent, because there are residual terms
in the coefficient of $\ln\Lambda$ that are not canceled by the PV
sector.  These are total derivatives, so the S-matrix of the regulated
theory is finite, but a field-dependent cut-off can induce finite
terms that are not invariant under the classical symmetries.  These
new contributions, which are anomalous only under T-duality
invariance, combine to form a supermultiplet provided the ultraviolet
cut-off has the field dependence given in \myref{uvcut}.  When this
contribution is combined with the PV contribution, we obtain results
in agreement with string-loop calculations of the anomalous
coefficients of the squared (nonanomalous) Yang-Mills and space-time curvatures.

It is well known that these contributions to the anomaly can be
canceled by a four-dimensional version of the Green-Schwarz mechanism.
We showed that the remaining contributions depend on the choice of PV
sector couplings.  Provided that the F-term contributions
factorize~\cite{ssanom}, they can be canceled by a generalization this
mechanism, corresponding to a generalization of the modified linearity
condition for the linear supermultiplet that is dual to the dilaton
chiral supermultiplet.  It may be that factorization is possible only
with constraints on the superpotential of the type discussed at the
beginning of \myapp{orbs}.  Such constraints would be somewhat
analogous to the constraint on gauge charges mentioned above, although
the cancellation of UV divergences by itself requires no constraint on
the K\"ahler potential. Such a factorization condition could be used
as a tool to probe higher order terms in the twisted sector K\"ahler
potential.  This is of importance for phenomenology if some fields
require large vacuum values as, for example, in \ux\, gauge symmetry
breaking at the string scale.  As an example, we found that if we
impose a K\"ahler potential of the form \myref{twisted} for the
twisted sector of the {$\mathbf{\mathbb{Z}_7}$} model of \myapp{fiqs}
one can find values of the parameters such that terms proprotional to
$\Xc\ga^n$, with coefficients both linear and quadratic in the modular
weights, either factorize or vanish, while the cancellation in
\myref{OXcancel} is unaffected.  This approach will be pursued further
elsewhere.

There are additional contributions to the conformal anomaly that may
require new counter-terms.  These include terms that are nonlinear in
the parameters of the anomalous transformations.  On the other hand it
is possible that some or all of these terms can be made to cancel. 

We briefly considered the possible connections between the 4-d
counterterms that we find and the 10-d GS term; establishing a clear
connection requires further work.


%% file: AppendixA.tex
\subsection{Linear divergences}\label{lindiv}
\setcounter{equation}{0} 
The one-loop contribution to the bosonic part of the effective action is
\bea 
\L_1 &=& {i\over 2}\Tr\ln(\hD_\Phi^2 + H_\Phi) 
-{i\over 2}\Tr\ln(-i\notD + M_\Theta)
\nonumber \\
& & + i\Tr\ln(D_{Gh}^2 + H_{Gh}) - i\Tr\ln(\hD_{gh}^2 + H_{gh})
\nnn &\equiv& {i\over2}\STr\ln(\hD^2 + H) -{i\over 2}T_-,
\label{boseL1}\eea
where $\Phi$ is a $2N+4N_G+10$ component boson, $\Theta$ is an $N +
N_G + 5$ component Majorana fermion, with $N$ is the number of chiral
multiplets and $N_G$ is the number of gauge multiplets, and the last
two terms are the (4-component fermion) ghostino and ($N_G + 4$
component boson) ghost contributions, respectively.  The last equality
is obtained after casting the fermionic determinant in the form
\begin{equation}
-{i\over 2}\Tr\ln(-i\notD + M_\Theta) =
-{i\over 8}\Tr\ln(\hD^2_\Theta + H_\Theta) -{i\over 2}T_-,
\label{fdet}\end{equation}
where in \myref{fdet} $\hD_\Theta$ and $H_\Theta$ are 
$8\times 8$ matrices in
Dirac space that act on the vector-valued 4-component fermion
$(\Theta_R,\Theta_L)$; for example $H_\Theta$ for PV the fields with
noninvariant masses is given in \myref{defH} below.  The first term is
the helicity averaged part of the fermion contribution, and~\cite{us2}
\beq T_- = \Tr\lbr-\[\notD^2 
+ i\notD M\]^{-1} i\notD\cN\rbr\label{tminus00}\eeq
is the helicity-odd part,\footnote{As discussed in~\cite{pv1} the
  separation of the fermion determinant into helicity odd and even
  parts is ambiguous; there is a unique choice that allows a PV
  regularization of the quadratic divergences. For the universal axion
  couplings to gauginos, this choice is also consistent with
  nonrenormalization~\cite{russ} of the $\theta$-parameter and with
  linearity constraints~\cite{tom} in the dual linear formulation for
  the dilaton supermultiplet.} where $\cN$ is the helicity-odd part of
$\displaystyle{-i\notD + M_\Theta}$.  $T_-$ is not invariant under chiral
transformations because, unlike the the other operators appearing in 
\myref{boseL1}, $\cN$ cannot be recast in invariant form; it explicitly
contains the axial current.  If the linear divergences in $T_-$ cancel,
the non-invariance can be shifted to the PV masses. This requires 
\beq \Tr\eta(\G\cdot\tilde\G\phi) = 0,\label{cacond1}\eeq
where the signature $\eta = 1$ for light fields, 
\beq \G_{\mu\nu} = [D_\mu,\D_\nu]  = G_{\mu\nu} + Z_{\mu\nu},\label{ggphi}\eeq
is the fermion field strength, the fermion covariant derivative is 
\beq D_\mu = \pp_\mu + Z_\mu + v_\mu + J_\mu\gamma_5 = d_\mu + Z_\mu,\qquad 
 G_{\mu\nu} = [d_\mu,d_\nu],\label{defdG}\eeq
with $Z_\mu$ the spin connection:
\beq Z_\mu = {1\over4}\gamma_\nu[\nabla_\mu,\gamma^\nu],\qquad
Z_{\mu\nu} = {1\over4}r_{\mu\nu\rho\sigma}\gamma^\sigma\gamma^\rho,
\label{defZs}\eeq
and
\beq \del J_\mu = - i\pp_\mu\phi \eeq
under a chiral transformation.  In fact, the trace in \myref{cacond1}
does not vanish because the gravitino connection contains the affine
connection as well as the spin connection, whereas the PV fields are
chiral fermions and (Abelian) gauginos, with only the spin connection
contributing to space-time curvature dependent terms.  In addition
there is an off-diagonal gaugino-gravitino connection that is not
reproduced in the PV sector.  This implies that there is a residual
linear divergence.  When the PV fields are included all the other
terms are reproduced, provided they satisfy condition \myref{cacond1}.

The first equality in \myref{2quadconds}, which follows from the
cancellation of on-shell quadratic divergences, ensures the equality
\beq \Tr\eta Z\cdot \tilde Z\phi = \half\(N + N' - N_G - N'_G - 3
- 2\sum_C\eta_C\alpha^C\)Z\cdot \tilde Z\im F = 0.\label{Trphi}\eeq

To implement the condition \myref{cacond1} for the chiral fermions it
is useful to use the relations
\bea 
\(G^+_{\mu\nu}\)^{\n}_{\m} &=& - K_{p\m}\(G^-_{\mu\nu}\)^p_q K^{q\n}
+ 2Z_{\mu\nu}\del^{\n}_{\m},\nnn
\(D^+_{\mu}\)^{\n}_{\m} &=& - K_{p\m}\(D^-_{\mu}\)^p_q K^{q\n} +
K^{q\n}\pp_\mu K_{q\m} + 2Z_\mu\del^{\n}_{\m},\label{chifs0} \eea
Writing
\beq  D^\pm_\mu = \pp_\mu + Z_\mu + J^\pm_\mu,\qquad (J_\mu)^i_j\equiv
(J_\mu^-)^i_j, \qquad (J_\mu)^{\m}_{\n} \equiv (J^+_\mu)^{\m}_{\n},\eeq
for a chiral superfield with a general K\"ahler metric ${\mathbf K}$:
${\mathbf K}_{i\m} = K_{i\m}$, the anomalous part of the one loop
action under a chiral transformation takes the form [see \myref{tminus}]
\bea &&\half\[\Tr\ln(-i\notD^+R) - \Tr\ln(-i\notD^-L)\] =
\half\[\Tr\ln(-i\notD^+R) - \Tr\ln(-i{\mathbf K}\notD^-
{\mathbf K}^{-1}L)\]\nnn&&\qquad\qquad\qquad=
\half\[\Tr\ln(-i\{\notd + \notZ + \notj^+\}R) - 
\Tr\ln(-i{\mathbf K}\{\notd + \notZ + \notj^-\}
{\mathbf K}^{-1}L)\]\nnn&&\qquad\qquad\qquad=
\half\[\Tr\ln(-i\{\notd + \notZ + \notj^+\}R) - 
\Tr\ln(-i\{\notd + \notZ - \notj^+\}L)\],\eea
where in the last equality we used \myref{chifs0}, so we can simply
take $J^+_\mu$ (or $- J^-_\mu$) to be the axial current: $J_\mu = J^\pm_\mu$.
Then
\beq G^V_{\mu\nu} = Z_{\mu\nu} + [J^\pm_\mu,J^\pm_\nu],\qquad
J_{\mu\nu} = \pp_\mu J^\pm_\nu - \pp_\nu J^\pm_\mu,\eeq
and the condition \myref{cacond1} reduces to
\beq 0 = \epsilon^{\mu\nu\rho\sigma}\Tr
\(\pp_\mu J^\pm_\nu\pp_\sigma J^\pm_\rho\phi\).\label{cacond2}\eeq
This implies, in particular,
\beq 0 = \Tr T^2_a T_X,\label{gcubic}\eeq
and
\bea 0 = \sum_C\eta_C\(-8\alpha^3_C + 12\alpha^2_C - 6\alpha_C\) 
+ N + N' - N_G - N'_G - 3 = -8\sum_C\eta_C\alpha^3_C - 8,\nonumber\eea
\beq \sum_C\eta_C\alpha^3_C =  - 1.\label{alph3}\eeq
where $\alpha_C$ is defined in \myref{basicf2} and we used the constraints
\myref{sigs}--\myref{condquad}.  There are also mixed terms in the gauge and
K\"ahler $U(1)$ connections; these require
\bea 0 &=& 2\sum_C\eta_C\alpha_C(T_a)_C^2 - \Tr T^2_a 
= 2\sum_C\eta_C\alpha_C(T_a)^2_C,\label{at2}\\
0 &=& 4\sum_C\eta_C\alpha_C^2(T_X)_C- 4\sum_C\eta_C\alpha_C(T_X)_C 
+ \Tr T_X = 4\sum_C\eta_C\alpha^2_C(T_X)_C,\label{a2t}\eea
since cancellation of quadratic divergences \myref{lquad} requires
$\Tr T_X = 0$, and cancellation of the logarithmic divergences
\myref{fops} requires $\Tr T^2_a = \sum_C\eta_C\alpha_C(T_X)_C = 0$ when PV
contributions are included in the trace. The contributions to \myref{at2}
and \myref{a2t} from $\hZ,\tY$ fermion loops are, respectively  
\beq 2\sum_\alpha\hat\eta_\alpha\(\alpha_\alpha^{\hZ} 
+ \alpha_\alpha^{\tY}\)C_a^M,\qquad
4\sum_\alpha\hat\eta_\alpha\[\(\alpha_\alpha^{\hZ}\)^2 -
\(\alpha_\alpha^{\tY}\)^2\](\Tr T_X)_M.\label{newyz}\eeq
Since $\sum_\alpha\hat\eta_\alpha=0,$ both terms in \myref{newyz} as well as
\myref{alphayz} vanish if
\beq \alpha_\alpha^{\hZ} = \alpha^{\hZ}, \qquad \alpha_\alpha^{\tY} = \alpha^{\tY},
\label{alphayz2}\eeq
independent of $\alpha$.  For example, if there are no threshold
corrections, $\mu_\alpha^{\hZ}$ and $\mu_\alpha^{\tY}$ are constants
in \myref{w1}, and modular covariance of $W^{\hZ},\;W^{\tY}$ requires
$\alpha^{\hZ} = \alpha^{\tY} = 1$.  More generally, \myref{alphayz2}
requires that $\mu_\alpha^{\hZ,\tY} = c_\alpha\mu^{\hZ,\tY}(T^i)$,
which simplifies the constraints in \mysec{quadm}. This constraint
\myref{alphayz2} also assures that there are no other contributions to
\myref{cacond1} from $\hY,\hZ,\tY,\tZ$ loops.  There is no
contribution to \myref{at2} and \myref{a2t} from the adjoint fermions
$\chi^a_\alpha = \l\Da\varphi^a\r$ which have vector gauge connections
and no $U(1)_X$ charges. Then these constraints are trivially satisfied
if the superfields $\phi_\gamma$ in \myref{kahl} carry no gauge
charge, which is natural since they were introduced to regulate
gravitational couplings.  The constraints involving the
parameters $\beta_C$ are, using \myref{cond1} and \myref{condquad},
\bea 0 = \sum_C\beta^2_C\(2\alpha_C - 1\) = 2\(\sum_C\beta^2_C\alpha_C - 1\)
= \sum_C\beta_C\(4\alpha_C^2 - 4\alpha_C + 1\) = 4\sum_C\beta_C\alpha_C^2,
\nonumber\eea\beq \sum_C\beta^2_C\alpha_C = 1,
\qquad \sum_C\beta_C\alpha_C^2 = 0, \qquad \sum_C\beta^3_C = 0.
\label{newbeta}\eeq
Finally, we consider the gravitino and gaugino sector.
As discussed in \myapp{chigauge}, cancellation of 
quadratic divergences arising from the axion connection
in the gaugino covariant derivative, which is defined as
\bea \D_\mu\lambda^a_L &=& \(\pp_\mu + Z_\mu -i\Gamma_\mu + \ell_\mu\)
\lambda^a_L + i(T_c)^a_b A^c_\mu
\lambda^b_L,\nnn \D_\mu\lambda^a_R &=&
\(\pp_\mu + Z_\mu +i\Gamma_\mu + \ell_\mu\)\lambda^a_R + 
i(T_c)^a_b A^c_\mu \lambda^b_R,\label{lamcon}\eea
where $\Gamma_\mu$ is the K\"ahler \uk\, connection defined in \myref{defJ},
and 
\beq \ell_\lambda = {1\over24}\epsilon^{\mu\nu\rho\sigma}\gamma_\mu
\gamma_\nu\gamma_\rho\gamma_\sigma{\pp_\lambda y\over2x}
\label{defell}\eeq
is the ``vector'' axion connection.  Note that although the gauge
representation is real, we have, instead of \myref{chifs0},
\bea 
\(G^+_{\mu\nu}\)^{a}_b &=& - K_{c b}\(G^-_{\mu\nu}\)^c_d K^{d a}
+ 2\(Z_{\mu\nu} + \ell_{\mu\nu}\)\del^{a}_b,\nnn
\(D^+_{\mu}\)^{a}_b &=& - K_{c b}\(D^-_{\mu}\)^c_d K^{d a} +
K^{d a}\pp_\mu K_{d b} + 2\(Z_\mu + \ell_\mu\)\del^{a}_b,
\label{chifs02} \eea
where $\ell_{\mu\nu}$ is the field strength associated with the connection
$\ell_\mu$, and $K_{a b} =  x\del_{a b}$, so 
\beq K_{b c}(T_e)^c_d K^{d a} = (T_e)^b_a = (T^T_e)^a_b = - (T_e)^a_b.\eeq
We therefore identify $J^\pm$ in \myref{cacond1} as $J^\pm_\mu = \pm
i\Gamma_\mu + iT\cdot A_\mu$. The contribution from $\Gamma_\mu$
alone to \myref{cacond1} is included in the sum rule
\myref{alph3}. It remains to cancel the gaugino contribution
proportional to $\Gamma_\mu C_a$, where $C_a$ is the adjoint Casimir.
From \myref{phiparams} and \myref{defJ} we have the following
contribution from $\varphi^a,\hph^a,\tph^a$:
\beq C_a\[\sum_\gamma\eta_\gamma^\varphi\(\Gamma_\mu - \pp_\mu y/2x\)
+ \sum_\gamma\eta_\gamma^{\hph}\(\pp_\mu y/2x - \Gamma_\mu\)
- \sum_\gamma\eta_\gamma^{\tph}\Gamma_\mu\],\eeq
which cancels the gaugino contribution provided
\beq \sum_\gamma\eta_\gamma^\varphi = \sum_\gamma\eta_\gamma^{\hph} =
\sum_\gamma\eta_\gamma^{\tph} = 1,\label{etavarphi}\eeq
in accordance with the overall constraint  
\beq \sum_\gamma\eta_\gamma^\varphi + \sum_\gamma\eta_\gamma^{\hph} +
\sum_\gamma\eta_\gamma^{\tph} = 3,\label{etavarphi2}\eeq
needed for the cancellation of gauge and gaugino one-loop logarthmic
divergences that arise from their gauge couplings~\cite{pv1,pvdil}.

Additional constraints on the PV sector required to cancel the
remaining contributions to \myref{cacond1} from PV chiral
supermultiplets that regulate chiral matter loops are discussed in
\myapp{orbs}.


%% file: AppendixB.tex
\subsection{Chiral Anomalies}\label{chiralan}
\setcounter{equation}{0} 
\subsubsection{Chiral anomaly analysis in the regulated theory}\label{gen}
The chiral anomaly arises from the helicity odd part of the fermion
determinant 
\beq T_- = {1\over 2}\[\Tr\ln\cM(\gamma_5) -
\Tr\ln\cM(-\gamma_5)\],\qquad \cM = -i\notD + M,\label{tminus} \eeq 
in which the axial connection appears explicitly.  In
\cite{pv1}--\cite{pvdil}, explicit cancellation of ultraviolet
quadratic and logarithmic UV divergences in \myref{boseL1} was shown
to be possible by the introduction of the PV fields discussed in
\mysec{pvreg}.  In the previous appendix we found additional
constraints that insure cancellation of most linear divergences.
We defer to \myapp{chigauge} the discussion of residual linear divergences 
that are associated with the gauge-gravity-dilaton sector.

In this section we outline a check that, in the absence of 
linear divergences, the anomaly appears only
through the noncovariance of the PV masses, and show that the
``consistent anomaly'' of YM theory is recovered under appropriate
assumptions.  Under an infinitesimal chiral transformation 
\beq
\Theta\to e^{-i\gamma_5\phi}\Theta,\qquad\del\Theta =
-i\gamma_5\phi\Theta,\label{theta}\eeq 
where $\Theta =
(\psi_\mu,\lambda^a,L\chi^p + R\chi^{\bar p},\alpha)$ is a $2N + 4N_G
+ 10$ component Majorana spinor\footnote{The chiral fermion $\alpha$
is an auxiliary field used to implement gravitino gauge
fixing~\cite{us}.}, we have $\cM\to \cM + \delta\cM,\; \delta\Tr\ln\cM
= \Tr\cM^{-1}\delta\cM$, and \myref{tminus} is shifted by 
\beq
\delta T_- = {1\over 2}\[\Tr\cM^{-1}(\gamma_5)\delta\cM(\gamma_5) -
\Tr\cM^{-1}(-\gamma_5)\delta\cM(-\gamma_5)\].\label{deltminus} \eeq 
Using the methods
developed in Appendix A of~\cite{us2} the corresponding shift in the
Lagrangian can be cast in the form 
\bea \del\L &\ni& -{i\over 2}\del
T_- = -i\int{d^4p\over4(2\pi)^4}\[T(\gamma_5) -
T(-\gamma_5)\],\nonumber \\ T(\gamma_5) &\equiv& T = \Tr\sum_{\ell =
0}^{\infty}(-\R)^{\ell}\delta\R,\label{delLchi} \eea 
where now the
trace is over only Dirac indices and internal quantum numbers, and
Lorentz indices for the gravitino, and 
\bea \R &=& \pr\[p^2 -
T^{\mu\nu}\Delta_\mu\Delta_\nu - {i\over2}\hat\G\cdot\sigma + X +
P_{\mu\nu}\(p^\nu + \G^\nu\)\hM^\mu \],\ddd \Delta_\mu = p_\mu +
\G_\mu + \del_\mu\nonumber \\ 
\delta\R &=& \pr P_{\mu\nu}\(p^\nu + \G^\nu\)\widehat{N}^\mu
= \pr p_\mu N^\mu + O\({\pp\over\pp p}\).\label{RdR}\eea 
The operators in these expression are given in \myapp{cdexp} 
as covariant derivative expansions.  Specifically,
$T^{\mu\nu}, X, P_{\mu\nu},\del_\mu$ and $\G_\mu,\hat{\G}_{\mu\nu}$ 
are expansions in covariant derivatives of, respectively, the 
space-time curvature tensor\footnote{There 
is a sign error in the third and second terms,
respectively, of the expressions for $T_{\mu\nu}$  and $X$ in 
Eq. (A.19) of~\cite{us2}.} and the field 
strength $\G_{\mu\nu}$ acting on fermions:
\beq \G_{\mu\nu}^{\pm} = [D^{\pm}_\mu,D^{\pm}_\nu] = G^{\pm}_{\mu\nu}
+ Z_{\mu\nu} + iG_{\mu\nu}^L\gamma_5, \qquad Z_{\mu\nu} = -
{1\over4}r_{\rho\sigma\mu\nu}\gamma^\rho\gamma^\sigma,
\label{defZ}\eeq
where $+(-)$ refers to right(left)-handed fermions, and
$G_{\mu\nu}^{\pm}$, $G_{\mu\nu}^L$ are unit matrices in Dirac space.
$G^L$ contributes only through loops in the gaugino-gravitino-dilatino
sector.  In addition we defined
\beq N_\mu = -\pmatrix{R\gamma_\mu i\delta\notD^+ R& - 
R\gamma_\mu \delta ML\cr 
- L\gamma_\mu \delta \M R & L\gamma_\mu i\delta\notD^- L\cr}, \;\;\;\;
M_\mu = \pmatrix{ 0 & R\gamma_\mu ML\cr 
L\gamma_\mu\M R & 0 \cr},\label{NM} \eeq
and $T(-\gamma_5)$ is obtained from $T(\gamma_5)$ by the
substitutions $ (D^+,D^-,M,\M) \to (D^-,D^+,\M,M),$ with
\beq  M = M_0 + M_{\mu\nu}\sigma^{\mu\nu}.\label{Mmu} \eeq
The matrix $M_{\mu\nu}$ has nonvanishing elements only between the
gauginos $\lambda$ and either the dilatino $\chi^S$ or the auxiliary
field $\alpha$ introduced to fix the gravitino
gauge~\cite{us,us2}.  Under the chiral transformation \myref{theta},
\beq \delta D^{\pm}_\mu = \pm i\[D^{\pm}_\mu,\phi\],\qquad
\del\M = i\{\M,\phi\},\qquad \del M = -i\{M,\phi\}.\eeq
In order to evaluate the light quark contribution to the chiral
anomaly, we must resum the derivative expansion of~\cite{us2}.
This resummation can be expressed in terms of the action of the
operators in (\ref{tminus}) on a function of momentum:
\bea \hat{F}f(p) &=& f(p - iD)F, \qquad
 \G_\mu f(p) = \int_0^1d\lambda\lambda{\pp\over\pp p_\rho}f(p - i\lambda D)
\G_{\rho\mu},\nonumber \\
 T_{\mu\nu}f(p) &=& g_{\mu\nu} - 2\int_0^1d\lambda\lambda
(1-\lambda){\pp^2\over\pp p_\rho\pp p_\sigma}f(p - i\lambda D)
r_{\mu\rho\sigma\nu} + O(r^2),\label{resum1} \eea
and, by partial integration
\bea f(p)\hat{F} &=& f(p + iD)F, \qquad 
f(p)\G_\mu = -\int_0^1d\lambda\lambda{\pp\over\pp p_\rho} 
f(p + i\lambda D)\G_{\rho\mu}, \nonumber \\
 T_{\mu\nu}f(p) &=& g_{\mu\nu} - 2\int_0^1d\lambda\lambda
(1-\lambda){\pp^2\over\pp p_\rho\pp p_\sigma}f(p + i\lambda D)
r_{\mu\rho\sigma\nu} + O(r^2),\label{resum2}\eea
where $F$ is any local field operator.
 
First neglecting $G_L$ and $M_{\mu\nu}$, the anomaly is mass-independent
and arises from terms of order $p^{-6}$ in \myref{tminus}.  Writing
\beq T_n = \half\int{d^4p\over(2\pi)^4}
\Tr\lbr\[\R(\gamma_5)\]^n\del\R(\gamma_5)
- \[\R(-\gamma_5)\]^n\del\R(-\gamma_5)\rbr 
= \int{d^4p\over(2\pi)^4}t_n,\eeq
we have $T_0=0$, and we only need to evaluate $t_1\sim p^{-3}$ and
$t_2\sim p^{-5}$.  Since $T_2$ is finite we can evaluate it directly
using \myref{resum1} and \myref{resum2}.  This is a local operator
which is bilinear in the field strength $\G_{\mu\nu}$.  Since it is
independent of momenta we can set one external momentum to zero.  That
is, in the product $\G_{\rho\sigma}\G'_{\mu\nu}$ we first set
$D_\lambda\G_{\rho\sigma}=0$ and then $D'_\lambda\G'_{\mu\nu}=0$, and
take the average.  This trick considerably simplifies the calculation
and, using standard Feynman parameterization techniques and the
Bianchi identities, and dropping total derivatives:
\beq \Tr(\G\G D\phi) \to - \Tr[\phi D(\G\G)],\eeq
we find that the various contributions cancel:
\beq T_2 = 0.\eeq
Since $T_1$ has logarithmically divergent contributions, we must explicitly
cancel them against one another before using integration by parts or
shifts in the integration variable.  To this end we first rewrite
\beq \Tr(\G D\phi) \to - \Tr(\phi D\G)\label{drop}\eeq
and
\beq [D^\nu,{}^*\hG_{\mu\nu}] =
i[\hG^\nu,{}^*\hG_{\mu\nu}] = {i\over2}\epsilon_{\mu\nu\rho\sigma}
[\hG^\nu,\hG^{\rho\sigma}] , \eeq 
where 
\beq \hG_\mu f(p) = \int_0^1d\lambda {\pp\over\pp p_\rho}f(p - i\lambda D)
G_{\rho\mu} ,\qquad
 f(p)\hG_\mu = -\int_0^1 d \lambda{\pp\over\pp p_\rho} 
f(p + i\lambda D)G_{\rho\mu}.\label{dgcomm}\eeq
Using this result in the expression for $T_1$ that contains
a factor $\sigma\cdot\hG$ one finds the contribution 
\beq -{i\over2}\delta T_- = + {i\over2}T_1 = 
{1\over16\pi^2}\(G\cdot\tG\phi\).\label{GGphi}\eeq
In the regulated theory, we have to subtract a
contribution with $-p^2\to -(p^2 - m^2)$ in the denominators in
\myref{RdR}, and drop the terms that vanish in the limit
$m^2\to\infty$.  Using
\beq G^{\rho\nu}\tG'_{\mu\nu} = \half G\cdot\tG' g^\rho_\mu - 
\tG_{\mu\nu}G'^{\rho\nu},
\qquad D_\mu\tG^{\mu\nu} =0,\label{rels}\eeq
and assuming $[m^2,G]=0$ we get a contribution
\bea t_1(m^2) - t_1(0) &=& -8\Tr\int_0^1d\lambda\[{p_\rho p^\mu
\over(p^2-m^2)^3}G^{\rho\nu}\tG_{\mu\nu} +
{\tG^{\mu\nu}G_{\rho\nu}\over(p^2-m^2)^2}\({p_\mu p^\rho\over p^2-m^2}
- {g_\mu^\rho\over2}\)\]\phi\label{t1m2}\eee
-4\Tr\int_0^1d\lambda\[\({p^2
\over(p^2-m^2)^3} - {1\over(p^2-m^2)^2}\)G\cdot\tG\phi\]
\eee -4\Tr\int_0^1d\lambda\({m^2
\over(p^2-m^2)^3}G\cdot\tG\phi\),\nnn
\int_p\[t_1(m^2) - t_1(0)\] &=&   
+ {i\over8\pi^2}\Tr\(G\cdot\tG\phi\),
\qquad \int_p \equiv \int{d^4p\over(2\pi)^4}\eea
and we recover \myref{GGphi}
\bea -{i\over2}\delta T_- = +{i\over2}\int_p\[t_1(0) - t_1(m^2)\] = 
{1\over16\pi^2}\(G\cdot\tG\phi\) = {i\over2}\int_p t_1(0),\eea 
which is the standard result 
for the contribution from renormalizable gauge couplings with
$G_{\mu\nu}^{\pm}\to\mp iT\cdot F_{\mu\nu}$.  However 
there are additional contributions
from the PV sector when $[m^2,G]\ne0$, 
as we will see in \mysec{pvchiral}.

The space-time curvature term in $T_1$ has two contributions.
Those arising from $Z_\mu Z^\mu$ and 
$(T^{\mu\nu} -g^{\mu\nu})\{p_\mu,Z_\nu\}$
are UV finite and may be straightforwardly evaluated as described above.
After dropping a total derivative as in \myref{drop}, the remaining
contribution reduces to 
\beq {1\over p^2}\Tr\(\{p^\mu,[D^\nu,Z_\mu]\}{\notp\over p^2}
\gamma_\nu\gamma_5\phi\).\label{uone}\eeq
To evaluate this we adopt the
convention that all $p$-derivatives act to the left so that
\beq \[{\pp\over\pp p_\alpha},p^\beta\] = -g^{\alpha\beta},\eeq
and we define
\beq C_{\mu\nu} = [D_\mu,D_\nu] = [\nabla_\mu,\nabla_\nu] + Z_{\mu\nu}.
\label{cmunu}\eeq
Then writing
\beq \{p^\mu,Z_\mu\} = 2p^\mu Z_\mu + \[Z_\mu,p^\mu\],\eeq
the first term drops out of (\ref{uone}) because 
\beq {p^\mu\over p^2}Z_\mu = {2p^\mu p^\rho - p^2\del^{\mu\rho}\over p^4}
\sum_{m=0}\int_0^1d\lambda{\lambda\over m!}
\(-i\lambda D\cdot{\pp\over\pp 
p}\)^m Z_{\rho\mu} = 0\eeq
by symmetry, and we obtain
\bea \[Z_\mu,p^\mu\] &=& i\int_0^1d\lambda\lambda
D^\mu Z_{\mu}(\lambda) + \int_0^1d\lambda\lambda\int_0^\lambda d\eta\eta
\[C^\mu(\eta),Z_\mu(\lambda)\],\eea
 where if $G = Z,C$
\bea  G_\mu(\lambda)f(p) &=& {\pp\over\pp
p_\rho}f(p -i\lambda D)G_{\rho\mu}, \nnn f(p)G_\mu(\lambda) &=&
-{\pp\over\pp p_\rho}f(p -i\lambda D)
G_{\rho\mu}.\eea
Finally, using \myref{dgcomm} we obtain
\bea [D_\nu,[Z_\mu,p^\mu]] &=& 
-\int_0^1d\lambda\lambda^2\int_0^\lambda d\eta
\[C_\nu(\eta),D^\mu Z_\mu(\lambda)\] + i\int_0^1d\lambda\lambda
D_\nu D^\mu Z_{\mu}(\lambda)\ddd
+ \int_0^1d\lambda\lambda\int_0^\lambda d\eta\eta\(
\[C^\mu(\eta),D_\nu Z_\mu(\lambda)\]
+ \[D_\nu C^\mu(\eta),Z_\mu(\lambda)\]\).\eea
Evaluating this contribution by the above procedure and combining
all three contributions gives
\bea \delta T &=& - \int_p t_1 = 
{i\over192\pi^2}r\cdot\tr\Tr\phi,\qquad
- {i\over2}\delta T = {1\over384\pi^2}r\cdot\tr\Tr\phi\nnn
r\cdot\tr &=& \half \epsilon^{\mu\nu\rho\sigma}r_{\rho\sigma\tau\lambda}
r^{\tau\lambda}_{\;\;\;\;\mu\nu}.\label{totr}\eea
It is straightforward to check that there is no contribution to this term
when $p^2\to p^2 -m^2$ in the denominators of $\R,\del\R$ in \myref{RdR}.

In the following section we will see that the remaining terms of the
``consistent anomaly''~\cite{zum} will emerge when we include all the
contributions from the PV sector.

\subsubsection{Full PV contribution}\label{pvchiral}
In this section we consider PV fields with no spin-dependent mass
terms.  Then the only relevant mass is the PV mass $m$, and $\cM_{PV}$
is given by (A.7) of~\cite{us2} with $L=0,\;M= 0,\;m=m_{PV}$. To
evaluate \myref{tminus} we follow (A.13) of~\cite{us2}, but take $\cM_0 =
\cM_{PV}(-m,-\vec\sigma)$. Then we have
\bea \R_{PV} &=& {1\over
-p^2 + \bfm^2}\[p^2 - \bfm^2 - T^{\mu\nu}\Delta_\mu\Delta_\nu + 
\hat \bfm^2 + \hat h + X -
i\widehat{\notD \bfm} \], \nonumber \\ \delta\R_{PV} &=& {1\over -p^2 +
\bfm^2}\(p_\mu N^\mu + N\) + O\({\pp\over\pp p}\),\label{RdR2}\eea 
with $\delta M\to\delta m$ in $N_\mu$, defined in \myref{NM}, and 
\bea \notD \bfm &=& \pmatrix{ 0 & R[\notD^+ m -
m \notD^-] L\cr L [\notD^-\m - \m \notD^+]R & 0 \cr} \equiv
\pmatrix{ 0 & R\notD m L\cr L \notD\m R & 0 \cr}, \label{Dm} \\ N
&=& -\pmatrix{R m\delta\m R & -iR m\delta\notD^- L\cr
-iL\m\delta\notD^+ R & L \m\delta m L\cr}, \;\;\;\; \bfm^2 =
\pmatrix{R m\m R & 0 \cr 0 & L \m m L\cr}. \eea
First consider chiral fermion contributions to 
the space-time curvature term.  The only contribution
from the PV sector that is not suppressed by $m^{-2}$ comes from
the replacement $\displaystyle{\notp i\del\notD\phi\to\bfm\del\bfm}$
in the term involving $Z_\mu Z^\mu$, giving
\bea \del T^{PV} &=& - \int_p t_1^{PV} =
{r\cdot\tr\over192\pi^2}\Tr\eta{1\over4}\(\m^{-1}\del\m -
 m^{-1}\del m\),\nnn
- {i\over2}\del T^{PV} &=&
- {r\cdot\tr\over384\pi^2}\Tr\eta{i\over4}\(\m^{-1}\del\m -
 m^{-1}\del m\),\eea
and the total contribution from light and heavy modes is
\beq - {i\over2}\(\del T + \del T^{PV}\) = -
     {r\cdot\tr\over16\pi^2}{1\over24}\Tr\[\half\(\phi^+ - \phi^-\) +
     {i\over4}\eta\(\m^{-1}\del\m -  m^{-1}\del m\)\],\eeq
where for a chiral transformation defined by \myref{theta} $\phi^+ = 
- \phi^-= -\phi$.  If the PV masses were modular covariant they 
would satisfy
\beq \del\m = i(\phi^-\m - \m\phi^+)\to i\{\m,\phi\},\qquad
\del m = i(\phi^+ m - m\phi^-)\to -i\{ m,\phi\},\eeq
and the regulated theory would be anomaly free, provided \myref{Trphi}
is satisfied.  If we define 
\beq d\m = \del\m - i\{\m,{\phi}\}, 
\qquad d m = \del m + i\{m,{\phi}\}, ,\eeq
the total contribution from light and heavy modes reads, setting $\eta=-1$,
\beq - {i\over2}\(\del T + \del T^{PV}\) = -
     {r\cdot\tr\over16\pi^2}{1\over24}\Tr\eta
     {i\over4}\(\m^{-1}d\m -  m^{-1}d m\).\label{deltZ}\eeq
In other words, we may write
\bea -{i\over2}\(\del T + \del T^{P V}\) &=& 
- {i\over2}\(\del T + \del T^{P V}_{m =0}\) -
{i\over2}\(\del T^{P V} - \del T^{P V}_{m =0}\)
\eee - {i\over2}\(\del T + \del T^{P V}_{m =0}\)
-{r\cdot\tr\over16\pi^2}{1\over24}\Tr\eta
     {i\over4}\(\m^{-1}d\m -  m^{-1}d m\)\label{deltZ2}.\eea
The first term on the right hand side of \myref{deltZ2} vanishes by
virtue of \myref{Trphi}, which is consistent with the absence of
linear divergences associated with the spin connection.

To evaluate the remaining contributions we have to take into account
that $m$ is {\it a priori} matrix-valued and field dependent.  Then
(\ref{t1m2}) is replaced by
\bea \tau_1 &=& - 
2\Tr\eta\({m^2\over(p^2 - m^2)^2}G^{\mu\nu}\tG_{\mu\nu}{1\over
p^2-m^2} + {1\over
p^2-m^2}\tG_{\mu\nu}G^{\mu\nu}{m^2\over(p^2 - m^2)^2}\)\phi,\nnn
\int_p\tau_1(p^2) &=& - 
{i\over16\pi^2}\int_0^\infty dp^2 p^2\tau_1(-p^2)\eee
{i\over8\pi^2}\Tr\eta\int_0^\infty dp^2\(\lbr G\cdot\tG,{m^2\over(p^2 +
  m^2)^2} \rbr - {m^2\over p^2 + m^2}\lbr G\cdot\tG,{1\over p^2 +
  m^2}\rbr {m^2\over p^2 + m^2}\)\phi\eee
{i\over8\pi^2}\Tr\eta\[2G\cdot\tG + \int_0^\infty dp^2{\pp\over\pp p^2}
\({m^2\over p^2 + m^2}G\cdot\tG {m^2\over p^2 + m^2}\)\]\phi
\eee {i\over8\pi^2}\Tr\eta\(G\cdot\tG\phi\),\eea
where here and below we use a shorthand notation with $\phi = -\phi^+ =
\phi^-$, and, for example, 
\beq \Tr\eta\(G\cdot\tG\phi\)\equiv \half\Tr\eta\[\(G^+\cdot\tG^+ + 
G^-\cdot\tG^-\)\phi\].\eeq
That is, we average over helicities with the convention that 
\beq O\phi \equiv -\half\(O^+\phi^+ - O^-\phi^-\) = \half
\(O^+ + O^-\)\phi.\eeq
If $\del m\ne0$, there is another contribution with
$\displaystyle{\notp i\del\notD\phi}\to m\del m$:
\bea \sigma'_2 &=& {1\over4}\Tr\eta\({\sigma^{\mu\nu}\over
p^2-m^2}G_{\mu\nu}\)^2{1\over p^2-m^2}m\del m\gamma_5
\eee
 2i\Tr\eta{1\over p^2-m^2}\tG_{\mu\nu}{1\over p^2-m^2} G^{\mu\nu}{1\over
p^2-m^2}m d m\ddd - 2\Tr\eta\lbr m,{1\over p^2-m^2}\tG_{\mu\nu}{1\over
p^2-m^2} G^{\mu\nu}{1\over p^2-m^2}\rbr m\phi.\eea
These are the only contributions if $[D_\mu,m]= 0$; in this case
$[G,m]=0$, and if also $dm=0$ this reduces to
\bea T_2^{PV} &\to& \int_p\sigma'_2 \to \int_p\tau_1 = T_1^{PV},
\qquad T_1 - T_2 \to 0, \eea
as expected.  The remaining contributions involve covariant
derivatives on the mass matrix. Writing
\beq T^{PV}_n = -\half\int_p\Tr\eta\lbr\[\R(\gamma_5)\]^n\del\R(\gamma_5)
- \[\R(-\gamma_5)\]^n\del\R(-\gamma_5)\rbr_{PV} = \int_p\(\tau_n +
\sigma_n\),\eeq
where $\tau_n$ and $\sigma_n$ are the contributions without and with,
respectively, a factor $\del m$.  They take the general form
\bea \tau_n &=& \Tr\eta\(O^\tau_n\notD\phi\) \to 
-\Tr\eta\(\[\notD,O^\tau_n\]\phi\),\nnn
\sigma_n &=& \Tr\eta\(O^\sigma_n\del m\) = \Tr\eta\(O^\sigma_n d m\) +
i\Tr\eta\(\{m,O^\sigma_n\}\phi\).\eea
Using relations such as 
\bea \notD^2m &=& D^2m - {i\over2}[\sigma\cdot G,m],\nnn
 {[{\notD,m^2}]} &=& \{m,\notD m\},\qquad [\notD,P] = \{m,P[\notD,m]P\},\nnn
P\ppd D m^2P &=& \ppd[D,P] + 2P[p\cdot D,P] = [D,P]\dpp -
2[p\cdot D,P]P,\nnn
\{m,P[G,m]P\} &=& [G,P],\qquad P = {1\over p^2 - m^2},\label{dp}\eea
it is possible to show that 
\beq \sum_n\tau_n + \sum_n\sigma_n = \sum_n\Tr\eta\(O^\sigma_n d m\).\eeq
As a check of our anomaly calculation, consider
the case with constant PV masses such that $[D,m^2] = [G,m^2] = \del m
= 0$.  Then $\sigma_i=0$ and
\bea \tau_2 &=& -\Tr\eta\[P i\widehat{\notD
m}\]^2P\notp[\notD\phi]\gamma_5 + {1\over2}\Tr\eta\lbr
P[\notD,m],P\sigma\cdot G\rbr P m[\notD,\phi]\gamma_5\eee
-2\Tr\eta{P^3 m^2}P[\tG^{\mu\nu},m] [G_{\mu\nu},m]\phi
- 2\Tr\eta P^3\lbr[\tG^{\mu\nu},m], G_{\mu\nu}\rbr m\phi\ddd
- 2\Tr\eta p^2P^4[\tG_{\mu\nu},[D^\mu,m]][D^\nu,m]\phi
+ 4\Tr\eta P^2\lbr[D^\mu,m],\tG_{\mu\nu}\rbr [D^\nu,m]\phi\ddd
+ 2\Tr\eta P^3[D^\mu,m][G^{\mu\nu},[D_\nu,m]]\( p^2P - 2\)
\phi,\nnn
\tau_3 &=& \Tr\eta\[P i\notD m\]^3Pm[\notD,\phi]\gamma_5 \eee
- 4\Tr\eta P^4\{[\tG^{\mu\nu}, m],D_\mu m D_\nu m\}
 m\phi
+ 4\Tr\eta P^4D_\mu m[\tG^{\mu\nu},m]D_\nu m m\phi\ddd
+ 4\epsilon^{\mu\nu\rho\sigma}\Tr\eta P^4D_\mu m D_\nu m D_\rho D_\sigma
 m\phi.\eea
The covariant derivative on chiral fermions is given in \myref{defdG}
with
\beq J^{\pm}_\mu = \Gamma^{\pm}_\mu \mp
i\Gamma_\mu \mp iT^{\pm}\cdot A_\mu,\qquad T^+ = (T^-)^T = (T^-)^*,
\label{defJ0}\eeq
where $A_\mu$ is a gauge field and  
\beq\Gamma_\mu = {i\over4}\(K_i\D_\mu z^i - K_{\m}\D_\mu\z^{\m}\), \qquad
\(\Gamma^-_\mu\)^p_q = \Gamma^p_{qk}\D_\mu z^k =
\[\(\Gamma^+_\mu\)^{\bar p}_{\bar q}\]^{\dag}.\label{defJ}\eeq
are the K\"ahler $U(1)$ and reparameterization connections,
respectively. It follows from gauge invariance of the K\"ahler potential
that 
\beq K_{i\m}(D^+_\mu)^{\m}_{\n}K^{\n j} = (\pp_\mu + Z_\mu)\del^j_i
- (J^-_\mu)^j_i.\eeq
For chiral fermions the field strength
$G_{\mu\nu}$ can be expressed in terms of the general two-forms
$T_{\mu\nu}$ defined\footnote{The two-forms $T_{\mu\nu}$ used here and
below should not confused with the tensor $T^{\mu\nu}$ introduced in
the derivative expansion in \myref{RdR} and defined
in \myref{Tmunu}.} in \myref{E2}; explicitly:
\bea \(G^-_{\mu\nu}\)^p_q &=& {1\over2}X_{\mu\nu}\delta^p_q +
iF^a_{\mu\nu}(T_a)_q^p - \Gamma^p_{q\mu\nu}, \nonumber \\
\(G^+_{\mu\nu}\)^{\n}_{\m} &=& - K_{p\m}K^{q\n}\(G^-_{\mu\nu}\)^p_q,
\quad X_{\mu\nu} = K_{\mu\nu}.\label{chifs} \eea 
For constant masses we have
\beq [D_\mu,m] = [J_\mu,m] = [a_\mu,m],\qquad
J_\mu = j_\mu + a_\mu = \pmatrix{J_\mu^+&0\cr0&J_\mu^-},
\label{defja0}\eeq 
where $a_\mu$ is the connection associated with
the anomalous symmetry:
\beq  [j_\mu,m] = \{a_\mu,m\} = 0, \qquad
G_{\mu\nu} = g_{\mu\nu} + a_{\mu\nu},\qquad [g_{\mu\nu},m] =
\{a_{\mu\nu},m\} = 0.\label{defja}\eeq
For the helicity components of these matrices, this gives 
\beq m a_{\mu\nu}^- = - a^+_{\mu\nu}m,\qquad 
a_{\mu\nu}^-\m = - \m a^+_{\mu\nu},\qquad m g^-_{\mu\nu} =
g^+_{\mu\nu}m,\qquad g_{\mu\nu}^-\m = \m
g_{\mu\nu}^+,\label{defja2}\eeq
and we obtain in this case
\bea T_2 &=&  {i\over12\pi^2}\Tr\eta\(\tilde a^{\mu\nu}a_{\mu\nu}\phi\)
\ddd + {i\over3\pi^2}\Tr\eta\[\(\{\tilde g^{\mu\nu},
a_\mu a_\nu\} + a_\mu{\tilde g}^{\mu\nu}a_\nu +
\{\tilde a^{\mu\nu},a_\mu a_\nu\} - a_\mu{\tilde a}^{\mu\nu}a_\nu\)\phi\]\nnn
T_3 &=&  {i\over3\pi^2}\Tr\eta\[\(\{\tilde a^{\mu\nu},a_\mu a_\nu\} -
a_\mu{\tilde a}^{\mu\nu}a_\nu\)\phi\]\ddd +
{2i\over3\pi^2}\epsilon^{\mu\nu\rho\sigma}\Tr\eta\( a_\mu a_\nu a_\rho
a_\sigma\phi\).
\eea
Terms linear and cubic in $a$ drop out of 
$\del T = \sum_{n=1}^3(-1)^nT_n$,
and, provided there is a PV particle with negative signature, $\eta=-1$,
for every light particle, we recover the standard~\cite{zum} result:
\bea -{i\over2}\delta T &=&  {1\over16\pi^2}\Tr\phi\(g\cdot{\tilde g} + 
{1\over3}a\cdot{\tilde a} + {r\cdot\tr\over24}\right.  \ddd \left. -
{8\over3}\[\{{\tilde g}_{\mu\nu},a^\mu a^\nu\} +
a^\mu\tilde g_{\mu\nu}a^\nu\] + {16\over3}\epsilon_{\mu\nu\rho\sigma}
a^\mu a^\nu a^\rho a^\sigma\).\label{chiral}\eea
\subsubsection{Gauge, gravity and dilaton sector: 
nonrenormalizable operators}\label{chigauge}  

In the preceeding subsections the cancellation of linear divergences
was implicitly assumed, that is, we assumed
\beq \del T = - \del T^{P
V}_{m=0},\qquad\Tr(\phi G\cdot\tG)_{\rm light} = - \Tr(\eta\phi
G\cdot\tG)_{\rm P V},\eeq 
However contributions from nonrenormalizable interactions to the
gaugino and gravitino connections, including an additional curvature
term in the gravitino connection and~\cite{us2} an off-diagonal
gravitino-gaugino connection, do not have counterparts in the PV
sector.  These have to be treated separately.  The contribution
from the gravitino field strength:
\beq \Tr\(g\cdot\tg\)_\psi \ni - r\tr\eeq
leads to the contribution \myref{affan} to the chiral anomaly.
There is a an additional contribution from the off-diagonal
gaugino-gravitino connection; 
the corresponding field strength $g_{\mu\nu}$
has terms linear and quadratic in the Yang-Mills field $F_{\mu\nu}$.
The terms quartic in $F_{\mu\nu}$ cancel between gaugino
and gravitino loops:
\beq g\cdot\tg\ni \pm{x^2\over2}\epsilon_{\mu\nu\rho\sigma}
F_a^{\alpha\rho}F^{a\;\;\;\nu}_{\beta}F_{b\alpha}^{\;\;\;\;\sigma}
F^{b\beta\mu} \quad {\rm for}\quad \cases{\lambda\cr\psi\cr}.
\eeq
The remaining contribution takes the form
\bea \Tr\(g\cdot\tg\)_{\lambda +\psi} &\ni&
4\[\D^\mu\(\sqrt{x}F_a^{\rho\nu}\)\D_\rho\(\sqrt{x}\tF_{\mu\nu}^a\)
- \D_\rho\(\sqrt{x}F_a^{\rho\nu}\)\D^\mu\(\sqrt{x}\tF_{\mu\nu}^a\)\]
 \nonumber \\ & & +
2x\tF_a^{\mu\nu}F^a_{\rho\sigma}r_{\mu\nu}^{\;\;\;\;\rho\sigma} - x r
F^a_{\mu\nu}\tF_a^{\mu\nu} - 4x
c_{abc}\tF^c_{\mu\nu}F^{a\rho\mu}F^{b\;\;\nu}_\rho\eee
4\D^\mu\(xF^{\rho\nu}\D_\rho\tF_{\mu\nu}\).\label{g+G}\eea
The second expression on the right hand side of \myref{g+G} is obtained 
from the first by using the identities \myref{rels} and~\cite{us2}
\beq \tF^a_{\mu\nu}[\D^\mu,\D_\rho]F_a^{\rho\nu} =
- F^a_{\mu\nu}[\D^\mu,\D_\rho]\tF_a^{\rho\nu} =
c_{abc}\tF^a_{\mu\nu}F^{b\mu\rho}F^{c\nu}_{\;\;\;\;\rho}
+ r^\mu_\nu\tF^a_{\mu\rho}F_a^{\nu\rho} - {1\over2}
r_{\mu\nu}^{\;\;\;\;\;\rho\sigma}\tF_a^{\mu\nu}F^a_{\rho\sigma},\eeq
The contribution from \myref{g+G} to the chiral anomaly is given
in \myref{offan}.  As described in \mysec{strat}, these anomalies arise
from uncanceled linear divergences that have counterparts
in uncanceled logarithmic divergences which are total divergences, and so
do not affect the finiteness of the S-matrix, but the resulting
anomalies form superfields provided the the cut-off has the form
\myref{uvcut}.

The field strength $G^L$ in \myref{defZ} arises from a term in the
gaugino connection, $i\gamma_5L_\mu = -i\gamma_5\pp_\mu y/2x$, that,
as was shown in~\cite{pv1}, must be defined as a ``vector'' (rather
than an ``axial vector'') connection, through the use of the identity
\beq \gamma_5 = (i/24)\epsilon^{\mu\nu\rho\sigma}\gamma_\mu
\gamma_\nu\gamma_\rho\gamma_\sigma, \eeq
in order to allow Pauli-Villars regularization of the quadratic
divergences.  This choice further insures that the
nonrenormalization~\cite{russ} of the topological charge, $\theta =
8\pi^2y$, is consistent with linear-chiral multiplet
duality~\cite{frad} for the dilaton supermultiplet, and
preserves~\cite{tom} the modified linearity condition in the linear
multiplet formulation.  Since BRST invariance requires (and
supersymmetry allows) the regulation of nonabelian multiplet gauge
loops by PV chiral multiplets, it is clear that only the chiral
multiplet axial connections can appear in the anomalies associated
with the regulated pure Yang-Mills sector.  As a result,
contributions from $\del L_\mu$ are absent\footnote{In Feynman diagram
language the gaugino-loop contribution to the $L_\mu$ anomaly arising from
a shift in the axion $y$ is canceled by
a gauge vector loop contribution~\cite{shif}.} from $\del\R$.  The operator $T_-$ is
defined in such a way that the contribution to the anomaly involving
only $G^L$, and $G$ is proportional to
\bea &&{1\over2}\Tr\lbr\[\([G^+ + i\gamma^5G^L]\cdot\sigma\)^2 + \([G^- +
i\gamma^5G^L]\cdot\sigma\)^2\]\gamma_5\phi\rbr\eee
\Tr\phi\lbr\[(g\cdot\sigma)^2 + (a\cdot\sigma)^2 -
(G^L\cdot\sigma)^2
\]\gamma_5 + i\{(g\cdot\sigma),(G^L\cdot\sigma)\}\rbr.\label{TrGg}\eea
From the expressions for the field strengths given in Eq. (C.18)
of~\cite{us2}, it is easy to see that the traces involving $G^L$
vanish identically in the above expression.  Those involving only 
$G$ have already been taken into account.  In particular,
the dilatino $\chi^S$, gauginos $\lambda$ and gravitino $\psi$
transform only under modular transformations:
\beq \chi^S\to e^{{i\over2}\im F}\chi^S, \quad \lambda\to
e^{-{i\over2}\im F} \lambda, \quad \psi\to e^{-{i\over2}\im
F}\psi.\label{chik}\eeq
 To fix the gravitino gauge, we follow \cite{us,us2} and introduce an
auxiliary field $\alpha$ that transforms like a chiral fermion:
$\alpha\to e^{{i\over2}\im F}\alpha$.  With this procedure, the
ghostino makes no contribution to the chiral anomaly, but we must
include the contribution of the auxiliary field.  

The fields in this sector are \ux-neutral, and $\lambda$ transforms
under modular transformations with the opposite phase from the phase
of $\chi^S,\alpha$.  Therefore the masses \myref{Mmu} are modular
invariant, and terms quadratic or quartic in $M\cdot\sigma$ and linear
in $\phi$ cannot contribute to (\ref{tminus}); the only nonvanishing
contributions involve the additional, modular invariant, connections
$i\gamma_5 L_\mu$ and $(\Gamma^\pm)^S_S$. 
The $\lambda$-$\alpha$,$\lambda$-$\chi_S$ matrix
elements \myref{Mmu} satisfy~\cite{us2}
\bea M_0 &=& M^T_0, \qquad  M_{\mu\nu} = - M_{\mu\nu}^T, \qquad
M^0_{\lambda\alpha}  = - \bM^0_{\lambda\alpha},\qquad
M_{\lambda\alpha}^{\mu\nu} = \bM_{\lambda\alpha}^{\mu\nu}\nnn
0 &=& M_{\lambda\chi}^{\mu\nu}\M^{\lambda\chi}_{\mu\nu}, \qquad
\tM_{\lambda\chi}^{\mu\nu} = iM_{\lambda\chi}^{\mu\nu},\qquad
{\tilde{\bM}}_{\lambda\chi}^{\mu\nu} = - i\bM_{\lambda\chi}^{\mu\nu},\qquad
M^0_{\lambda\chi}  = - \bM^0_{\lambda\chi}.
\label{mrels}\eea
The contributions to the anomaly that may arise from these masses can
be determined by identifying local operators that could be obtained
from the terms of order $p^{-6}$ in the expansion \myref{tminus}. The 
masses in \myref{mrels} are gauge invariant and modular covariant: 
\beq \del\M = i\(\M\phi + \phi\M\) = 0\eeq
because $\phi_\chi = \phi_\alpha = - \phi_\lambda = \half\im F$.
Then since only even powers of $M$ can occur, the {\it a priori}
possibilities are $T_{2,3,4}$  with 
\bea T_i &=& \half\int{d^4p\over(2\pi)^4}\Tr\(t_i\hc\),\nnn
t_2 &=& - {1\over p^2}\(\notp + \notG\)\hM{1\over p^2}\(\notp + \notG\)
\widehat{\M}{1\over\notp}\notD\phi\gamma_5 \equiv t_{\infty} + t'_2,
\qquad t_{\infty} = - {1\over p^2}\notp\hM{1\over p^2}\notp
\widehat{\M}{1\over\notp}\notD\phi\gamma_5,\nnn
t_3 &=&  - \[{1\over p^2}\(\{\G_\mu,p^\mu\} + {i\over2}\hat\G\cdot\sigma\)
{1\over\notp}\hM{1\over\notp}\widehat{\M}
+ {\rm permutations}\]{1\over\notp}\notD\phi\gamma_5,\nnn
t_4 &=& - \({1\over\notp}\hM{1\over\notp}\widehat{\M}\)^2
{1\over\notp}\notD\phi\gamma_5.\label{p5}\eea
Because $M_{a\alpha(\chi)}^{\mu\nu} = c_{\alpha(\chi)}F_a^{\mu\nu}$,
no Lorentz and gauge invariant operator can be constructed from
$M_{\mu\nu}^2G_{\rho\sigma}\phi$ or from three factors of $M_{\mu\nu}$
in the trace. The only invariant involving the space-time Riemann
tensor has two factors of $M^{\mu\nu}$ and vanishes due to \myref{mrels}.
Then from power counting in $p_\mu\sim D_\mu$, we 
can get the following local operators:
\bea T_4 &\propto&\Tr\(M\M M\M\hc\)\phi\gamma_5 = 0,\nnn
T_3 &\sim& t'_2\sim\Tr\(\G^+_{\mu\nu}M^{\mu\nu}\M_0 + \G^-_{\mu\nu}
\M^{\mu\nu}M_0 + {\rm permutations}\hc\)\phi
%
.\label{tops}\eea
Terms even $M_{\mu\nu}$ cancel between $\lambda$ and $\alpha,\chi_S$.
For the odd terms, from \myref{mrels} we have 
\beq M_{\lambda\alpha}^{\mu\nu}\M_0^{\alpha\lambda} +
\M_{\lambda\alpha}^{\mu\nu}M_0^{\alpha\lambda} = 0,\eeq
and
\beq \(M^{\mu\nu}\M_0 + M_0\M^{\mu\nu}\)_{\lambda\lambda}\hc =
\(M^{\mu\nu}\M_0 + M_0\M^{\mu\nu}\)_{\chi\chi}\hc = 0.\label{Mm=0}\eeq
It follows that $T_4$ vanishes, as does any term with $\G_{\mu\nu}$
real.  A part that is not real is $X_{\mu\nu}$ which cancels between
$\lambda$ and $\alpha,\chi_S$.  There are also terms with
$G_\lambda^{\pm}\to i\gamma_5G^L$, $G^\pm_\chi\to \mp i G^L,$ where
\beq G^L_{\mu\nu} = {1\over2x}\(\pp_\mu x\pp_\nu y - 
\pp_\nu x\pp_\mu y\).\eeq
Since for these contributions
\beq G_\chi^+R = - G_\lambda^+R,\quad G_\chi^-L = - G_\lambda^-L,\eeq
if we keep explicit the helicity projection operators in the expressions
\myref{RdR}, it becomes clear that these also cancel.  The divergent
piece $t_{\infty}$ in \myref{p5} is proportional to a total derivative, 
as we will explicitly display below.\footnote{See \myref{MM0td}
for $m^2=0$; \myref{TMT2} vanishes identically in this case.}

Next consider the PV sector.  The $\lambda$-$\theta$ PV mass that
is generated by the supersymmetric Higgs mechanism, satisfies
\beq \del m = 0,\qquad \phi_{\lambda} = - \phi_{\chi} 
= - {1\over2}\im F,\qquad 
dm = - i\{\phi,m\} = 0,\label{phi0}\eeq
and there are no linear divergences or residual
anomalies generated by $\lambda$,$\theta$ loops. The divergent
contribution $t_{\infty}$ is regulated by the coupling of $\lambda_0$
to $\hph$ defined by \myref{kin1}, giving rise to masses $M_{\lambda^0\hph^a_-}$
of the form \myref{Mmu} which have the same properties \myref{mrels}
as $M_{\lambda\chi}$.  $N_\mu$ is still defined as in \myref{RdR2} since
$\del M = 0$, but $R_{PV}$ contains an additional term
\beq \lbr\hat\bfm + \(p_\nu + \G_\nu\)P^{\mu\nu}\gamma_\mu\rbr\hM\eeq
inside the square brackets. Under a chiral modular transformation the 
phases satisfy \myref{phi0} provided the PV masses $m$ are constant.  
Since $M\bfm M=\del m =0$, for constant masses the only potentially 
nonvanishing contribution analogous to \myref{p5} is
\bea t_2 &=& -{1\over p^2 - m^2}\(\notp\widehat{\M}{\notp\over p^2 - m^2}
\hM\notp + \m{\hM}{\notp\over p^2 - m^2}\widehat{\M}m\)
{1\over p^2 - m^2}\notD\phi\gamma_5.\label{p5PV}\eea
Terms in \myref{p5PV} with no factor of $\sigma\cdot M$ vanish
identically; evaluation of the terms with one of $\sigma\cdot M$ gives
\bea T_2 &=& 4\int{d^4p\over(2\pi)^4}{p^2\over(p^2 - m^2)}\[
i\Tr\eta\(D_\mu\tM^{\mu\nu}{1\over p^2 - m^2}\M_0 
+ \tM^{\mu\nu}{1\over p^2 - m^2}D_\mu\M_0 \right.\right.\mmm \l\l 
- D_\mu M_0{1\over p^2 - m^2}\tbM^{\mu\nu} 
- M_0{1\over p^2 - m^2}D_\mu\tbM^{\mu\nu}\)\hc\]\D_\nu\im F,
\label{TPV2}\eea
%
\bea T_2 &\ni& 4\int{d^4p\over(2\pi)^4}{p^2\over p^2 - m^2}D_\mu\[
i\Tr\eta\(\tM^{\mu\nu}{1\over p^2 - m^2}\M_0 - M_0{1\over p^2 -
m^2}\tbM^{\mu\nu}\)\hc\]\D_\nu\im F\eee {\rm total\; derivative},
\label{MM0td}\eea
since 
\beq 2D_\nu D_\mu\(M^{\mu\nu}\bM_0\) =
\[G^+_{\nu\mu},M^{\mu\nu}\M_0\] = 0.\eeq 
Note that the coefficient
of the divergent integral in \myref{TPV2} is just the variation of the
first line of the expression for Tr$\R^2\R_5$ in (B.29) of~\cite{us2}.
The condition on $\sum\eta e^2$ in \myref{cond0} assures that the
divergence is canceled.  Finally, taking into account the symmetry
properties \myref{mrels}, terms containing two factors of $\sigma\cdot
M$ vanish in the limit of equal PV masses: $m^2_{\lambda_0\theta_0}
= m^2_{\chi\hat\chi}$, and we get a contribution
\beq -{i\over2}T_2 = {\im F\over16\pi^2}f(r)
\D^\mu\(xF^{\rho\nu}\D_\rho\tF_{\mu\nu}\),
\qquad r = m^2_{\lambda_0\theta_0}/m^2_{\chi\hat\chi},\qquad f(1) = 0.
\label{TMT2}\eeq
The function $f(r)$ takes all possible values $-\infty \le f(r)\le
\infty$ over the range $0\le r\le \infty$; therefore the PV mass ratio
can be chosen so that \myref{TMT2} exactly cancels the contribution
\myref{g+G} from the off-diagonal gaugino-gravitino connection.  Since
the PV regularization procedure respects supersymmetry, the conformal
anomaly necessarily includes the supersymmetric completion,
proportional to \myref{mixan}, of \myref{TMT2}, and the contribution
\myref{LMx} can thus be cancelled.  Therefore in the following section
we include only PV fields that have noninvariant PV masses through
superpotential couplings to one another.

%% file: AppendixC.tex
\subsection{The full anomaly}\label{fullan}
\setcounter{equation}{0} 

In this appendix we calculate the full contribution to the anomaly
that is generated by superpotential couplings of PV chiral superfields
resulting in noninvariant masses.

The one-loop effective action from chiral multiplet loops is given by
\beq S_1 = {i\over 2}\Tr\eta\ln\(\hD_\Phi^2 + H_\Phi(\bfm)\) 
-{i\over 2}\Tr\eta\ln\[-i\notD_\Theta + M_\Theta(\bfm)\],\eeq
where the subscripts $\Phi$ and $\Theta$ denote scalar and fermion
loops, respectively.  We require that under an infinitesimal
transformation on superfields $Z(\theta)\to Z'(\theta') =
g(\theta)Z(\theta)$, such that $\delta\Phi = \phi_\Phi\Phi,\;
\delta\Theta = \phi_\Theta\Theta,$
\beq \delta S_1 = S_1(\bfm + d\bfm) - S_1(\bfm) = {i\over2}\STr\eta [1
+ (1 + \R)^{-1}d\R] = {i\over2}\STr\eta\sum_{n=0}^4(-\R)^nd\R,
\label{PVsum}\eeq
where
\bea d\bfm_\Theta &=& e^{-\phi_\Theta}\bfm(z',\vx')e^{\phi_\Theta} -
\bfm(z,\vx) = \delta\bfm - [\phi_\Theta,m] = \tilde\bfm - \bfm
\nonumber \\ &=& \pmatrix{0&(\delta m - \phi_\Theta^+m +
m\phi_\Theta^-)R\cr(\delta\m - \phi_\Theta^-\m +
\m\phi_\Theta^+)L&0\cr} = \pmatrix{0&dm L\cr d\m R&0\cr}\\
d\bfm_\Phi &=& e^{-\phi_\Phi}\bfm(z',\vx')e^{\phi_\Phi} -
\bfm(z,\vx) = \delta\bfm - [\phi_\Phi,m]
\nonumber \\ &=& \pmatrix{0&(\delta m - \phi_\Phi^+m +
m\phi_\Phi^-)\cr(\delta\m - \phi_\Phi^-\m +
\m\phi_\Phi^+)&0\cr}. \eea
Here we define, using the notation introduced in \myref{RdR} and \myref{RdR2}, with
now $\displaystyle{\del\notD\to0,\; \del m\to d m}$,
\bea \STr Fd\R &=& \Tr(Fd\R)_\Phi - \Tr(Fd\R)_\Theta,\label{StrFdR} \\
\R_{\Phi,\Theta} &=& {1\over p^2 -
\bfm^2}\[T^{\mu\nu}\Delta_\mu\Delta_\nu - \hat H - X - p^2 +
\bfm^2\]_{\Phi,\Theta}, \\ d\R_\Theta &=& - {1\over p^2 -
\bfm^2}\[-\notp + M_\Theta(\bfm)\]d\bfm, \\ d\R_{\Phi} &=& - {1\over p^2 -
\bfm^2}\[e^{-\phi_\Phi}H_\Phi(z',\vx') e^{\phi_{\Phi}} -
H_\Phi(z,\vx)\] .\label{defs1} \eea
We argued in \mysec{strat} that fields $\Phi^P$ with couplings in
$W_2$, or with K\"ahler curvature terms $R_{P\m Q\n}\ne0$, must have
modular and \ux\, invariant masses. Then $M_\Theta(\bfm) = \bfm,$ and
$\hat H,\G_\mu$ are given by the derivative expansions of Appendix A
of \cite{us2} in terms of the following operators:\footnote{In the
interest of simplification we have evaluated the component Lagrangian
in WZ gauge for $\vx$: $\l\vx\r = F_X = 0$, since the regulated
theory, including counterterms, must be \ux\ gauge invariant.  However
we have left explicit the $F_X$ term in \myref{defh} because it gives
a contribution to $d\bfm_\Phi$ through $F_X\to F_\Lambda = -
{1\over4}\D^2\Lambda$ if $\vx\to\Lambda + \bar\Lambda$.  This
term arises from mixed $F^P F_X$ terms in the Lagrangian when
K\"ahler potential terms take the form
$e^{q^P_X\vx}f(Z,\Z)|\Phi^P|^2$.}
\bea H_\Phi &=& \pmatrix{\h + \Delta\H & h'\cr \h' & h + \Delta H
\cr}, \qquad H_\Theta = \pmatrix{\h R & 0\cr0 &hL\cr} -
{i\over2}\sigma\cdot \G_\Theta - i[\notD,\bfm],\label{defH}  \\ (\Delta
H)^Q_P &=& \delta^Q_P\(\hV + M^2\) + R^Q_{P\m k} \(e^{-K}\A^kA^{\m} +
\D_\mu z^k\D^\mu \z^{\m}\) + {1\over x}\D_aD_P(T^az)^Q,\label{defDelH}\\
h &=& \m m, \qquad m_{PQ'} = \delta_{PQ}e^{K/2}\mu_P,
\\ h'_{PQ} &=& e^{-K}\(\A^kD_k - \A\)e^{K/2}m_{PQ} 
- (q^P_X + q^Q_X)F_X m_{PQ},\qquad F_X = 
- \l{1\over4}\D^2\vx\r\label{defh},\\ 
D_\mu m_{PP'} &=& \D_\mu z^q\[m_{P P'}\(K_q+
\pp_q\ln\mu_P\) - m_{Q P'}\Gamma^Q_{Pq} - m_{P Q'}\Gamma^{Q'}_{P'q}\]
\mmm + iA^X_\mu\(q^A_X + q^{P'}_X\)m_{P P'}, \\ {[{\notD,\bfm}]} &=&
\pmatrix{0&\notD m L\cr \notD \m R&0\cr}, \qquad \G_{\mu\nu}^\Theta =
G_{\mu\nu}^\Phi + Z_{\mu\nu} - \gamma_5\Gamma_{\mu\nu},
\label{defs2}\eea
where indices are raised and lowered with the K\"ahler metric, the
operator $A$ and its covariant derivatives are defined in
\mysec{uvdiv} [see \myref{kr}], $Z_{\mu\nu}$ is defined in
\myref{defZ}, and

\bea \Gamma_{\mu\nu} &=& {1\over2}(\D_\nu\z^{\m}\D_\mu z^i -
\D_\mu\z^{\m}\D_\nu z^i)K_{i\m} = {1\over2}X_{\mu\nu},\label{kconn} \\
\(G^{\Phi-}_{\mu\nu}\)^P_Q &=& iF^a_{\mu\nu}(T_a)^P_Q -
\Gamma^P_{Q\mu\nu} = - K^{P\M} \(G^{\Phi+}_{\mu\nu}\)^{\N}_{\M}K_{\N
Q}, \qquad a\ne X,\label{conns}\eea
are the field strengths associated with the K\"ahler and
reparameterization + gauge connections, respectively.  The expression for
$d\R_\Theta$ is obtained using the methods of Appendix A
of~\cite{us2}, with $\R_5\to d\R,$ $\cM_0\to \cM_4(-M,-\vec\sigma)$.
Since $d\R$ appears only on the far right in (\ref{PVsum}), we can
drop all momentum derivatives in the resulting operator.

For a gauge transformation $\phi^{\pm}_\Phi = \phi^{\pm}_\Theta.$ Now
consider a K\"ahler transformation that is induced by a chiral field
redefinition: 
\beq Z'^p(\theta) = f^p[Z(\theta')], \qquad W' = e^{-F}W, \qquad K' = K
+ F + \bF, \qquad \theta' = e^{{i\over2}\im F}\theta
.\label{kahlt2}\eeq
For an infinitesimal transformation
\beq d  Z'^p = {\pp f^p\over\pp Z^q}d Z^q \approx d Z^p +
(\phi^-)^p_q d Z^q.\eeq
Then for example if $K(Z^p)$ is the light field K\"ahler potential,
the corresponding light loop divergences can be canceled by PV fields
$Z^P$ with K\"ahler potential \beq K_{PV} = Z^P\Z^{\Q}K_{p\q},\eeq
which is modular invariant provided
\beq \delta Z^P = (\phi^-)^p_qZ^Q \eeq
under \ref{kahlt2}.  This gives 
\beq \delta z^P =
(\phi^-)^p_qz^Q, \qquad \delta \chi^P = (\phi^-)^p_q\chi^Q +
{i\over2}\im F\chi^P + (\pp_k\phi^-)^p_q z^Q\chi^k.\eeq 
Since the last term does not contribute to the effective bosonic
Lagrangian that we are evaluating, from now on we will set
\beq \phi_\Theta = \phi_\Phi - {i\over2}\gamma_5\im F =
\phi,\label{delphi} \qquad \[\phi^+_\Phi,m\m\] =
\[\phi^+_\Theta,m\m\].\eeq
Then, since $m\m = \m m \equiv m^2$ for the chiral PV fields that 
contribute to the anomaly, we obtain
\bea d\R_\Phi &=& - {1\over p^2-m^2}\pmatrix{d\h_\Phi&dh'\cr d\h'&
dh_\Phi\cr}, \qquad d\R_\Theta = - {1\over p^2-m^2}\pmatrix{d\h_\Theta
R&-\notp dm L \cr-\notp d\m R& dh_\Theta L\cr}, \\ dh_\Theta
&=& \m dm, \qquad dh_\Phi = d\m m + \m dm, \\ dh'^{\bar P}_Q &=&
\[\lbr e^{-K}\(\A^kD_k - \A\)e^{K/2} - (q^P_X + q^Q_X)F_X\rbr dm\]^{\bar P}_Q 
- (q^P_X + q^Q_X)F_\Lambda m^{\bar P}_Q\eee 
\tph^+h'^{\bar P}_Q -\[F^k D_k\tph^+ - (q^P_X + q^Q_X)F_\Lambda\]m^{\bar P}_Q,
\qquad F_\Lambda
= - \l{1\over4}\D^2\Lambda\r, \label{defs3}\eea
where $F^k = e^{-K/2}\A^k = -{1\over4}\D^2Z^k$ at lowest order in the
loop expansion.

{\it A priori} the expansion (\ref{PVsum}) contains ultraviolet
divergences which must vanish if the regulated theory is truly finite.
The quadratic divergences are contained in
\beq \STr d\R = -\Tr\eta{1\over p^2-m^2}\(dh_\Phi - 2dh_\Theta\) + {\rm h.c.} = 0,\eeq
where the trace is over internal indices only.  The logarithmic
divergences occur in 
\bea \STr \R d\R &=& \Tr\eta{1\over p^2- m^2}\[-i\widehat{\notD\m}
 {1\over p^2 - m^2}\notp dm + \(
T^{\mu\nu}\G^-_{\Theta\mu}\G^-_{\Theta\nu} - {\hat r\over4}\) {1\over
p^2 - m^2}dh_\Theta\]L \nonumber\\ & & + \Tr\eta {1\over p^2 -
m^2}\[\(\widehat{\Delta H} -
T^{\mu\nu}\G^-_{\Phi\mu}\G^-_{\Phi\nu}\){1\over p^2 - m^2}dh_\Phi +
\hat{\h}'{1\over p^2 - m^2}dh'_\Phi\]
\nonumber \\ & & + \Tr\eta{1\over p^2-m^2}\(\hat h - m^2\)
{1\over p^2-m^2} \(dh_\Phi - dh_\Theta L\) + {\rm
h.c.},\label{rdr}\eea
where the traces are over both Dirac and internal indices.  Since PV
fields with $dm\ne0$ have no superpotential couplings and have
$K^{PV}_{P Q}=0$, there are no terms odd in $m$, and the
logarithmically divergent part of this expression is proportional to
\beq \Tr\eta\[{r\over2}\m dm - \(d\m m + \m
dm\)\Delta H - \h'dh' - {1\over2}D^2\m dm\] + {\rm
h.c.}\label{divrdr}\eeq
where here the traces are over internal indices only.  The
expression \myref{divrdr} is just the variation of 
the logarithmically divergent
$m$-dependent part of the one loop action,\footnote{There is no
quadratically divergent $m$-dependent contribution to the one loop
action $S_1$.} which, under the above assumptions, is proportional
to~\cite{pvcan}
\beq \Tr\eta\[\{\m m,\Delta H\} - {r\over2}\m m - {1\over2}D_\mu
mD^\mu\m + h'\h'\].\label{m2logdiv}\eeq
The ultraviolet divergent terms that are independent of $m$ have been
constructed to cancel the light loop divergences.  We also require
that the logarithmically divergent terms proportional to $m^2$ vanish,
which means that for a given functional form of $m_C^2(z,\z,\cx)$,
where $c_X = \l\vx\r$,\footnote{Here and throughout this appendix
$\l\Phi\r$ is the $\theta=\bth=0$ component of the superfield $\Phi$
with all fermion fields also set to zero.} we
have to introduce a set of PV fields with masses
\beq m^2_{C_\gamma} = \rho^C_\gamma m_C^2(z,\z,\cx),\qquad
\sum_\gamma\eta^C_\gamma\rho^C_\gamma=0.\label{pvmcond}\eeq
This condition assures the vanishing of \myref{m2logdiv} and
\myref{divrdr} as well as the finite terms proportional to $m^2$ 
that arise from ${i\over2}\Tr\R^2d\R$.  For the masses that are
not modular and \ux\, invariant, we need also to eliminate the 
residual finite terms proportional to $m^2\ln m^2$, which
requires the additional constraint
\beq \sum_\gamma\eta^C_\gamma\rho^C_\gamma\ln(\rho^C_\gamma)=0.\eeq
for these masses.  Both the PV masses and the functions $H$ are
controlled by the choice of K\"ahler metric for the PV fields, which
is constrained by the requirement of cancellation of quadratic
divergences.  Assuming that $D^{2n}m^2$ commutes with other operators
if $dm\ne0$, the last line drops out in
\myref{rdr}.  Dropping terms that vanish as $m^2\to\infty$, the finite
part of \myref{rdr} is
\bea - {i\over2}\l\STr\eta\R d\R\r_{\rm finite} &=&
- \Tr\eta{1\over192\pi^2m^2}\Bigg\{{1\over2}\(G^-_\Phi\cdot
G^-_\Phi\) \(d\m m - \m dm\) - G^-_{\mu\nu}X^{\mu\nu}\m dm\ddd\qquad 
- \[D^4\m - {1\over2}G_{\mu\nu}G^{\mu\nu}\m -
{2\over3}\([D_\mu,G^{\mu\nu}] + D_\mu r^{\mu\nu}\)D_\nu\m\]dm\ddd\qquad +
{1\over8}\(r_{\mu\nu\rho\sigma}r^{\mu\nu\rho\sigma} - ir\cdot\tr +
2X_{\mu\nu}X^{\mu\nu} - 4\nabla^2r\)\m dm \ddd\qquad +
\(D^2\Delta H\)\(d\m m + \m dm\) + \(D^2\h'\)dh'\Bigg\}\hc,
\label{finrdr}\eea
where we define
\beq D_\mu\m = D_\mu^-\m - \m D_\mu^+, \qquad G_{\mu\nu}\m =
[D_\mu,D_\nu]\m = G_{\mu\nu}^-\m - \m G_{\mu\nu}^+, \qquad etc.,\label{defDm}\eeq
and $G_{\mu\nu}^{\pm} = (G^{\pm}_\Phi)_{\mu\nu} \mp\half X_{\mu\nu}$ 
is defined as in \myref{defZ}.
If $\L_m$ is the part of the Lagrangian that contains the mass matrix
$m$ for PV chiral fermions $f^P=f_L^P$:
\beq m^{\bar{Q}}_{P'} = - K^{\bar{Q}R}(\L_m)_{RP'}, \qquad
(\L_m)_{RP'} = {\pp^2\over\pp f^R\pp f^{P'}}\L_m, \eeq
under a transformation $Z^p \to Z'^p$, $\phi^P \to g\phi^P$,
$f^P \to e^{i\alpha}gf^P$, we have
\beq m'^{\bar{Q}}_{P'} = -
e^{-i(\alpha + \alpha')}g^{\bar{Q}}_{\bar{M}}K^{\bar{M}R}(\L'_m)_{RN'}
(g^{-1})^{N'}_{P'},\label{delmgen} \eeq
since $K'_{PV} = K_{PV}$ by construction.  
For superpotential fermion mass terms $\L_m =
- e^{K/2}\mu_{PQ'}\chi^P\chi^{Q'}$, $\alpha = {1\over2}\im F$, and if
under \myref{kahlt2} $\mu_{PQ}(Z') = e^{\omega_i^{PQ}F^i}\mu_{PQ}(Z)$, we
have
\bea \L'_m &=& - e^{K'/2 + i\im
F}g^P_{R}g^{M'}_{Q'}\mu'_{PM'}\chi^{R}\chi^{Q'} = - e^{K/2 + F +
\omega_i^{P M'}F^i}g^P_{R}g^{M'}_{Q'}\mu_{PM'}\chi^{R}\chi^{Q'}, \nonumber \\
m'^{\bar{Q}}_{P'} &=& - e^{-i\im
F}g^{\bar{Q}}_{\bar{N}}K^{\bar{N}R}(\L'_m)_{R M'}(g^{-1})_{P'}^{M'}
= e^{\re F +
\omega_i^{P M'}F^i}g^{\bar{Q}}_{\bar{N}}K^{\bar{N}R}g_R^{M}K_{M\S}
m^{\bar{S}}_{P'}, \nonumber \\ \tm^{\bar{Q}}_{P'} &=& e^{i\im
F}(g^{-1})^{\bar Q}_{\N}m'^{\N}_{M'}g^{M'}_{P'} = e^{F +
\omega_i^{P M'}F^i}
K^{\bar{Q}R}g_R^{M}K_{M\S}m^{\bar{S}}_{N'}g^{N'}_{P'},\label{dmgen}\eea
For the chiral PV fields $\Psi^P_\gamma,\Psi'^Q_\gamma$, $\Psi =
U,V,\Phi$, with couplings defined by \myref{kpsi}--\myref{psi} 
in \myapp{orbs}, we have
\bea \mu^\gamma_{P Q'} &=& \delta_{P Q}\mu^P_\gamma(T), \qquad 
m^{\bar P}_{Q'} =e^{K/2}K^{\bar P Q}\mu^P_\gamma(T), 
\nnn K_{P\Q} &=& \del_{P
Q}f^P_\gamma,\qquad K_{P'\Q'} = \del_{P Q}f^{P'}_\gamma,
\nnn
(g_\gamma)^R_Q &=& \del^R_Q e^{\phi^-_{R_\gamma}},\qquad
(g_\gamma)^{R'}_{Q'} = \del^R_Q e^{\phi^-_{{R'}_\gamma}}.
\eea
Explicitly
\bea \ln f^P_\gamma &=& \sum_n q^{P_\gamma}_n g^n + q^{P_\gamma}_X\cx,
\qquad \phi^-_{P_\gamma} = - \sum_i q^{P_\gamma}_n F^n -
q^{P_\gamma}_X\lambda, \nnn
\cx &=&\l\vx\r, \qquad \lambda = \l\Lambda\r.\label{kpq}
\eea
where $\Lambda$ is the \ux\, gauge transformation superfield introduced
in \myref{gaugetr} and \myref{uxpv}, and $q_n$ and $q_X$ are modular
weights and \ux\, charges, respectively.
Then we obtain
\beq \sum_R g^R_Qg^{R'}_{P'} = e^{\varphi^-_{P_\gamma} +
\varphi^-_{P'_\gamma}}\delta_{PQ}, \qquad
\tm^P_\gamma = e^{\sum_i F^i(1 + \omega_i^{P_\gamma}) +
\varphi^-_{P_\gamma} + 
\varphi^-_{P'_\gamma}}m_\gamma\equiv e^{\tph^+_{P_\gamma}}m^P_\gamma,
\label{dmp}\eeq
To evaluate \myref{PVsum} we need only consider fields with
noninvariant masses: $\tm\ne m$. In addition to the PV superfields
$U,V,\Phi,$ considered above, these include the gauge singlets
$\phi_\gamma$ for which we take
\beq \ln f^{\phi}_\gamma = \alpha^{\phi}_\gamma K,
\qquad \varphi_{\phi_\gamma}^- = - \alpha^{\phi}_\gamma F, \qquad
\tph_{\phi_\gamma}^+ = \sum_i\(1 - 2\alpha^{\phi}_\gamma
+ \omega_i^\gamma\)F^i,
\qquad \mu^{\phi}_{\gamma\gamma'}\ne0\label{tphph},\eeq
and the adjoint chiral multiplet $\tph^a_\alpha$ with
\beq f^{\tph}_\alpha = 1,\qquad \varphi_\gamma^- = 0, \qquad
\tph_\gamma^+ = \sum_i\(1 + \omega_i^\gamma\)F^i.\eeq
Let us first consider the coefficient of $r\cdot\tr$ in
\myref{finrdr}.  To evaluate this we use the conditions \myref{sigs},
\myref{cond1}, \myref{sumqpsi} and \myref{condqx}.  These include
contributions from chiral PV fields that have covariant masses: $\tm =
m,\; \tph^{\pm} = 0$; we can also include them in the sums here, since
their net contribution vanishes. However we must exclude the fields
$\theta^\gamma$; their masses arise from D-terms rather than
F-terms. Each pair $\Phi^P,\Phi'^{P'}$ gives gives an identical
contribution, and we remove from $\Tr\eta = N'$ the contribution $N'_G
= \sum_\gamma \eta^\theta_\gamma$.  On the other hand, the
moduli-dependent prefactors in \myref{kpq}, with, referring to
\myref{kpsi},
\beq q_n^{U^A_\gamma} = \alpha^A_\gamma q^A_n,\qquad etc.,
\label{qnphi}\eeq
that have been chosen to cancel loop contributions from the PV fields
$\dY$ are not included; to include these we use \myref{corb121}--\myref{corb12} 
and the identifications \myref{qnphi}, giving the result in
\myref{psitphi}. However we also 
have to exclude the net contribution of all the 
fields\footnote{Equivalently, we remove $N_\Psi$ which is already
included in \myref{psitphi}, and include the reparameterization 
+ threshold 
contributions from ${\dZ},{\dY}$  which exactly cancels their K\"ahler
connection contributions $N_{\dZ} + N_{\dY}$.} $\dZ,\dY,\Psi$
from $N'$ where $N_\Psi = - N_{\dY} = - N_{\dZ} = N + 2$. That
is, we have to add a factor $N + 2$ to $N'$, giving
\bea {16\pi^2\over\sqrt{g}}\del S_r &=& -{i\over96}r\cdot\tr\Tr\eta(\tph^-
- \tph^+) = - {1\over24}r\cdot\tr\Tr\eta\phi_{P V} = {1\over24}r\cdot\tr\Tr\phi\eee 
- {1\over48}r\cdot\tr\l\Bigg[\(N' - N'_G  - 2\alpha_1
  \)\im F + 2\Tr T_X\aax + \sum_n\(\omega_n + 2\sum_p q^p_n\)\im F^n\]\eee 
 {1\over48}r\cdot\tr\[\sum_n\(N
  - N_G - 3 - \omega_n - 2\sum_p q^p_n\)\im F^n - 2\Tr
  T_X\aax\],\label{delsr} \eea
where [see \myref{cond1}], 
\beq \alpha_1 = \sum_C\eta_C\alpha^C = - 10,\qquad
\omega_n = \sum_C\eta_C \omega_n^C,\qquad \aax = \im\lambda.
\label{newdefs}\eeq 
Apart from the ``threshold corrections'' $\omega_n(T^i)$, this is
precisely the result in \myref{totr} (with the identification
$\phi = -\eta\phi_{P V} = \eta\phi^+_{P V} = -{i\over2}\eta\tph^-$) 
for $N$ chiral fermions with
$\phi^p = \sum_n(\half - q^p_n)\im F^n - q^p_X\aax$, the auxiliary
fermion needed for gravitino gauge fixing~\cite{us} with $\phi^\alpha
= \half\sum_n\im F^n$, and $N_G + 4$ gaugino and gravitino degrees of
freedom with $\phi = - \half\sum_n\im F^n$.

To evaluate the full anomaly, we can simplify further by setting
\beq \mu^\phi_{\alpha\beta} = \mu_\alpha^\phi\del_{\alpha\beta}
\label{diagmass}\eeq
in \myref{w1}, and imposing \myref{neworb0} and \myref{hfeta} with
$q_X^{U^A} = q_X^{U_A}$.

Then for the fields with noninvariant PV masses introduced in
\myapp{anmass} we have
\bea \hf^0,\hf^\pm:&&\quad \ln f_{\hf_\gamma} =
\hat\alpha_\gamma^{\hf}K, \qquad \hat\alpha^0_\gamma =
\hat\alpha^+_\gamma = 0,\qquad \hat\alpha^-_\gamma = 1,\\
\Psi^P:&&\quad \ln f_{\Psi^P} = \sum_n q^P_n g^n + q^P_X\vx ,\qquad
q^P_n = 1 - q^p_n + \dot\omega_n,\qquad q^N_X = q^I_X = 0,\\
\tph^a:&&\quad f^{\tph^a} = 1,\eea

and the $q^A_X$ are subject to the constraints given in \myref{neworb3},
\myref{neworb6} and \myref{neworb4}.  Here $P = N,I,A$, 
and we identify [see \myref{kpsi}]
\beq \Psi^N = \Phi^n,\qquad \Psi^I = \del^I_N\Phi^N,\qquad
\Psi^A = U^A,U_A,V^A,\qquad q^n_m = 0,\qquad q^i_n = 2\del^i_n.\eeq
The PV superfields $\Psi^P,\tph^a$ and 
$\hf_\gamma,\;\gamma = 1,\ldots,5,$ have $\eta = +1$ and
$\hf^{\pm}_\gamma,\; \gamma = 1,\ldots 2N -4$ has $\eta = -1$.
Using \myref{diagmass},
and denoting the fields $U,V,\Phi,\hf,\tph$ collectively by $\Phi^C$,
the covariant derivative in \myref{defs2} reduces to
\bea D_\mu m_{CC'} &=& \lbr\D_\mu z^i\[K_i + \pp_i\ln(\mu_C -
2\ln f_C)\] -  2q_X(i A^X_\mu + \pp_\mu\cx)\rbr m_{CC'}\nnn &\equiv&
\half\(V_\mu^C + iA_\mu^C\)m_{CC'}\equiv V^+_\mu m_{CC'} 
= (V^-_\mu)^\dag m_{CC'}.\label{dmass}\eea
Since the K\"ahler metric is covariantly constant: $D_q K^{C\bar C} =
0$, we also have 
\bea D_\mu m^{\bar C}_{C'} &=& K^{\bar C C}\D_\mu
m_{C C'} ,\qquad \pp_i\ln\mu_C = \del_{i t^n}\omega^C_n\zeta(t^n),
\qquad \zeta(t) = \pp_t\eta(t)/\eta(t)\label{dmass2}.\eea 
The vectors $V_\mu,A_\mu$ satisfy 
\bea i A^C_{\mu\nu} &=& i\(\D_\mu
A^{C}_\nu - \D_\nu A^{C}_\mu\) = 2 f^C_{\mu\nu} - X_{\mu\nu},\qquad
\D_\mu V^{C}_\nu - \D_\nu V^{C}_\mu = 0,
%
\nnn f^C_{\mu\nu} &=& \pp_i\pp_{\m}\ln f_C\(\D_\mu z^i\D_\nu\z^{\m} -
\D_\nu z^i\D_\mu\z^{\m}\) - i F^a_{\mu\nu}(T_a z)^i D_i\ln f_C
- 2iq^C_X F^X_{\mu\nu},\label{c53}\eea 
and from \myref{defs2}--\myref{conns} and \myref{defDm} we have
\bea \(G_\Phi^\pm\)^C_{\mu\nu} &=& \pm\(f^C_{\mu\nu} - iT^C\cdot F_{\mu\nu}\),
\qquad  F_{\mu\nu} \ne F^X_{\mu\nu},\\
\(G_\Theta^\pm\)^C_{\mu\nu} &\equiv& (G_\Phi^\pm)^C_{\mu\nu} \mp
\half X_{\mu\nu} = \pm i\(\half A^C_{\mu\nu} - T^C\cdot F_{\mu\nu}\),
\qquad F_{\mu\nu} \ne F_{\mu\nu}^X\\
G_{\mu\nu} m_C &=& iA^C_{\mu\nu} m_C,\qquad 
G_{\mu\nu}\m_C = - iA^C_{\mu\nu}\m_C, \qquad G_{\mu\nu}\m m = 0\eea
We are interested only in the variation
of the on shell action.  Therefore we may drop terms proportional to 
\beq g^{-{1\over2}}g_{\mu\nu}{\pp\L_{tree}\over\pp g_{\mu\nu}} =
{r\over2} - 2V + \D_\mu\z^{m}\D^\mu z^iK_{i\m} = {r\over2} -
\l\half\D^\alpha X_\alpha\r - 3\(\hV + M^2\).\label{gdldg} \eeq 
Then, defining 
\bea f_\alpha &=& \cP\D_\alpha\ln f = -{1\over8}\chiproj\D_\alpha\ln f,\nnn
f_\alpha^{P} &=& \sum_n q_n^{P}g^n_\alpha + q^{P}_X W^X_\alpha,\qquad
f_\alpha^{\hf_\gamma} = \alpha^{\hf}_\gamma K_\alpha,\qquad
f^{\tph^a}_\alpha = 0,\eea 
 we have
\bea \Delta H^C_D &=& \del^C_D\(\hV + M^2 + \half\l\D^\alpha f^C_\alpha\r\)
+ {1\over x}\D_a(T^a)^C_D\eee \del^C_D\[{1\over6}\(r -\DX\) + 
\half\l\D^\alpha f^C_\alpha\r\]+ {1\over x}\D_a(T^a)^C_D,
\nnn a&\ne& X,\qquad (\Tr T_a)_C = 0,\label{defDH}\\ 
\(m^{-1}h'\)^{\bar C'}_C &=&
e^{-K/2}\[\A^k\(K_k - 2\pp_k\ln f_C + \pp_k\ln\mu_C\) - \A\] - 2q^C F_X
\eee - F^i\D_i\ln m^2 + 2q_X F_X - \bM = {1\over4}\l\(\D^2\ln\cM^2
- 8\bR\)\r \equiv \y ,\label{defby}
\\ (m^{-1}d h')_C^{\bar C'} &=& \y\tph^+ + 
\sum_n\(1 - 2q_n^C + \omega_n^C\)e^{-K/2}\A^k
F^n_k- 2q^C F_\Lambda\eee \y\tph^+ + F_{\tph^+_C}
\qquad\qquad F_{\tph^+_C} =- {1\over4}\l\D^2\tph^+_C\r,\label{defFph}\eea
where
\beq   \cM^2 =  e^{K - 2f}|\mu|^2, \qquad \l\cM^2\r = m^2,\label{defcM}\eeq
is a real superfield, and the identifications on the right in \myref{defby}
and \myref{defFph} hold to the order we are working in because to that
order the auxiliary fields can be interchanged with their tree-level 
values\footnote{Our normalization of the auxiliary field $M$ differs by a 
factor $-3$ from that of Bin\'etruy, Girardi and Grimm~\cite{bggm}:
$M = - {1\over3}M_{\rm B G G}$.}
\beq F^k = - e^{-K/2}\A^k, \qquad M = e^{K/2}W = e^{-K/2}A = 2\l R\r.\eeq
Since the relevant PV masses have a diagonal K\"ahler metric, it
follows from \myref{chifs} that $G^+ = - G^-$, so the first term in
\myref{finrdr} drops out, and, from, \myref{defDm}, $G\m = \{G^-,\m\},$
etc.  Then writing
\beq T = \sum_k(-)^k T_k,\qquad T_k 
= {i\over2}\l\STr\eta\R^k d\R\r_{\rm finite}, \eeq
we obtain, imposing \myref{pvmcond},
\bea T_1 &=& + {\sqrt{g}\over192\pi^2}\Tr\eta\Bigg[\tph^+\Bigg\{{1\over8}\(
r_{\mu\nu\rho\sigma}r^{\mu\nu\rho\sigma} - i r\cdot\tr + 
2X_{\mu\nu}X^{\mu\nu} + 4iA_{\mu\nu}X^{\mu\nu}\) 
- \(\nabla^\mu V_\mu^- + V_-^\mu V_\mu^-\)^2
\mmm\qquad + \Box\(\Df - {1\over3}\DX - \nabla^\mu V_\mu^- - V_-^\mu V_\mu^- 
- {1\over6}r\) \mmm\qquad
- 2V_\nu^-\nabla^\nu\[\(\nabla_\mu + V_\mu^-\)V^\mu_-\]  -
\half A_{\mu\nu}A^{\mu\nu} 
- {2i\over3}V_\nu^-\D_\mu A^{\mu\nu}\mmm\qquad
+ {2\over3}\nabla_\mu\(r^{\mu\nu}V^-_\nu\) 
+ {2\over3}r^{\mu\nu}V^-_\mu V^-_\nu\Bigg\}\mmm  + \(\y\tph^+
+ F_{\tphi^+}\)\(\Box y + 2V^-_\mu\pp^\mu y + y\nabla^\mu V^-_\mu
+ y V^-_\mu V^\mu_-\)\Bigg]\hc
\label{finrdr2}\eea
\bea T_4 &=& - {\sqrt{g}\over32\pi^2}\Tr\eta\tph^+\[{1\over6}
(y\y)^2 - \half y\y V_\mu V^\mu + {1\over3}\(V_\mu V_-^\mu\)^2 
- {1\over6}\(V_\mu^-V^\mu_-\)^2 + {1\over3}V_\mu^-V^\mu_-V_\nu V^\nu\]
\ddd \hc, \eea
The expressions for $T_2$ and $T_3$ are more complicated; they
combine to give
\bea T_2 - T_3 &=& - {\sqrt{g}\over32\pi^2}\Tr\eta
\[\tph^+\lbr\(\half\Df - {1\over6}\DX + {1\over6}r\)^2  
- {1\over3}V_\mu \pp^\mu\(\half\Df - {1\over6}\DX + {1\over6}r\)
\right.\right.\mmm \l\l
- {1\over3}\(\nabla_\mu V^\mu + 3y\y + r\)
\(\half\Df - {1\over6}\DX + {1\over6}r\)
\right.\right.\mmm \l\l + {1\over12}\nabla^\mu\(V_\mu r\) + 
{r^2\over48} + {4i\over9}V_-^\nu\D^\mu A_{\mu\nu}
\right.\right.
\mmm\l\l - {1\over 6}\[y\Box\y + \y\Box y + \pp_\mu y\pp^\mu\y 
+ \y(V^-_\mu - V_\mu)\pp^\mu y 
+ y(V^+_\mu - V_\mu)\pp^\mu\y\right.\right.\right.\mmm\l\l\l\qquad\qquad
- y\y\(\nabla_\mu V^\mu + 3V_\mu V^\mu + V^+_\mu V^\mu_-\)\]
\right.\right.\mmm\l\l
+\[{1\over x^2}D^a D^b - {1\over2}\(F^a\cdot F^b - i\tF^a\cdot F^b\)\]T_a T_b
+ {1\over24}\(A\cdot A + i\tilde A\cdot A\)\right.\right.\mmm \l\l 
\right.\right.\mmm \l\l 
+ {1\over6}\[\Box\(V_-^\mu V^-_\mu\)  + 
V^\mu\nabla_\mu\nabla^\nu V_\nu^- 
+ \(\nabla^\mu V_\mu\)\(\nabla^\nu V_\nu^-\) 
- \(\nabla^\mu V_-^\nu\)\nabla_\mu V_\nu^-\]
\right.\right.\mmm\l\l
+ {1\over6}\nabla^\mu\(V_\mu V^-_\nu V_-^\nu +
V^-_\mu V_-^\nu V^-_\nu\)
\right.\right.\mmm\l\l
- {1\over3}\[V_\mu V^\mu V_-^\nu V^-_\nu + (V^-_\mu V^\mu)^2 
- (V^-_\mu V^\mu_-)^2\]
\right.\right.\mmm\l\l - {1\over36}\[10V_\mu^-V_\nu^-r^{\mu\nu} 
+ \nabla^\mu\(3r V_\mu^- -  2r_{\mu\nu}V^\nu_-\) + 3r V_\mu^-V^\mu_-\]
+ {1\over6}r y\y\rbr\right.\mmm\l
- {1\over6}\(y^2\y + V_\mu\pp^\mu y + V_\mu V^\mu_-y 
+ y\nabla^\mu V_\mu + y\DX - 3y\Df\)\(F_{\tphi^+} + \y\tph^+\)\]
\ddd \hc,\label{T2T3}\eea
In writing \myref{T2T3} we used \myref{c53}, the Bianchi identities:
\beq 0 = D^\mu\tG_{\mu\nu} = 2\nabla^\mu r_{\mu\nu} -
\nabla_\nu r = \epsilon^{\lambda\mu\nu\rho}r_{\mu\nu\rho\sigma},\qquad
V^-_\mu V_\nu\nabla^\mu V^\nu_- = V_\mu V^-_\nu\nabla^\mu V^\nu_- +
i A^{\mu\nu}V^+_\mu V^-_\nu, \eeq
the identities
\beq 0 = [D_\mu,D_\nu]V_\rho  + r^\sigma_{\rho\nu\mu}V_\sigma 
= \([D_\mu,D_\nu]V^+_\rho + r^\sigma_{\rho\nu\mu}V^+_\sigma 
- iA_{\mu\nu}V^+_\rho\)m
= \([D_\mu,D_\nu] - iA_{\mu\nu}\)m,\eeq
and the conjugate relations.
Combining these contributions using, from \myref{c53}
\bea A_{\mu\nu}A^{\mu\nu} &=& 2i\D_\mu(V^-_\nu A^{\mu\nu}) 
- 2iV^-_\nu\D_\mu A^{\mu\nu}, \nnn V_-^\mu\Box V^-_\mu &=&
\half V_-^\mu\{\nabla_\mu,\nabla^\nu\}V^-_\nu 
- \half V_\nu^- V^-_\mu r^{\mu\nu} - {i}V^-_\nu\D_\mu A^{\mu\nu}.\eea
 gives
\bea {i\over2}T &=& {\sqrt{g}\over64\pi^2}\Tr\eta\Bigg[\tph^+\Bigg\{
\(F^a\cdot F^b - i\tF^a\cdot F^b - {2\over x^2}D^a D^b\)T_a T_b
+ {1\over12}\(3A\cdot A - i\tilde A\cdot A\)\mmm\qquad  
- {1\over24}\(r_{\mu\nu\rho\sigma}r^{\mu\nu\rho\sigma} 
- i r\cdot\tr + 2X_{\mu\nu}X^{\mu\nu}\) + {r^2\over72} 
- {i\over6}A_{\mu\nu}X^{\mu\nu}
\mmm\qquad + {1\over3}\nabla^\mu\(\nabla_\mu - V_\mu\)\({1\over3}\DX 
- \Df + {1\over6}r + \nabla^\nu V^-_\nu + V^\nu_-V^-_\nu\)\mmm\qquad
- \half\(\Df - {1\over3}\DX\)^2 
+ {1\over3}\nabla^\mu\[y\(\pp_\mu - V^-_\mu\)\y 
- \nabla^\nu\(V_\mu^-V^-_\nu\)\]\mmm
\qquad + {2\over3}\nabla^\mu\[V^-_\mu\nabla^\nu V_\nu^- 
+ \half V^-_\mu V_-^\nu V_\nu^-\]
+ {1\over6}\nabla^\mu\(r V_\mu^- -  2r_{\mu\nu}V^\nu_-\)\Bigg\}\mmm
- {1\over3}F_{\tph^+}\lbr\Box y - \D_\mu y V^\mu 
+ y\(\D^\mu V^-_\mu + V_-^\mu V^-_\mu - \D_\mu V^\mu - V^\mu V_\mu^-\)
\right.\mmm\qquad\l + 2V^-_\mu\D^\mu y - y^2\y + y\(3\Dc f_\alpha
- \Dc X_\alpha\)\rbr\Bigg]\hc \label{totanom}\eea 
This expression is the bosonic part of the superfield expression
\bea \del\L_1 &=&
{1\over8\pi^2}\superint\Tr\[\eta\widetilde\Phi^+(T^n,\Lambda_X)\Omega_1\]
\hc\label{tot2}\\
\Omega_1 &=& - {1\over48}\Omega_m + {1\over3}\Omega_W + \Omega^0_{\rm YM} -
{1\over36}\Omega_{X^m},\qquad \Omega_m = \Omega_D - 4\Omega_G + 8\Omega_R,\label{Om2}\eea
where the operators in \myref{Om2} are defined in 
\myref{OYM} and \myref{defxm}--\myref{Gm}, $\Omega^0_{Y M}$ is the Chern-Simons
superfield for the nonanomalous gauge group, and
$\l\widetilde\Phi^+\r = \tph^+$. 
The expression \myref{tot2} is the result of an infinitesimal
transformation, and must be integrated to give the expression for a
finite transformation; in particular the modular transformations are
discrete and therefore finite.  This is possible if the coefficient of
$\widetilde\Phi^+$ contains only 1) operators such as
$\Omega_W,\;\Omega_{\rm YM}$ and $\Omega_{X^m}$ whose chiral
projections are invariant under \ux\, and modular transformations, 2)
linear multiplets that drop out of the superspace integral and 3)
derivatives of $\ln\cM$.  That this is indeed the case is shown in
\mysec{anomsf}.

%% file: AppendixD.tex
\subsection{Orbifold compactification: PV sector for matter}\label{orbs}
\setcounter{equation}{0} 
We argued in \mysec{strat} that PV mass terms must have well-defined
modular weights, with invariant masses for those fields that
contribute to the renormalization of the K\"ahler potential.  One way
to achieve this would be to couple in the PV mass superpotential all
PV fields $Z^\sigma$ with metric $K^Z_{\sigma\brh}$ to fields
$Y_\sigma$ with a metric proportional to its inverse, as in
\myref{diagm2}.  However each $Z,Y$ pair gives no contribution linear
in the scalar curvature, and doubles the contribution quadratic in
scalar curvature.  This requires an even number of pairs with
signatures that sum to zero, and the introduction of other fields that
reproduce the curvature terms from the light fields.  This was done
in~\cite{pvdil} for the untwisted sector with K\"ahler potential:
\bea G_u &=& \sum_n G^n(Z^i_n), \quad G^n = - \del^n_i\ln\(T^i + \T^i
- \sum_{a=1}^{N_n-1}|\Phi^{ai}|^2\) = \del^n_i\[g^i - \ln(1 -
e^{g^i}\sum_{a=1}^{N_n-1}|\Phi^{ai}|^2)\], \nnn g^i &=& - \ln(T^i +
\T^i),\label{untw2}\eea
where the special property $G^n_{ij} = G^n_iG^n_j$ was exploited to
mimic their curvature terms by other fields.  The twisted sector
K\"ahler potential is not known beyond leading order:
\beq G = G_u + f(Z,\Z), \quad f = X^a + O(\Phi^3), \quad
X^a = e^{g^a}|\Phi^a|^2, \quad g^a = \sum_i q^a_i g^i. \eeq
Under a nonlinear transformation $Z^i_n\to
Z'^i_n(Z^j_n)$ such that
\beq G_u(Z') = G_u(Z) + F + \bF, \quad G^n(Z'_n) = 
G^n(Z_n) + F^n + \bF^n, \quad F = \sum_n F^n,
\eeq
$X^a$ is invariant provided 
\beq\Phi^a\to \Phi'^a = e^{-F^a}\Phi^a,\quad F^a = \sum_n 
q^a_n F^n.\label{modphi}\eeq
To use the same trick including the twisted sector fields
requires a constraint on the overall K\"ahler potential
analogous to the constraint on gauge charges discussed in
\mysec{gauge}.  For example this is possible with 
a K\"ahler potential of the form
\beq G = g + f(X^a) + H(Z) + \H(\Z), \qquad g = \sum_n g^n(Z^i_n),
\quad g^a = \sum_n q^a_n g^n,\label{orbk}\eeq
where $g^n$ transforms like $G^n$ under a modular transformation, for
example $g^n = \del^n_i g^i$ or $g^n = G^n$.  The modular invariant
holomorphic function H is constructed from operators of the form
\beq \prod_a[\Phi^a\prod_{i=1}^3\eta^{2q^a_{n_i}}(iT^i)], \label{other}\eeq
and the fields $\Phi^a$ are separated into groups $\Phi_m\ni
\Phi^a_m,\;a=1,\ldots N_m$ with the function $f(X^a)$ in
\myref{orbk} restricted to the form
\beq f = \sum_m k^m, \quad k^m = - \lambda^{-1}_m\ln\(1 - \lambda_m
\sum_{a=1}^{N_m}X^a_m\).\label{twisted}\eeq
This is option 1) of \mysec{pvmass}; its implementation requires the
introduction of a number of additional PV fields.  Here we will
focus on option 2) which entails no restriction on the twisted sector K\"ahler
potential. In \mysec{orb3} we will consider a hybrid case where option 1) is
implemented for the untwisted sector in order to include the
possibility of maximal Heisenberg invariance.

In order to make the PV K\"ahler potential and superpotential fully
modular invariant we introduced a set of PV fields $\dZ^\sigma =
\dZ^N,\dZ^P$, with negative signature, that regulate the 
UV divergent contributions to the renormalization of the light
field K\"ahler potential and to the operator $\phi_3$ in 
\myref{dops}.  Their K\"ahler potential is given by \myref{pvz0}
and \myref{xkahl2} with $\zeta_{m\n}=\del_{m\n}$:
\bea K(\dZ^\sigma) &=& K^{\dZ} + K(\dZ^N) =
K^{\dZ}_{\rho\bsig}\dZ^\rho\dbZ^{\bsig} +
{1\over2}\(K_{\sigma\rho}\dZ^\sigma\dZ^\rho \hc\),\nnn
\sigma &=& P,N,\qquad P = I,A,\label{kalph}\eea
which is modular invariant provided $\dZ^\sigma$ transforms as in
\myref{xkahl} under \myref{genkahl}. These fields couple in the PV 
superpotential $W^{\dZ}_1$ given in \myref{wzi2} to superfields
$\dY_\sigma = \dY_N,\dY_P,$ with the K\"ahler potential given in
\myref{kdy}. To calculate the contributions of these fields to
$\L_Q,\Phi_1,\Phi_2$, we need the affine connection derived from the
PV K\"ahler metric.  For $\dZ^\alpha$ we have
\bea \dot\Gamma^P_{Nr} &=& K^{p\q}\pp_r\bc^{\n}_{\q}, \quad
\dot\Gamma^P_{Q r} = \Gamma^p_{q r} + \chi^n_q\dot\Gamma^P_{Nr}, \quad
\dot\Gamma^N_{N' r} =  - \chi^n_q\dot\Gamma^Q_{N'r},\nnn
\dot\Gamma^N_{P r} &=&  + D_r\chi^n_p 
- \chi^n_q\chi^{n'}_p\dot\Gamma^Q_{N'r}, \eea
If $\chi^n_p = \dot a\pp_p h^n(Z)$, with $h^n(Z)$ holomorphic, $\pp_r
\h(\z)^{\n}_{\q} = \dot\Gamma^P_{N r} = 0$, $\chi^n_p$ drops out of 
the traces of products of $\dot\Gamma$ and its derivatives.  In addition,
since $(T_a)_p^q\chi^n_q=0,$ it also drops out of 
the traces of $\dot\Gamma$ with 
gauge generators. Alternatively, since
$\chi^n_p$ is proportional to the parameter $\dot a$ by virtue of the
condition \myref{xkahl2}, we can cancel the UV divergent terms
involving $\chi^n_p$ by including several copies of the
$\dZ^\sigma\to\dZ^\sigma_\lambda$, $\lambda = 1,\ldots, 2n_{\dZ}+1$
with signatures $\eta_\lambda^Z$ and parameters $\dot a_\lambda^Z$
such that
\beq\dot n = \sum_\lambda\eta_\lambda^{\dZ} = -1, \qquad
\sum_\lambda\eta_\lambda^{\dZ} \dot a^2_\lambda 
= \sum_\lambda\eta_\lambda^{\dZ} \dot a^4_\lambda = 0.
\label{azconds}\eeq  
We also need to impose the condition \myref{cacond1}. This is
again automatically satisfied if $\chi^n$ is holomorphic because it
also drops out of the trace of products of $J_\mu^\pm$ with $\phi$
because $\phi_N^\rho= 0$, where
\beq (J^-_\mu)^\rho_\sigma = \dot\Gamma^\rho_{\sigma r}\D_\mu z^r +
iT^\rho_\sigma\cdot A_\mu + i\del^\rho_\sigma\Gamma_\mu = 
\[(J^+_\mu)^\rho_\sigma\]^\dag, \qquad \Gamma_\mu = {i\over4}
\(K_i\D_\mu z^i - K_{\m}\D_\mu\z^{\m}\). \eeq 
For the more general case, \myref{cacond1} requires the conditions 
on $\dot a^2$ in \myref{azconds} and the additional constraint
\beq \sum_\lambda\eta_\lambda^{\dZ} \dot a^6_\lambda =
0.\label{azconds2}\eeq 
Only one set of the $\dZ_\lambda$, say $\dZ_0$, with negative
signature, $\eta^{\dZ}_0 = -1$, need have couplings in the
superpotential \myref{pvz} and in the additional terms
$K_{\sigma\rho}$ in the K\"ahler potential \myref{kalph}.  Gauge and
modular invariant mass terms for all the above fields may be
constructed as in \myref{wzi2} by introducing a superfield
$\dY_\sigma^\lambda$, with the K\"ahler potential \myref{kdy}, for
each $\dZ^\sigma_\lambda$.

Rather than use the most general parameterization, we will simply set 
\beq \chi^n_p = \pp_p\chi^n,\label{simp}\eeq 
and consider three choices for the pair of functions $g^n$ in
\myref{kdy} and $\chi^n$ that we take to be the same in \myref{kalph}
and in \myref{kdy}.  In our second and third scenarios we make some
assumptions on the form of the K\"ahler potential for the light fields.

\subsubsection{$\chi^n$ holomorphic}\label{orb1}
We set
\bea g^n &=& \del^n_i g^n(T^i +\T^{\ibar}) 
= -\del^n_i\ln(T^i+\T^{\ibar}),\qquad i,n = 1,2,3,\nnn
\chi^n &=& 2\dot a\del^n_i\ln\eta(T^i),
\qquad \chi^n_i = 2\dot a{\pp\eta(T^i)\over\pp T^i}\del^n_i
\equiv 2\dot a\zeta(T^i)\del^n_i,\label{o1}
\eea
in \myref{xkahl2} and \myref{wzi2}.  Then for $\dZ^\sigma$:
\beq \dot\Gamma^P_{Q r} = \dot\Gamma^p_{q r},\qquad \dot\Gamma^N_{Q r} =
D_r\chi^n_q,\qquad
\dot\Gamma^P_{N r} = \dot\Gamma^N_{N r} = 0,\label{zo1}\eeq
and the UV divergent contributions from 
$\dZ$ simply cancel the UV divergences from the light fields $Z$.  
The affine connections for $\dY_{I,N}$ are
\beq \dot\Gamma^I_{I j} = G^N_j - 2g^n_j,\qquad \dot\Gamma^N_{I j} =
-\(\pp_j - 2g^n_i\)\chi^n_i,\qquad
\dot\Gamma^I_{N j} = 0,\qquad \dot\Gamma^N_{N j} = G_j^N,\label{zo2}\eeq
and the UV divergent contributions from $\dY_{I,N}$, with
$\sum_\lambda\dot\eta_\lambda = -1$, involve the operators
\bea \sum_\lambda\dot\eta_\lambda(\Gamma^n_{\dY})^\sigma_{\sigma\alpha} &=& 
-2\(G^N_\alpha - g^n_\alpha\), \label{orbops0}\\
\sum_\lambda\dot\eta_\lambda(\Gamma^n_{\dY})^\sigma_{\rho\alpha}
(\Gamma^n_{\dY})^\rho_{\sigma\alpha} &=& - 
\(G^N_\alpha - 2g^n_\alpha\)\(G^N_\beta - 2g^n_\beta\) - G^N_\alpha
G^N_\beta.\label{orbops}\eea
All the UV divergent contributions from $\dY$ can be canceled by
additional PV fields $\Psi= U,V,\Phi$ with K\"ahler and superpotential
\bea K(\Psi) &=& \sum_\gamma\[\sum_A\(e^{\alpha^A_\gamma G^A}|U^A_\gamma|^2
+ e^{\alpha^A_\gamma G^A}|U^\gamma_A|^2 + e^{\gamma^A_\gamma
G^A}|V^A_\gamma|^2\)\right.\ddd\qquad\left. + \sum_{N=n}
\(e^{\delta^N_\gamma(G^N
- 2g^n)}|\Phi_\gamma^N|^2 + e^{\epsilon^n_\gamma
G^N}|\Phi_\gamma^n|^2\)\],\label{kpsi}\\
W(\Psi) &=& \sum_{\gamma}\[\sum_A\mu^U_{\gamma}U^A_\gamma U_A^\gamma +
\half\(\sum_A\mu^V_\gamma(V^A_\gamma)^2 +
\sum_N\mu^N_\gamma(\Phi^N_\gamma)^2 
+ \sum_n\mu^n_\gamma(\Phi^n_\gamma)^2\)\],\label{wpsi}\\
T_a(U^A_\gamma) &=& - T_a^T(U_A^\gamma),\quad a\ne X;\qquad
T_a(V) = - T^T_a(V),
\quad T_a(\Phi) = 0 \quad\forall\;\; a,\label{psi}\eea
if we require 
\bea \sum_{\Psi,\gamma}\eta^\Psi_\gamma &=& 6 + d_M = N + 2,\label{corb121}\\
C^a_M &=& \sum_{\gamma}\lbr\eta^U_\gamma\[(\Tr T_a^2)_{U_\gamma}
+ (\Tr T_a^2)_{U^\gamma}\] + \eta^V_\gamma(\Tr T_a^2)_{V_\gamma}\rbr,
\label{corb122}\\ 1 &=& \sum_{\gamma}\eta^N_\gamma\del^N_\gamma 
=\sum_{\gamma}\eta^n_\gamma\epsilon^n_\gamma =
\sum_{\gamma}\eta^N_\gamma(\del^N_\gamma)^2=
\sum_{\gamma}\eta^n_\gamma(\epsilon^n_\gamma)^2,\qquad \label{corb123} \\
(\Tr T_X)_M &=& - \sum_{\gamma}\eta^U_\gamma\[(\Tr T_X)_{U_\gamma}
+ (\Tr T_X)_{U^\gamma}\],\label{corb124}  \\
(\Tr T_b)_{M^a} &=& - 
\sum_{\gamma}\eta^{U^A}_\gamma\alpha^A_\gamma\[(\Tr T_b)_{U^A_\gamma}
+ (\Tr T_b)_{U_A^\gamma}\],
\label{corb125}\\
d_{M^a} &=& \sum_{\gamma}\(2\eta^{U^A}_\gamma\alpha^A_\gamma
d_{U^A_\gamma} + \eta^{V^A_\gamma}\delta^A_\gamma
d_{V^A\gamma}\) \label{corb126}\\ &=&
\sum_{\gamma}\[2\eta^{U^A}_\gamma(\alpha^A_\gamma)^2
d_{U^A_\gamma} + \eta^{V^A}_\gamma(\delta^A_\gamma)^2
d_{V^A_\gamma}\],\label{corb12} \eea
where the subscript $M$ stands for the gauge-charged light sector, and
$d_{M^a}$ is the dimension of the (generally reducible) gauge group
representation in the matter sector with modular weights $q^a_n$.  
In addition to \myref{azconds} and
\myref{azconds2}, the constraint \myref{cacond1} 
imposes further conditions on the parameters in $K(\Psi)$. 
The $\dY$ contributions to the first condition in \myref{cacond1},
namely
\beq \sum_{\lambda,A}\(\sum_i\pp_\mu t^i G^A_i 
+ i T_A\cdot A_\mu + i\Gamma_\mu\)
\(\sum_j\pp_\nu t^j G^A_j + i T_A\cdot A_\nu +i\Gamma_\nu\)q_X^A\im\lambda,
\label{op1} \eeq
and
\bea &&\im F^m(t^m)\[\sum_{\lambda,A}\(\sum_i\pp_\mu t^i G^A_i 
+ i T_A\cdot A_\mu + i\Gamma_\mu\)
\(\sum_j\pp_\nu t^j G^A_j 
+ i T_A\cdot A_\nu +i\Gamma_\nu\)q^A_m \right. \mmm \l+
\sum_{\lambda,N}\(\sum_j\pp_\mu t^j G_j^N - 2\pp_\mu t^n g^n_n + i\Gamma_\mu\)
\cdot\(\sum_k\pp_\mu t^k G_k^N - 2\pp_\mu t^n g^n_n + i\Gamma_\mu\)\right. 
\mmm \l
+ \sum_{\lambda,N}\(\sum_j\pp_\mu t^j G^N_j + i\Gamma_\mu\)
\(\sum_k\pp_\nu t^k G_k^N q^N_m + i\Gamma_\mu\)\],\label{op2prime}\eea
require
\bea (\Tr T_a^2T_X)_{M} &=& -\sum_{\gamma}\eta^U_\gamma
\[(\Tr T_a^2 T_X)_{U_\gamma}
+ (\Tr T_a^2 T_X)_{U^\gamma}\] \label{corb131}\\
1 &=&  \sum_{\gamma}\eta^N_\gamma(\del^N_\gamma)^3 = 
\sum_{\gamma}\eta^n_\gamma(\epsilon^n_\gamma)^3, \label{corb133}\\ 
(\Tr T_b T_c)_{M^a} &=& 
\sum_{\gamma}\lbr\eta^{U^A_\gamma}\alpha^A_\gamma
\[(\Tr T_b T_c)_{U^A_\gamma} +
(\Tr T_b T_c)_{U_A^\gamma}\] + \eta^{V^A_\gamma}
(\Tr T_b T_c)_{V^A_\gamma}\delta^A_\gamma\rbr,\label{corb134}\\
(\Tr T_b)_{M^a} &=& - 
\sum_{\gamma}\eta^{U^A}_\gamma(\alpha^A_\gamma)^2
\[(\Tr T_b)_{U^A_\gamma} + (\Tr T_b)_{U_A^\gamma}\],\label{corb135}\\
d_{M^a} &=& \sum_{\gamma}\[2\eta^{U^A}_\gamma(\alpha^A_\gamma)^3
d_{U^A_\gamma} + \eta^{V^A}_\gamma(\delta^A_\gamma)^3
d_{V^A_\gamma}\rbr.\label{corb13} \eea
Note that, for example, since the right
hand side of \myref{orbops} is equal to $ -2\(G^N_\alpha -
g^n_\alpha\)\(G^N_\beta - g^n_\beta\) - 2g^n_\alpha g^n_\beta$, we can
cancel \myref{orbops0} and \myref{orbops} if we replace the last term in
\myref{kpsi} and the condition \myref{corb123} by, respectively
\beq \sum_{N=n}\(e^{\delta^N_\gamma(G^N
- g^n)}|\Phi_\gamma^N|^2 + e^{\epsilon^n_\gamma
g^n}|\Phi_\gamma^n|^2\)\eeq
and 
\beq 2 = \sum_{\gamma}\eta^N_\gamma\del^N_\gamma =
\sum_{\gamma}\eta^N_\gamma(\del^N_\gamma)^2=
\sum_{\gamma}\eta^n_\gamma(\epsilon^n_\gamma)^2,\qquad
0 =\sum_{\gamma}\eta^n_\gamma\epsilon^n_\gamma.\eeq
Imposing the additional constraints \myref{cacond1} removes this 
ambiguity.  However we will see in \myapp{orb3} that there is 
at least one other choice of PV sector with this (and the following)
choice of $\dZ,\dY$ K\"ahler potentials.

A simple solution to the constraints \myref{corb121}--\myref{corb12}
and \myref{corb131}--\myref{corb13} is\footnote{The constraints \myref{corb125},
and \myref{corb135} are nontrivial only if
$T_{b}$ is a \uo\, generator; $V^A$ is a \uo\, singlet.}
\beq \del^N = \epsilon^n = 
\alpha^A = \delta^A = 1,\label{neworb0}\eeq 
in which case they reduce to
\bea 1 &=& \sum_\gamma\eta_\gamma^N = \sum_\gamma\eta_\gamma^n, \label{neworb1} \\
(\Tr T_b)_{M^a} &=& - \sum_\gamma\eta^{U_\gamma^A}
\[(\Tr T_b)_{U^A_\gamma} + (\Tr T_b)_{U_A^\gamma}\],\label{neworb3}\\
(\Tr T_b T_c)_{M^a} &=& \sum_\gamma\lbr\eta^{U^A_\gamma}
\[(\Tr T_b T_c)_{U_\gamma^A} + (\Tr T_b T_c)_{U_A^\gamma}\] +
\eta^{V^A_\gamma}(\Tr T_b T_c)_{V_\gamma^A}\rbr,\label{neworb6}\\
d_{M^a} &=& \sum_\gamma\[2\eta^{U_\gamma^A}d_{U_\gamma^A} + 
\eta^{V^A_\gamma}d_{V^A\gamma}\],\label{neworb5}\\
(\Tr T_a^2T_X)_{M} &=& -\sum_{A,\gamma}\eta^{U^A\gamma}
\[(\Tr T_a^2 T_X)_{U^A_\gamma} + (\Tr T_a^2 T_X)_{U_A^\gamma}\].
\label{neworb4} \eea
Note that \myref{neworb1} and \myref{neworb5} are equivalent to
\myref{corb121}.

\subsubsection{Preserving shift symmetry}\label{orb2}
The shift symmetries $\im t^I\to \im t^I + \alpha^I$, $\im s\to \im s
+ \beta$, $\alpha^I,\beta\in\R$, of the classical K\"ahler
potential are preserved if $K = K[(T^I + \T^I),(S + \S),\Phi^a$].
Then shift symmetry and modular covariance of the light particle
K\"ahler potential is preserved if it takes the form
\beq K = k(S + \S) + g + f(X^a), \quad X^a =
e^{g^a}|\Phi^a|^2, \quad g^a = \sum_n q^a_n g^n,\quad
f(X) = \sum_a X^a + O(X^2).\label{orbk2}\eeq
The metric is  
\bea K_{a\bar{b}} &=&
e^{g^a}\delta_{ab}f_a + \phi^b\bph^{\bar{a}}f_{ab}e^{g^a + g^b},\quad
K_{a\ibar} = \sum_bg^b_{\ibar}\bph^{\bb}K_{a\bb},\nonumber \\
K_{j\ibar} &=& g_{j\ibar}f_{n_i} +
\sum_{a,b}g^a_{j}g^b_{\ibar}\phi^a\bph^{\bb}K_{a\bb},
\quad f_{n_i} = \(1 +
\sum_aq_{n_i}^aX^af_a\),\label{orbmet}\eea
where 
\beq f_a = {\pp f\over \pp X^a}, \qquad f_{a b} 
= {\pp f\over\pp X^a\pp X^b}. \eeq
\beq \chi^n = \dot a g^n = \dot a\del^n_ig(T^i+\T^i).\eeq
If we define 
\beq k_{a\bb} = \pp_a\pp_{\bb} f(X^a) =
K_{a\bb}, \quad  k_{a\bb}k^{\bb c} = \delta^a_c, 
\quad \tg_{i\bj} = f_{n_i}g_{i\bj} = f_{n_i}g_i^2\del_{i j}, 
\quad \tg^{k\bj}\tg_{i\bj} =
\delta^k_i,\label{orbdefs}\eeq
the inverse metric is:
\beq K^{a\bar{b}} = k^{a\bb} + \phi^a\bph^{\bb}\sum_{\ibar j}
g_j^ag_{\ibar}^b\tg^{\ibar j}, \quad K^{j\ibar} = \tg^{\ibar j}
 = f^{-1}_{n_i}g_i^{-2}\del^{i j},\quad
K^{a\ibar} = - \phi^a\sum_j g^a_j\tg^{j\ibar}. \label{invorbk}\eeq
Using the properties $K^{\ibar j} = \del^{i j},\;g_i = g_{\ibar} = e^{g_{n_i}}$
and $(T_a)^i_j g_i = 0$, one can see that only \myref{azconds} is required for
\myref{cacond1} to be satisfied in this case; however a minimum of 3 sets of the
$\dZ,\dY$ is still  required.

To maintain invariance of the PV K\"ahler potential we take
\beq \chi^n = \dot a g^n = \dot a\del^n_ig(T^i+\T^i).\label{gchi2}\eeq
For $\dY$ we now have
\beq \Gamma^I_{N j} = - \dot a\del^i_j\del^i_n, \qquad 
\Gamma^I_{N \alpha} = {\dot a\over8}\del^i_n\chiproj\D_\alpha T^i = 0,\eeq
because $T^i$ is gauge invariant, and once \myref{azconds} is imposed
the contributions from $\dY$ involve the same operators as in 
\myref{orbops}, and are canceled by contributions from the fields
$\Psi = U,V,\Phi$ introduced in \myref{kpsi}--\myref{psi},
subject to the conditions \myref{corb121}--\myref{corb12}
and \myref{corb131}--\myref{corb13}.

\subsubsection{Heisenberg invariant untwisted sector K\"ahler
potential}\label{orb3} 
Here we set $\chi^n = \dot a g^n =\dot a
G^n$, where $G^n$ is defined in \myref{untw2}.  Once the conditions
\myref{azconds} and \myref{azconds2} are imposed, 
the $\dZ$ cancel the UV divergences from the light
fields $Z^p$. The contributions from $\dY_I$ decouple from those from
$\dY_N$ in the relevant sums.  Those from $\dY_I$ reduce to those from
fields with metric equal to the inverse of the untwisted sector metric
derived from \myref{untw2}, with an extra overall factor $e^{G^N}$,
and those from $\dY_N$ are the same as the contributions from a field
with metric $e^{G^N}$.  We may cancel their contributions to the UV
divergences with the method used in~\cite{pvdil} 
to cancel the UV divergences from the
light untwisted sector.  That is, the $\dY$ contributions to
the UV divergences can be canceled with $\Psi =
U,V,\Phi^N,\Phi^n_I,\Phi_n^I$ where $U,V,$ have the same K\"ahler
potential as in \myref{kpsi}. The $\Phi^N$, which are no longer all
gauge singlets, have K\"ahler potential
\beq K(\Phi^N) = \sum_\gamma\sum_{N=n}e^{\delta^N_\gamma(G^N - g^n)}
|\Phi^N_\gamma|^2,\label{kphin}\eeq
and the constraints \myref{corb121}--\myref{corb12} are 
appropriately modified. The fields
$\Phi^n_I$, $I = 0,\ldots,N_n$, have a K\"ahler potential of the same
form as $\dY_{I,N}$, but without the prefactor $e^{G^N}$ in
\myref{kdy}, and the fields $\Phi^I$ have the inverse K\"ahler metric:
\beq K^{\Phi^I}_{I\J}K_{\Phi_I}^{\J K} = \del^K_J,\qquad
I,J,K = 0,\ldots,N_n.\label{orthog}\eeq
This set cancels the remaining $\dY_{I,N}$ UV contributions if we impose
\beq \eta^{\Phi^I}_\gamma = \eta^{\Phi_I}_\gamma\equiv \eta^\Phi_\gamma, 
\qquad\sum_\gamma\eta^\Phi_\gamma = 0,
\qquad\sum_\gamma\eta^\Phi_\gamma(a^{\Phi}_\gamma)^2 
= - \sum_\gamma\eta^\Phi_\gamma(a^{\Phi}_\gamma)^4 = \half.
\label{lastphiconds}\eeq
However, it is not possible to cancel all the contributions
from $\dY$ to \myref{cacond1} because terms odd in
$(\Gamma_\mu^{\Phi_\gamma^n})^I_J$, $(T^{\Phi_\gamma^n}_a)^I_J$ and
$(\phi^{\Phi_\gamma^n})^I_J$ cancel between $\Phi^I$ and $\Phi_I$.
Terms proportional to $\Tr\(\dot\Gamma_\mu^{Y} \dot\phi^{Y}\)$ are
indeed canceled by virtue of \myref{lastphiconds}, but there is no
contribution from the $\Phi$ sector to cancel terms like $G_i\D_\mu
z^j(\dot T^Y_a\dot\phi^Y)^i_j$, for example.  These terms could be
made to vanish with $\sum_\lambda\eta^{\dZ}_\lambda\dot
a_\lambda^2 = -1$, but this contradicts the condition in
\myref{azconds}.  So in this case we need to modify 
\myref{kdy}--\myref{wzi2} as follows:
\beq K^{\dY} = \sum_Ae^{G^A}|\dY_A|^2 + \sum_N e^{G^N}\[g_n^{i\bj}\(\dY_I
- b G^n_i\dY_N\)\(\dbY_{\J} - b G^n_{\bj}\dbY_{\N}\) +
|\dY_N|^2\],\label{kdy3}\eeq
which is modular invariant provided under (\ref{modt})
\beq \dY'_A = e^{-F^A}\dY_A, \qquad\dY'_I = e^{-F^N}m_i^j\(\dY_J +
b F^n_j\dY_N\), \quad \dY'_N = e^{-F^{N}}\dY_N, \quad F^{N,A} =
\sum_m q^{N,A}_m F^m,\label{trdy3} \eeq
and the superpotential is given by \myref{wzi2}, except that
\bea W_n = \dot\mu_n(T^i)\(\sum_{P\in n}\dZ^P\dY_P + \dot a^{-1}b\dZ^N\dY_N\).
\label{wzi3}\eea
Now we have (see Appendix D of~\cite{pvdil}) 
\bea
\sum_\lambda\dot\eta_\lambda(\Gamma^n_{\dY})^\sigma_{\sigma\alpha} &=&
- (N_n + 1)\(G^N_\alpha - G^n_\alpha\), \label{gcond1}\\
\sum_\lambda\dot\eta_\lambda(\Gamma^n_{\dY})^\sigma_{\rho\beta}
(\Gamma^n_{\dY})^\rho_{\sigma\alpha} &=& - (N_n + 1)\(G^N_\alpha -
G^n_\alpha\)\(G^N_\beta - G^n_\beta\)+ 2(b_2 +  b_4)G^{n\alpha}_i\Z^i_\alpha
\ddd - (1 +  2b_2 + b_4)
\[G^n_\alpha G^n_\beta + (\hG^n)^i_{j\alpha}(\hG^n)^j_{i\beta}\],\label{gcond2}\\
\sum_\lambda\dot\eta_\lambda(\Gamma^n_{\dY})^J_{I\alpha}(T^a)^I_J &=&
(\Tr T^a)_{Z_n}\(G^N_\alpha - G^n_\alpha\) - (1 +  b_2)
(T^a)_i^j(\hG^n)^i_{j\alpha},\label{gcond3}\\ (\hG^n)^i_{j\alpha} &=&
\cP\(G^n_j\D_\alpha Z^i_n\), \qquad (\hG^n)_{i\alpha} =
\cP\(G^n_iG^n_j\D_\alpha Z^j_n\),\label{orbops3}\\
Z^i_\alpha &=& \cP\D_\alpha Z^i,\qquad\cP = -{1\over8}\chiproj,
\qquad b_p = -\sum_\lambda\dot\eta_\lambda
b^p_\lambda\label{Pbdefs}\eea 
If $b_p = 0$, we recover the inverse of the untwisted sector K\"ahler
potential.  However if 
\beq b_4 = - b_2 = 1,\label{bconds}\eeq
we eliminate both the terms arising from this contribution that cannot
be eliminated by $U,V,\Phi^I_N$, with
\beq K(\Phi^I_N) = \sum_{\gamma,N}\sum_{I=0}^{N_n}e^{\del^N_\gamma}
(G^N - G^n)|\Phi^I_N|^2, \eeq
as well as the new term involving the
last two operators in \myref{orbops3}.  A simple solution to \myref{bconds}
with three sets $\dY_\gamma$ is, for example
\beq (\eta,b) = (1,3), (-1,2), (-1,2).\label{bcondsol} \eeq
The constraints \myref{corb121}--\myref{corb12} now read
\bea \sum_{\Psi,\gamma}\eta^\Psi_\gamma &=& 3 + \sum_n N_n +
d_{M^T} = N + 2,\label{corb1}\\ C^a_{M} &=&
\sum_{\gamma}\lbr\eta^U_\gamma\[(\Tr T_a^2)_{U_\gamma} + (\Tr
T_a^2)_{U^\gamma}\] + \eta^V_\gamma(\Tr T_a^2)_{V_\gamma} +
\sum_N\eta^N_\gamma(\Tr T_a^2)_{\Phi^N_\gamma}\rbr, 
\label{corb2}\\ 1 &=& \sum_{\gamma}\eta^N_\gamma\del^N_\gamma =
\sum_{\gamma}\eta^N_\gamma(\del^N_\gamma)^2,\qquad (\Tr T_X)_{Z_n}  =
-\sum_{\gamma}\eta^N_\gamma\del^N_\gamma T^{N_\gamma}_X,\label{corb3}\\ 
(\Tr T_X)_{M} &=& -\sum_{\gamma}\(\eta^U_\gamma\[(\Tr T_X)_{U_\gamma}
+ (\Tr T_X)_{U^\gamma}\] + \eta^N_\gamma(T_X)_{\Phi^N_\gamma}\),
\label{corb4}\\  
(\Tr T_b)_{M^a} &=& - 
\sum_{\gamma}\eta^{U^A}_\gamma\alpha^A_\gamma\[(\Tr T_b)_{U^A_\gamma} +
(\Tr T_b)_{U_A^\gamma}\],\label{corb5}\\
d_{M^a} &=& \sum_{\gamma}\(2\eta^{U^A}_\gamma\alpha^A_\gamma
d_{U^A_\gamma} + \eta^{V^A}_\gamma\delta^A_\gamma
d_{V^A_\gamma}\) \eee
\sum_{\gamma}\[2\eta^{U^A}_\gamma(\alpha^A_\gamma)^2
d_{U^A_\gamma} + \eta^{V^A}_\gamma(\delta^A_\gamma)^2
d_{V^A_\gamma}\],\label{corb6} \eea
where $M^a$ now includes only twisted sector matter.  

For the relevant $\dY$ contributions to \myref{cacond1} we have 
\bea
\sum_\lambda\dot\eta_\lambda\Tr\[(\Gamma^n_{\dY})_{\mu\nu}\Phi_{\dY}\] &=&
- \(G^N_{\mu\nu} - G^n_{\mu\nu}\)\[(N_n + 1)(F - F_N) + \sum_i q^i_X\lambda_X\]
\ddd + (1 + b_2)Y_{\mu\nu},\label{Gcond1}\\
\sum_\lambda\dot\eta_\lambda\Tr\(\tilde\Gamma^n_{\dY}\cdot
\Gamma^n_{\dY}\Phi_{\dY}\)
 &=& - \(\tG^N - \tG^n\)\cdot\(G^N - G^n\)\[(N_n + 1)(F - F^N)
+ \sum_i q^i_X\lambda_X\]\ddd +  (1 +  b_2)Y + (b_2 + b_4)Z,\\
\sum_\lambda\dot\eta_\lambda\Tr\[(\Gamma^n_{\dY})_{\mu\nu}(T^a)_{\dY}\Phi_{\dY}\] 
&=&\(G^N_{\mu\nu} - G^n_{\mu\nu}\)\[(\Tr T^a)_{Z_n}(F - F^N) + (\Tr T^a T_X)_{Z_n}
\lambda_X\] \ddd+ (1 + b_2)Y^a_{\mu\nu},\\
\sum_\lambda\dot\eta_\lambda\Tr\[(T^a)_{\dY}(\Gamma^n_{\dY})_{\mu\nu}\Phi_{\dY}\] 
&=&\(G^N_{\mu\nu} - G^n_{\mu\nu}\)\[(\Tr T^a)_{Z_n}(F - F^N) + (\Tr T^a T_X)_{Z_n}
\lambda_X\]\ddd + (1 + b_2)Z^a_{\mu\nu}.\label{Gcondl3}\eea
The precise form of the various operators $Y,Z$ in the above is unimportant,
since they drop out when \myref{bconds} is imposed, and the remaining
contribution is canceled by $U,V,\Phi^N$, with the additional conditions
\bea (\Tr T_a^2T_X)_{M} &=& -\sum_{\gamma}\eta^U_\gamma
\[(\Tr T_a^2 T_X)_{U_\gamma} + (\Tr T_a^2 T_X)_{U^\gamma}\] 
-\sum_{N,\gamma}\eta^N_\gamma(\Tr T_a^2 T_X)_{\Phi^N_\gamma}
\label{cGorb131}\\
(\Tr T_X)_{Z_n} &=& \sum_{\gamma}\eta^N_\gamma(\Tr T_X)_{\Phi^N_\gamma}
(\delta^N_\gamma)^2,\qquad (\Tr T_b T_c)_{Z_n} = 
\sum_{\gamma}\eta^N_\gamma(\Tr T_b T_c)_{\Phi^N_\gamma}
\delta^N_\gamma, \label{cGorb133}\\ 
(\Tr T_b T_c)_{M^a} &=& \sum_{\gamma}\lbr\eta^{U^A_\gamma}
\alpha^A_\gamma\[(\Tr T_b T_c)_{U^A_\gamma} +
(\Tr T_b T_c)_{U_A^\gamma}\] + \eta^{V^A_\gamma}
(\Tr T_b T_c)_{V^A_\gamma}\delta^A_\gamma\rbr,\label{cGorb134}\\
(\Tr T_b)_{M^a} &=& - \sum_{\gamma}\eta^{U^A}_\gamma(\alpha^A_\gamma)^2
\[(\Tr T_b)_{U^A_\gamma} + (\Tr T_b)_{U_A^\gamma}\],\label{cGorb135}\\
1 &=&  \sum_{\gamma}\eta^N_\gamma(\del^N_\gamma)^3,\label{cGorb136}
 \\
d_{M^a} &=& \sum_{\gamma}\[2\eta^{U^A}_\gamma(\alpha^A_\gamma)^3 +
\eta^{V^A}_\gamma(\delta^A_\gamma)^3
d_{V^A_\gamma}\].\label{cGorb13} \eea
As before these constraints have a simple straightforward solution:
\bea 1 &=& \alpha^A = \gamma^N = \delta^N, \qquad 1
= \sum_{\gamma}\eta^N_\gamma, \label{corb7}\\ 
C^a_{M_b} &=&
\sum_{\gamma}\lbr\eta^{U^B}_\gamma\[(\Tr T_a^2)_{U^B_\gamma} + (\Tr
T_a^2)_{U^\gamma_B}\] + \eta^{V^B}_\gamma(\Tr T_a^2)_{V^B_\gamma}\rbr,
\label{corb8}\\ 
(\Tr T_b)_{M_a} &=& -\sum_{\gamma}\eta^{U^A}_\gamma
\[(\Tr T_b)_{U^A_\gamma}
+ (\Tr T_b)_{U^\gamma_A}\]\qquad (\Tr T_b)_{Z_n}  =
-\sum_{\gamma}\eta^N_\gamma T^{N_\gamma}_b,\label{corb9}\\  
d_{M^a} &=& \sum_{\gamma}\(2\eta^{U^A}_\gamma d_{U^A_\gamma} 
+ \eta^{V^A}_\gamma d_{V^A_\gamma}\), \qquad C^a_{Z_n} = 
\sum_\gamma\eta^N_\gamma(\Tr T_a^2)_{N_\gamma}.\label{corb10} \eea
Note that the K\"ahler potential $g^n$ used in the previous two
subsections is just the limiting case $\Phi^A_n\to 0$ of $G^n$,
so this gives an alternative PV sector for those cases.
They differ in the resulting sum rules that are quadratic and
cubic in the modular weights.

\subsubsection{Anomalous masses}\label{anmass}
The conditions \myref{corb121}--\myref{corb12} and 
\myref{corb131}--\myref{corb13} imply that the
the $\Psi$ masses are not modular invariant (at least in the absence
of moduli-dependent mass factors) and are not in general \ux\, invariant.
Their K\"ahler potential is invariant under K\"ahler
transformations arising from nonlinear transformations on the light
fields: 
\beq Z^p\to Z'^p(Z), \quad dZ'^p = M^p_qdZ^p, \quad K(Z') =
K(Z) + F(Z), \quad W(Z') = e^{-F(Z)}W(Z), \eeq 
provided 
\bea U'^A_\gamma
&=& e^{-\alpha_\gamma^A F^A}U^A_\gamma, \qquad U'^\gamma_A
= e^{-\alpha_\gamma^A F^A}U_A^\gamma, \qquad V'^A_\gamma
= e^{-\gamma_\gamma^A F^A}V^A_\gamma, \label{UVtr}\\
\Phi'^N_\gamma
&=& e^{-\delta_\gamma^N(F^N - F^n)}\Phi^N_\gamma, \qquad
\Phi'^n_\gamma = e^{-\epsilon_\gamma^n F^n}\Phi^n_\gamma.
\label{Phitr}\eea 
%
The K\"ahler potential is also gauge invariant; in Yang-Mills
superspace~\cite{bggm} the superfields $Z^p$ are defined to be
covariantly chiral. In particular, under
\ux\,
\beq Z^a\to g^{q_a}Z^a, \qquad Z^{\bar a}\to
g^{-q_a}Z^{\bar a}, \qquad \cA_M\to \cA_M + g^{-1}\D_M g\label{uxtr} \eeq
with the gauge covariant superspace derivative $\D_M$ given (neglecting
other connections) by
\beq \D_M Z^p = (D_M + q^p\cA_M)Z^p.\eeq
For the PV fields $\Psi^A = U^A,U_A$ we take instead, with \myref{gaugetr}, 
\bea \Psi^A &\to& e^{-q^A\Lambda}\Psi^A, \qquad
\bar\Psi^{\bar A}\to e^{-q^A\bL}\bar\Psi^{\bar A}.\label{uxpv}\eea
It is necessary to introduce the vector superfield $V_X$ in order that
the \ux-violating terms in the PV superpotential remain holomorphic
under a \ux\, transformation. In addition, in the regulated theory
noninvariance under \ux arises from the PV masses; this must cancel
the noninvariance of the GS term \myref{lgs} that explicitly involves
$\vx$.  Working in ``partial'' \ux\, superspace allows us to meet
these conditions while keeping the notation from becoming too
cumbersome.  With this modification the K\"ahler potential in \myref{psi}
for $U$ is replaced by 
\beq K(U) = \sum_{\gamma,A}\(e^{q^A_\gamma\vx
+ \alpha^A_\gamma G^A}|U^A_\gamma|^2
+ e^{p^A_\gamma\vx + \alpha^A_\gamma G^A}|U^\gamma_A|^2\).
\label{psi2}\eeq
%
%
%
%
%
In this way we have put all the anomaly associated with the T-moduli
and gauge charged chiral fields in the superpotential
$W(\Psi)$ in \myref{wpsi}.  We define 
\beq \dot\omega_m^n = q_m^N - 1,
\quad \dot\omega_m^a = q_m^A + q_m^a - 1,\label{omegadot}\eeq 
as the (negative of the) modular weights of $\dot\mu_{n,a}$ in the
superpotential \myref{wzi2}. If, for illustration, we use the PV sector
of \myapp{orb1} or \myapp{orb2}, the $\Psi$ have modular weights 
\bea q^{\Phi^N_\gamma}_m &=& q^N_m - 2\del^n_m =
\dot\omega_m^n - 2\del^n_m + 1, 
\qquad q^{\Phi^n_\gamma}_m = q^N_m =
\dot\omega^n_m + 1 ,\label{qmPhi}\\
q^{U^A_\gamma}_m &=& q^A_m = \dot\omega_m^a - q_m^a + 1, \qquad q^{U_A^\gamma}_m =
q^A_m = \dot\omega_m^a - q_m^a + 1, \label{qmU}\\
q^{V^A_\gamma}_m &=& q^A_m = \dot\omega_m^a - q_m^a + 1\label{qmV},\eea
where we have assumed\footnote{The assumption does not affect the results
below, except for the case with $p=4$ in \myref{MPsip}.} \myref{neworb0}.
Then, if we further assume that $\omega^{\Psi_\gamma}_n$ is independent
of $\gamma$, using the sum rules \myref{corb121}--\myref{corb12} we have
\bea \sum_{\Psi,\gamma}\eta^\Psi_\gamma q_m^{\Psi_\gamma}  
&=& \sum_a\(\dot\omega^a_m - q^a_m + 1\) + 2\sum_n
\(\dot\omega^n_m - \del^n_m + 1\)\eee
N + 2 - \sum_\lambda\dot\eta_\lambda\(2\sum_n
\dot\omega^n_m + 
\sum_a\dot\omega^a_m\) - \sum_{p} q^{p}_m,\label{sumqpsi}\eea
where $p$ refers to $t$-moduli and all gauge-charged matter, with, for
the untwisted sector charged matter $q^{a i}_m = \del^i_m$, and
for $t$-moduli $q^i_m = 2\del^i_m$.  In writing the last line
of \myref{sumqpsi} we used $\sum_\lambda\dot\eta_\lambda =-1$ and, 
including the dilaton, 
\beq N = d_M + 4 = 
\sum_{\Psi,\gamma}\eta^\Psi_\gamma - 2.\label{condqx}\eeq
Then $\tilde m^\Psi_\gamma$ is given by \myref{dmp} with, using also
the constraint \myref{corb124} on $(\Tr T_X)_\Psi$,
\bea \sum_{\Psi,\gamma}\eta^\Psi_\gamma\tph^+_{\Psi_\gamma} &=&
\sum_{\Psi,\gamma}\eta^\Psi_\gamma\[\sum_m\(1 - 2q^{\Psi_\gamma}_m
+ \omega^{\Psi_\gamma}_m\)F^m - 2q^{\Psi_\gamma}_X\lambda\]\eee
\sum_m\(2\sum_{p}q^{p}_m - N - 2 + \sum_{P = \Psi,\dY,\dZ}
\eta^P\omega^P_m\)F^m + 2\Tr T_X\lambda.\label{psitphi}\eea
When combined with other PV loop contributions, we get the general
result \myref{delsr}.  Similarly,  there is a contribution proportional
to (for $a\ne X$)  
\bea\sum_C\eta^C\tph^+_C(T^a)^2_C\hc &=& 
\sum_{\Psi,\gamma}\eta^\Psi_\gamma\tph^+_{\Psi_\gamma}C^a_{\Psi_\gamma}
+ \sum_{\varphi_\gamma}\eta^{\varphi_\gamma}\tph^+_{\varphi_\gamma}C_a
\hc\eee \sum_n\[C_a\(1 + \sum_{\gamma}\tilde\omega_n^{\tph^a_\gamma}\)
- \sum_bC^a_{M^b}\(1 - 2 q^b_n - \sum_{P = \Psi,\dY,\dZ}
\eta^{P^B}\omega^{P^B}_n\)\]F^n\ddd 
+ 2\Tr(T^a)^2T_X\lambda\hc,\label{tphita2}\eea
where we used \myref{corb131}, \myref{corb134} and \myref{etavarphi}.
This result is again completely general, and agrees with \myref{chiral}
when multiplied by $1/64\pi^2$ with $\phi^b = i\(q^b_n\im F^n
- \half\im F + q^b_X\im\lambda\)$.  The conditions that the anomalies
be canceled places constraints on the threshold factors as 
functions of charges; the coefficients of $F^n$
must be independent of $n$ in \myref{delsr}, and 
independent of $n$ and $a$ in \myref{tphita2}. To insure this, we set
\beq \tilde\omega_n^{\tph^a} \equiv 
\sum_\gamma\tilde\omega_n^{\tph^a_\gamma} = {1\over C_a}\[C_{G S}
+ \sum_bC^a_{M^b}\(1 - 2 q^b_n - \sum_{P = \Psi,\dY,\dZ}
\eta^{P^B}\omega^{P^B}_n\)\] - 1,\eeq
where $C_{G S}= 8\pi^2b$ is the Green-Schwarz coefficient, and
and we identify the coefficients $b^a_n$ of the threshold corrections
as
\beq b^a_n = C_a\tilde\omega^{\tph^a}_n + 
\sum_b C^a_{M^b}\sum_{P}\eta^{P^B}
\omega_n^{P^B} = C_{G S} - C_a + \sum_b C^a_{M^b}\(1 - 2q^b_n\).
\label{orbcond}\eeq
For the anomalous coefficient of $r\cdot\tr$, the expression in square 
brackets in \myref{delsr} is also equal to
the explicit contribution from just the fields with noninvariant 
masses:
\bea & & - \sum_m\[2\sum_{p}q^{p}_m - N - 2 + N_G + \sum_{P = \Psi,\dY,\dZ}
\eta^P\omega^P_m + \sum_a\tilde\omega^{\tph^a}_m
+ \sum_{\gamma}\hat\eta_\gamma\(1 - 2
\hat \alpha_\gamma + \hat\omega_m^{\hat\phi_\gamma}\)\]F^m\mmm
- 2\Tr T_X\lambda = -\sum_C\eta_C\tph^+_C = 
-24\(C_{G S}F + C'_{G S}\lambda\) - 24F,\label{psitphi3}\eea
where $\hat\phi_\gamma$ is the subset of neutral chiral multiplets
$\phi_\gamma$ that has noninvariant masses, and 
the coefficient $C'_{G S}$ of the
\ux\, GS term is defined in \myref{5.5} and \myref{uxconds}.  
In writing the last
equality in \myref{psitphi3} we used string theory results for the
anomalous coefficient of $\Omega_W$ in the full result \myref{LW},
including the contributions \myref{affan} and \myref{LGBanon} that are
not included in the PV contribution. String results require that
the coefficient of $F^m$ in \myref{psitphi3} be independent of $m$.

If we calculate the coefficient of $X_{\mu\nu}\tilde X^{\mu\nu}$
following the arguments leading to \myref{delsr}, using 
\myref{condquad} and \myref{alph3} as well as \myref{sigs} and
\myref{cond1}, we obtain $-2$ times the RHS of \myref{delsr} with $\omega_n$
in the expression in brackets replaced
by\footnote{The superpotential mass term $\mu^\varphi$ is constant
with $\omega^\varphi=0$ for $\varphi^a,\hph^a$ as required by modular
covariance for this term.}
\beq \sum_C\omega_n^C\eta^C(1-\alpha_C)^2 = \sum_{P = \Psi,\dY,\dZ}
\eta^P\omega^P_n + \sum_a\tilde\omega^{\tph^a}_n + \sum_\gamma
\eta^\phi_\gamma\omega_n^{\phi_\gamma}(1 - \alpha_\gamma)^2.
\label{altth1}\eeq
Alternatively, if we calculate directly using only the noninvariant
PV fields, we obtain ($- 2$ times) \myref{psitphi3} with the last term replaced
by 
\beq   \sum_{\gamma,m}\hat\eta_{\gamma}\[(1 - 2\hat\alpha_\gamma)^3
+ (1 - 2\hat\alpha_\gamma)^2\hat\omega_m^{\hat\phi_\gamma}\]F^m.
\label{altth2}\eeq
Comparing \myref{psitphi3} with\myref{delsr} and \myref{altth2} with
\myref{altth1}, consistency requires\footnote{The constraints \myref{sigs}
and \myref{alphayz2} assure that $\omega^{\tZ,\hY}$ drops out of 
these equations.}
\bea\sum_\gamma\hat\eta_{\gamma}\(1 - 2\hat\alpha_\gamma +
\hat\omega_m^{\hat\phi_\gamma}\) &=& 5 +
\sum_\gamma\eta^\phi_\gamma\omega_m^{\phi_\gamma},\label{D86}\\ 
\sum_\gamma\hat\eta_{\gamma}\[(1 - 2\hat\alpha_\gamma)^3 + (1 -
2\hat\alpha_\gamma)^2\hat\omega_m^{\hat\phi_\gamma}\] &=& 5 +
\sum_\gamma\eta^\phi_\gamma(1 - 2\alpha_\gamma)^2\omega_m^{\phi_\gamma}
\label{D87},\eea 
in conformity with the sum rules
\bea \sum\eta^\phi_\gamma &=& N' - N'_G + N + 2 - 3N_G = - 15 - 2N_G, 
\label{phicondN}\\
\sum\eta_\gamma^\phi\alpha_\gamma &=& - 10 - N_G,\label{phiconda1}\\
\sum\eta_\gamma^\phi\alpha_\gamma^2 &=& - 4 - N_G,\label{phiconda2}\\
\sum\eta_\gamma^\phi\alpha_\gamma^3 &=& - 1 - N_G,\label{phiconda3}\eea
which are obtained by subtracting the contributions of $\theta_\gamma,
\dZ,\varphi^a,\hph^a,\tph^a$ from $N'$ and subtracting
 the contribution of $\varphi^a$, with $\alpha^{\varphi^a} = 1$ from
the sum rules for $\alpha^n$. The constraints 
\myref{phicondN}--\myref{phiconda3} imply in particular
\beq \sum_\gamma\eta^\phi_{\gamma}\(1 - 2\alpha_\gamma\) = 
\sum_\gamma\eta^\phi_{\gamma}\(1 - 2\alpha_\gamma\)^3 = 5,\label{chph13}\eeq
and we obtain for the coefficient of $\Xc\Xa$ in $\Phi_L$ [see (E.9)]
\bea & & \sum_m\[2\sum_{p}q^{p}_m - N + 3 + N_G + \sum_{P = \Psi,\dY,\dZ}
\eta^P\omega^P_m + \sum_a\tilde\omega^{\tph^a}_m
+ \sum_{\gamma}\hat\eta_\gamma\(1 - 2\hat \alpha_\gamma\)^2
\hat\omega_m^{\hat\phi_\gamma}\]F^m+ 2\Tr T_X\lambda\mmm\qquad
= \sum_C\eta_C\tph^+_C(1 - 2\alpha_C)^2\mmm\qquad
= 24\(C_{G S}F + C'_{G S}\lambda\) + 24F +  
\sum_{\gamma,m}\hat\eta_\gamma\[\(1 - 2\hat \alpha_\gamma\)^2 - 1\]
\hat\omega_m^{\hat\phi_\gamma}F^m.\label{psitphi4}\eea
The last term in\myref{psitphi4} vanishes in the absence of
threshold corrections, or more generally if we impose
\bea  \sum_\gamma\hat\eta_{\gamma}\(1 - 2\hat\alpha_\gamma\)^2
\hat\omega_m^{\hat\phi_\gamma} = 
\sum_\gamma\hat\eta_{\gamma}\hat\omega_m^{\hat\phi_\gamma}\label{d97}.\eea
If we define $\phi_\gamma = \hf_\gamma,\tphi_\gamma$, where $\tphi_\gamma$
has a modular invariant mass term: $1 - 2\tilde\alpha_\gamma +
\tilde\omega^{\tphi_\gamma}_m = 0$, so that 
\beq \tilde\omega^{\tphi_\gamma}_m \equiv \tilde\omega^{\tphi}_\gamma\eeq
is independent of $m$, it is straightforward to show that
\myref{chph13} is equivalent to  \myref{D86}--\myref{D87}. A third 
constraint reads
\beq \sum_\gamma\eta^\phi_{\gamma}\(1 - 2\alpha_\gamma\)^2 = 
\sum_\gamma\hat\eta_{\gamma}\(1 - 2\hat\alpha_\gamma\)^2 +
\sum_\gamma\tilde\eta_{\gamma}\(\tilde\omega^{\tphi}_\gamma\)^2 = - 2N_G + 9,
\label{chph2}\eeq
A simple solution to these constraints, that satisfies \myref{d97} is to set 
$\tilde\omega^{\tphi}_\gamma 
= (1- 2\tilde\alpha_\gamma)=0$ with
\beq \tilde\eta_\gamma = -1,\qquad \sum_\gamma\tilde\eta_\gamma = -24, \qquad
\tilde\alpha_\gamma = \half,\label{feta}\eeq
and to take $\hf_\gamma = (\hf^0,\hf^\pm)$ with
\beq \hat\alpha^0_\gamma = 0,
\qquad \sum_\gamma\hat\eta^0_\gamma = 5,\qquad
1-2\hat\alpha_\gamma^{\pm} = \pm 1,
\qquad \sum_\gamma\hat\eta^{\pm}_\gamma = -N_G + 2.\label{hfeta}\eeq
In this case the coefficient of $F^m$ in \myref{psitphi3} is 
independent of $m$ provided $\hat\omega_n^{\hat\phi_\gamma}$ satisfies 
\bea \hat\omega_n\equiv
\sum_\gamma\hat\omega_n^{\hat\phi_\gamma}&=&
24C_{G S} - \sum_{P = \Psi,\dY,\dZ} \eta^{P}\omega^{P}_n
-\sum_a\tilde\omega_n^{\tph^a} + N - N_G + 21 - 2\sum_p q^p_n,
\label{homegan}\eea 
where
\beq \omega^p_n = \sum_{P = \Psi,\dY,\dZ}\eta^{P^p}\omega^{P^p}_n
= \sum_\lambda\eta^{\Psi^p}_\lambda\omega^{\Psi^p}_n - 2\dot\omega^p_n,
\qquad \omega_n = \sum_{P = \Psi,\dY,\dZ}\eta^{P}\omega^{P}_n =
\sum_\lambda\eta^{\Psi}_\lambda\omega^{\Psi}_n- 2\sum_p\dot\omega^p_n.\eeq
We identify the corresponding threshold corrections as 
\beq b_n^r = \sum_P\omega_n^P +
\sum_a\tilde\omega^{\tph^a}_n + \hat\omega_n =24C_{G S} - N_G
+ N + 21 - 2\sum_p q^p_n.\label{D98}\eeq 
The ``F-term operator $\Omega_{X^m}$ and the ``D-term'' 
operator $\Omega_m$ also contain terms linear 
in the PV sector parameters; the linear terms in \myref{E5}--\myref{E14}
have a single factor
\bea 
\ln\cM &=& \half\(K - 2\ln f + \ln|\mu^2|\).\label{lnM}\eea 
The sum rule \myref{chph2} assures consistency between the coefficient of
derivatives of $K$ as calculated by summing over all PV states as in \myref{delsr}
and \myref{psitphi3}, and by summing over only states with noninvariant
masses as in \myref{altth1} and \myref{altth2}. However in the former case,
when using the sum rules \myref{cond1} and \myref{condquad}, we have to subtract
a spurious contribution\footnote{The actual contribution from these fields
is 2$\sum_a(1 - \alpha^{\varphi^a} - \alpha^{\hph^a})^2 = 0$.}
\beq \sum_a\[(1 - 2\alpha^{\varphi^a})^2 + (1  - 2\alpha^{\hph^a})^2\] = 2N_G,
\qquad \alpha^{\varphi^a} = 1, \qquad \alpha^{\hph^a} = 0,\eeq
giving 
\bea && \sum_n\(N + N_G - 7 - \sum_{P = \Psi,\dY,\dZ}
\eta^P\omega^P_n - \sum_a\tilde\omega^{\tph^a}_n - \sum_\gamma
\eta^\phi_\gamma\omega_n^{\phi_\gamma}(1 - 2\alpha_\gamma)
 - 2\sum_p q^p_n\)F^n - 2\Tr T_X\lambda\eee
- \sum_m\[2\sum_{p}q^{p}_m - N - 2 + N_G + \sum_{P = \Psi,\dY,\dZ}
\eta^P\omega^P_m + \sum_a\tilde\omega^{\tph^a}_m\right.\mmm\l\qquad
+ \sum_{\gamma}\hat\eta_\gamma\(1 - 2\hat \alpha_\gamma\)\(1 - 2
\hat \alpha_\gamma + \hat\omega_m^{\hat\phi_\gamma}\)\]F^m
- 2\Tr T_X\lambda \eee -
\sum_C\eta_C\tph^+_C(1 - 2\alpha_C),\label{ak1}\eea
\nnn
which is consistent with \myref{chph2}.  

The chiral projection $\Phi_L$ of $\Omega_L$ is given by
\beq \Phi_L = \chiproj\Omega_L = 16\(\Xc - 2\fc)(\Xa - 2\fa\).\eeq
Using the above results we obtain
\beq {1\over192}\Tr\(\eta\tilde\Phi^+\Phi_L\) =
{1\over12}\sum_C\eta_C\Phi^+_C
(1 - 2\alpha_C)^2\Xc\Xa + {1\over3}\sum_{\Psi,\gamma}\eta^{\Psi}_{\gamma}
\Phi^+_{\Psi_\gamma}\fc_{\Psi_\gamma}\(\fa^{\Psi_\gamma} - \Xa\)
\label{sumF1}\eeq
where $\Phi^+_C$ is a chiral superfield with 
$\l\Phi^+_C\r = \tph^+_C$,
\bea \fa^n &=& \sum_m(1 + \dot\omega^n_m)g^m_\alpha - \ell^n_\alpha
,\qquad \fa^P = \sum_n(1 + \dot\omega^p_n - q^p_n)g^n_\alpha - q^p_X\Wa^X - 
\ell^P_\alpha\\
\Phi^+_n &=& - \sum_m(1 - \omega^n_m)F^m,\qquad \Phi^+_p = 
- \sum_n(1 - \omega^p_n - 2q^p_n)F^n + 2q^p_X\Lambda_X.
\label{TrOL0}\eea
The term in brackets can be evaluated using the conditions 
\myref{omegadot}--\myref{qmV}, together with the relation
\beq C_{G S} = \sum_p(\omega^p_l + 2q^p_l- 1)(q^p_X)^2\qquad
\forall\quad l,\label{D117}\eeq
which, like \myref{orbcond}, is known to be satisfied in orbifold
compactifications of the heterotic string.  If we take, again for
illustration, 
\beq \ell^C(Z,\Z) = 0,\qquad C = n,P,\label{ellp1}\eeq
when \myref{sumF1} is
combined with the other terms in $\Omega'_L$, defined in
\myref{OprimeL}, we obtain the result given in \myref{TrOL} with
\bea A_{m n l} &=& \sum_k(\omega^k_l - 1)
(1 + \dot\omega^k_n)(1 + \dot\omega^k_m)\ddd
+ \sum_p(\omega^p_l + 2q^p_l- 1)(1 + \dot\omega^p_n - q^p_n)
(1 + \dot\omega^p_m - q^p_m)
\label{D110}\\
A_{n l} &=& \sum_k(\omega^k_l - 1)(1 + \dot\omega^k_n)
+ \sum_p(\omega^p_l + 2q^p_l- 1)(1 + \dot\omega^p_n - q^p_n)
\\
A_l &=& 
\sum_p(1 - \omega^p_l - 2q^p_l)q^p_X\\
B_{n l} &=& \sum_p(1 - \omega^p_l - 2q^p_l)
(1 + \dot\omega^p_n - q^p_n)q^p_X\\
 a_{m n} &=& 2\sum_p q^p_X(1 + \dot\omega^p_n - q^p_n)
(1 + \dot\omega^p_m - q^p_m)\\
 a_n &=& 2\sum_p q^p_X(1 + \dot\omega^p_n - q^p_n)\\
a &=& - 2\sum_p(q^p_X)^2 = - 2C^M_X,\\
 b_n &=& - 2\sum_p(q^p_X)^2(1 + \dot\omega^p_n - q^p_n).
\label{D116}\eea

The anomalous part of the Lagrangian also contains terms quartic in
the PV parameters; these are not constrained by the requirement of
finiteness alone.   As an example we can assume the simple solutions 
defined by \myref{neworb0}, \myref{neworb1}, 
and \myref{feta}--\myref{hfeta}.   
If we further assume $\hat\omega^{\hf^{0,\pm}_\gamma}_n 
= \hat\omega^{0,\pm}_n$ is 
independent of $\gamma$, The real superfields $\cM_{\hf_C}$ take the form 
\beq \cM^{0,\pm}_\gamma = e^{K/2}\mu^{0,\pm}_\gamma
\prod_n|\eta(T^n)|^{2\hat\omega^{0,\pm}_n}.\eeq
The constant parameters $\ln\mu$ drop out in the variation of 
the action and in derivatives of $\ln\cM$, and we obtain  
\bea&&\sum_\gamma\hat\eta_\gamma\(\cO\ln\cM_{\hf_\gamma}\)^p \to
{5\over2^p}\(\cO K + 4\sum_n\hat\omega^0_n\cO|\eta(T^n)|\)^p\mmm
+ {2 - N_G\over2^p}\[\(\cO K + 4\sum_n\hat\omega^+_n\cO|\eta(T^n)|\)^p +
\(4\sum_n\hat\omega^-_n\cO|\eta(T^n)| - \cO K\)^p\]
\label{D119},\\
&&\sum_a\eta_{\tph^a}(\cO\ln\cM_{\tph^a})^p \to {N_G\over2^p}
\(\cO K + 4\sum_n\tilde\omega^{\tph^a}_n\cO|\eta(T^n)|\)^p\eea
where $\cO$ stands for any operator, and 
\beq \hat\omega_n = 5\hat\omega^0 + (2 - N_G)\(\hat\omega^+
+ \hat\omega^-\),\qquad \hat b_n^\phi = (N_G - 2)\hat\omega^-,\eeq
with $\hat\omega^n$ given by \myref{homegan}.

If in addition to \myref{neworb1}--\myref{neworb4} we assume 
\bea (\Tr T_b T_c T_d)_{M^a} &=& -\sum_\gamma\eta^{U^A_\gamma}
\[(\Tr T_b T_c T_d)_{U_\gamma^A} + (\Tr T_b T_c T_d)_{U_A^\gamma}\]\ddd +
\eta^{V^A_\gamma}(\Tr T_b T_c T_d)_{V_\gamma^A},\label{neworb7}\\
(\Tr T_a^2T^2_X)_{M} &=& \sum_{A,\gamma}\eta^{U^A\gamma}
\[(\Tr T_a^2 T^2_X)_{U^A_\gamma} + (\Tr T_a^2 T^2_X)_{U_A^\gamma}\],
\label{neworb8} \eea
we obtain 
\beq \sum_P\eta^\Psi_P(\cO\ln\cM_{\Psi^P})^p \to \sum_{n=1}^3 
(\cO\ln\cM_n)^p + \sum_{q = 1}^{N-1}(\cO\ln\cM_q)^p,\label{MPsip}\eeq
with 
\beq \ln\cM_n = {K/2 - g - \sum_m\dot\omega_m^n g^m},\qquad
\ln\cM_p = {K/2 - g + \sum_n(q^p_n - \dot\omega_n^p)g^n + q^p_X\vx},\eeq
where $q = {m = 1,2,3; a}$ with $q^m_n = 2\del^m_n,\;q^m_X = 0$
for the moduli, and $a$ denotes the gauge-charged matter fields.

As mentioned above, the terms quadratic and higher in weights are
modified if use instead the PV sector of \myapp{orb3}.  They are also
modified if $\ell^p \ne 0$; a natural choice might be
\beq \ell^P = K - g.\label{ellp2}\eeq

\subsubsection{Orbifold models}\label{fiqs}

Here we list the quantum numbers needed to evaluate the anomaly
coefficients for three orbifold compactification models for which all
the modular weights and gauge charges are known. These are $Z_7$ and
$Z_3$ models with no string threshold corrections: $b^{a,r}_n =
\omega^C_n = 0$, so the conditions \myref{orbcond} and \myref{D98}
reduce to \myref{wthconds}.  One can check that these are
satisfied\footnote{They are also satisfied for $T_a\in U(1)_a$ for the
nonanomalous \uo\, factors whose charges are not listed.  The fields
in a given set are not degenerate under under these \uo's.}  from the
tables given below, where we group the chiral multiplets according to
their representation (rep) under the semi-simple part of the gauge
group, and their modular weights $q_n$, and give the multiplicity
($n_s$) of each set with fixed quantum numbers, as well as the total
multiplicity $n$ for each set: $N = \sum n$. For the two $Z_3$ models
the superfields $T^{i\bj}$ are the K\"ahler moduli of which only the
diagonal elements $T^i = T^{i\ibar}$ are assumed to have nonvanishing
vacuum values.  In all the tables $S$ is the dilaton superfield, and the
indices $i,j = 1,2,3$.  The first two models were studied at the
string level in~\cite{ssanom}. They have no Wilson lines and no
anomalous \uo\, so they have 
\beq 8\pi^2b = C_{G S} = 30, \qquad \dx = C'_{G S} = 0,\eeq 
and all chiral multiplets are $E_8$ singlets.
For the third (FIQS) model,\footnote{See the last reference
in~\cite{iban}.} which is a $Z_3$ orbifold model with Wilson lines, we
also list the \ux\, charges $q_X$.

{$\mathbf{\mathbb{Z}_3}$}: The gauge group is 
$E_8\otimes E_6\otimes SU(3)$ with $N_G = 334$.

\begin{center}
\begin{tabular}{|l|r|c|r|r|}\hline
name & rep & $q_n$ & $n_s$ & $n/3$ \\ \hline
$U^i$ & (27,3) & $\del^i_n$ & 1 & 81\\
$T$ & (27,1) & ${2\over3}$ & 27 & 243\\
$Y^i$ & (1,$\bar3$) & ${2\over3} + \del^i_n$ & 27 & 81\\
$T^{i\bj}$ & (1,1) & $\del^i_n + \del^j_n$ & 1 & 3\\
$S$ & (1,1) & 0 & 1 & ${1\over3}$\\
\hline\end{tabular}
\end{center}
From the table we have
\beq N = 1225,\qquad \sum_p q^p_n = 816,\eeq  
and
\bea \sum_p q^p_n q^p_m &=& 542 + 168\del_{m n}, \\
\sum_p q^p_n q^p_m q^p_l &=& 396 + 56\(\del_{m n} + \del_{n l}
+ \del_{l m}\) + 168\del_{m n}\del_{n l}.\eea

{$\mathbf{\mathbb{Z}_7}$}: The gauge group is $E_8\otimes E_6\otimes U(1)^2$
with $N_G = 328$.

\begin{center}
\begin{tabular}{|l|r|c|r|r|}\hline
name & rep & $q_n$ & $n_s$ & $n/3$ \\ \hline
$U^i_1$ & 27 & $\del^i_n$ & 1 & 27\\
$U^i_2$ & 1 & $\del^i_n$ & 1 & 1\\
$t^i$ & 27 & $q^i_n$ & 7 & 189\\
$Y_1^{i}$ & 1 & $q^i_n + \del^i_n$ & 7 & 7\\
$Y_2^{i}$ & 1 & $q^i_n + 2\del^i_n$ & 7 & 7\\
$Y_3^{i}$ & 1 & $q^i_n + 4\del^i_n$ & 7 & 7\\
$Y_4^{i}$ & 1 & $q^i_n + \del^{M^i}_n$ & 7 & 7\\
$Y_5^i$ & 1 & $q^i_n + \del^{m^i}_n$ & 7 & 7\\
$Y_6^i$ & 1 & $q^i_n + 2\del^{M^i}_n$ & 7 & 7\\
$Y_7^i$ & 1 & $q^i_n + 2\del^i_n + \del^{M^i}_n$ & 7 & 7\\
$Y_8^i$ & 1 & $q^i_n + \del^i_n - \del^{m^i}_n $ & 7 & 7\\
$T^i$ & 1 & $2\del^i_n$ & 1 & 1\\
$S$ & 1 & 0 & 1 & ${1\over3}$\\
\hline\end{tabular}
\end{center}
where\footnote{These are respectively $\rho^1,\rho^4,\rho^2$ 
of~\cite{ssanom}.}
\beq q^1_n = \({6\over7},{5\over7},{3\over7}\), \qquad q^2_n =
\({3\over7},{6\over7},{5\over7}\),\qquad q^3_n =
\({5\over7},{3\over7},{6\over7}\),\label{rhoz7}\eeq
and
\beq M^i = i + 1\quad {\rm mod}\;3,\qquad m^i = i - 1\quad {\rm mod}\;3.
\eeq
We have
\beq N = 823,\qquad \sum_p q^p_n = 618,\eeq  
and
\bea \sum_p q^p_n q^p_m &=& 438 + 342\del_{m n}, \\
\sum_p q^p_n q^p_m q^p_l &=& 291 + 1425\del_{m n}\del_{n l}
+ 279\Del^+_{n m l} + 191\Del^-_{n m l},\eea
where
\bea \Del^+_{i j k} &=& \del_{i j}\(\del^1_i\del^2_k + \del^2_i\del^3_k
+ \del^3_i\del^1_k\) + {\rm cyclic}(i j k),\nnn \Del^-_{i j k} &=&
\del_{i j}\(\del^2_i\del^1_k + \del^3_i\del^2_k + \del^1_i\del^3_k\)
 + {\rm cyclic}(i j k).\eea

{\bf The FIQS model}:  The gauge group is 
$SO(10)\otimes SU(3)\otimes SU(2)\otimes[U(1)]^7\otimes U(1)_X$ and 
$N_G = 64$.

\begin{center}
\begin{tabular}{|l|r|r|c|r|r|}\hline
name & rep & $\sqrt{6}q_X$ & $q_n$ & $n_s$ & $n/3$ \\ \hline
$Q^i$ & (1,3,2) & $0$ & $\del^i_n$ & 1 & 6\\
$u^i$ & (1,$\bar3$,1) & $0$ & $\del^i_n$ & 1 & 3\\
$L^i$ & (1,1,2) & $0$ & $\del^i_n$ & 1 & 2\\
$\Phi$ & (16,1,1)  & ${3\over2}$ & $\del^i_n$ & 1 & 16\\
$D$ & (1,3,1) & ${2\over3}$ & ${2\over3}$ & 12 & 12\\
$d$ & (1,$\bar3$,1) & ${2\over3}$ & ${2\over3}$ & 15 & 15\\
$L_1$ & (1,1,2) & ${2\over3}$ & ${2\over3}$ & 33 & 22\\
$L_2$ & (1,1,2) & $-{4\over3}$ & ${2\over3}$ & 3 & 2\\
$T_1$ & (1,1,1) & ${2\over3}$ & ${2\over3}$ & 114 & 38\\
$T_2$ & (1,1,1) & $-{4\over3}$ & ${2\over3}$ & 30 & 10\\
$Y^i$ & (1,1,1) & ${2\over3}$ & ${2\over3}+\del^i_n$ & 9 & 9\\
$T^{i\bj}$ & (1,1,1) & $0$ & $\del^i_n + \del^j_n$ & 1 & 3\\
$S$ & (1,1,1) & $0$ & 0 & 1 & ${1\over3}$\\
\hline\end{tabular}
\end{center}

We have
\beq N = 415,\qquad \sum_p q^p_n = 258,\eeq  
and
\bea \sum_p q^p_n q^p_m &=& 158 + 42\del_{m n}, \qquad
\sum_p(q^p_X)^2 = 50,\qquad
\sum_p q^p_n q^p_X = 21\sqrt{6},\\ 
\sum_p q^p_n q^p_m q^p_X &=& \sqrt{6}\(12 + 5\del_{m n}\),\qquad
\sum_p(q^p_X)^2q^p_n = 28,\nnn
\sum_p q^p_n q^p_m q^p_l &=& 108 + 8\(\del_{m n} + \del_{n l}
+ \del_{l m}\) + 42\del_{m n}\del_{n l}.\eea

%% file: AppendixE.tex
\subsection{Construction of the GS Term}\label{gsterm}
\setcounter{equation}{0} 
In this appendix we detail the steps in the construction of
the GS term.  As in \myapp{fullan}, the notation $\l\Phi\r$ will
denote the bosonic part of the lowest component $\phi$ of the superfield
$\Phi$.

\subsubsection{Anomaly superfields}\label{anomsf}
Part of the anomaly can be expressed in term of 
supersymmetric field operators of the form\footnote{The sign on the RHS of
(A12) in I is incorrect.  As a consequece the signs of the $L_\alpha$ terms
should be flipped in (A7) and in the expressions for $\tL_0$ and $\tL_G$
in (2.11).  The coefficient of $\L_\alpha$ in (2.11) should be $- 1/18$, and
the coefficient of $\Xc\Xa$ in (A8) should be $+1/6$.}
\bea L(T,T',H) &=& 
{1\over2}\int d^4\theta {E\over R}T^\alpha T'_\alpha H(Z) 
+ {\rm h.c.}\nonumber \\
&=&  {\sqrt{g}\over2}H(z)\l\Da T'_\beta\D^\alpha T^\beta\r + {\rm h.c.} + 
{\rm fermions} \nonumber \\ 
&=&  \sqrt{g}\(2\re H T_0T'_0 + \re H T'_{\mu\nu}T^{\mu\nu} + \im H
{\tilde T}'_{\mu\nu}T^{\mu\nu}\) + {\rm fermions},\eea
where $H(Z)$ is a holomorphic function of the chiral fields, 
with $T_\alpha$ defined by (\ref{tft}) and \cite{pvdil}:
\bea T_0 &=& \half\l\D^\alpha T_\alpha\r = - D_{\m}T_i\(e^{-K}
\A^i A^{\m} + \D_\mu z^i\D^\mu\z^{\m}\) + x^{-1}\D_a T_i(T^a z)^i,
\nnn T_{\mu\nu} &=& \[\(\D_\mu z^i\D_\nu\z^{\m} - 
\D_\nu z^i\D_\mu\z^{\m}\)D_{\m} - iF^a_{\mu\nu}(T_az)^i\]T_i.
\label{E2}\eea
In particular we have contributions with $T_\alpha = \Xa$ and $\Xa^m$;
$\Xa^m$ is defined in \myref{defxm}.
In addition we have the supersymmetric operators
\bea L({\rm YM},H) &=& {1\over2}\int d^4\theta{E\over R}
H(Z)\WaWa + {\rm h.c.}\nonumber
\\ &=& {\sqrt{g}\over2}\l H(Z)\Da W_\beta^a\D^\alpha
W^\beta_a\r + {\rm h.c.} + {\rm fermions} \nonumber \\ &=&
- \sqrt{g}\[\re H\(F_a^{\mu\nu}F^a_{\mu\nu} - {2\over x^2}\D_a\D^a\) 
+ \im H F_a\cdot\tF^a\] + {\rm fermions}.\label{dYM}\eea
and ($X_{\mu\nu}= K_{\mu\nu})$
\bea L(W,H) &=& {1\over2}\int d^4\theta{E\over R}
H(Z)W^{\alpha\beta\gamma} W_{\alpha\beta\gamma} + {\rm h.c.}\nonumber
\\ &=& {\sqrt{g}\over2}\l H(Z)\Da W_{\beta\gamma\delta}\D^\alpha
W^{\beta\gamma\delta}\r + {\rm h.c.} + {\rm fermions} \nonumber \\ &=&
{\sqrt{g}\over8}\[\re H\(r^{\mu\nu\rho\sigma}r_{\mu\nu\rho\sigma} 
- 2r_{\mu\nu}r^{\mu\nu} + {1\over3}r^2\) + \im H r\cdot\tr\] 
\nonumber \\ & & + {\sqrt{g}\over12}\(\re H X_{\mu\nu}X^{\mu\nu} + 
\im H\tX_{\mu\nu}X^{\mu\nu}\) + {\rm fermions}.\label{d3}\eea

For the full cancellation of the anomalies we will also need the operators
introduced in \mysec{strat}, equations \myref{omR} and \myref{Gm}.  
The ``D-term'' combination $\Omega_m$ defined in \myref{Om2} can be
written 
\bea \Omega_m &=& - \({1\over4}\bD^2\D^2\ln\cM +
[\D^2,\Dd]\ln\cM\Db\ln\cM
- \bD^2\ln\cM\Dc\ln\cM\Da\ln\cM\hc\)
\ddd + {3\over2}\D^2\ln\cM\bD^2\ln\cM 
- \half\{\Da,\Dd\}\ln\cM \{\Dc,\Db\}\ln\cM
+ 2\Dc\ln\cM\Db\ln\cM[\Da,\Dd]\ln\cM\ddd
+ \(4R\Dc\ln\cM\Da\ln\cM - 2R\D^2\ln\cM - 4\Dc\ln\cM\Da R + \D^2R\hc\)
\ddd - 8R\bR + 2G_{\alpha\dot\beta}
\(2\Dc\ln\cM\Db\ln\cM - [\Dc,\Db]\ln\cM + G^{\alpha\dot\beta}\)\eee
- \Omega_L - 2\Omega_{L X}- 8R\bR + \D^2R + \bD^2\bR 
+ 2G_{\alpha\dot\beta}G^{\alpha\dot\beta}
+ \ln\cM\(\half\Dc\La + 2\ln\cM\Dc\Xa\)
\ddd - {\partial\over\pp\ln\cM}\[{1\over4}\(\D^2\ln\cM\Dd\ln\cM\Db\ln\cM\hc\)
- 2G_{\alpha\dot\beta}\Dc\ln\cM\Db\ln\cM
\right.\mmm\qquad\l + \(\ln\cM\lbr{1\over8}\bD^2\D^2\ln\cM 
+ \Dc(R\Da\ln\cM)\rbr
\hc\)\right.\mmm\qquad\l + \half\Dc\ln\cM\Da\ln\cM\Dd\ln\cM\Db\ln\cM\]
\nnn &\equiv& - \Omega_L - 2\Omega_{L X}- 8R\bR + \D^2R + \bD^2\bR 
+ 2G_{\alpha\dot\beta}G^{\alpha\dot\beta} - {\pp\over\pp\ln\cM}f(\ln\cM).
\label{E5}\eea 
The result \myref{E5} is determined only up to a linear supermultiplet $L_0$
\beq \chiproj L_0 = \bchiproj L_0 = 0.\label{E6}\eeq
which does not contribute to the variation of the action, 
and the argument of the $\ln\cM$ derivative is determined up to a total
spinorial derivative. The real superfields
\bea \Omega_L &=& \Lc\Da\ln\cM + \Dd(\ln\cM\Lb)
= \Ld\Db\ln\cM + \Dc(\ln\cM\La),\label{E7}\\
\Omega_{L X} &=& \Xc\Da\ln\cM + \Dd(\ln\cM\Xb)
= \Xd\Db\ln\cM + \Dc(\ln\cM\Xa)\eea
are the Chern-Simons superfields~\cite{cfv}, respectively, 
for the chiral superfields
\beq \Phi_L = \Lc\La, \qquad\Phi_{L X} = \Xc\La, 
\qquad \La = \chiproj\Da\ln\cM, \eeq
as discussed in the next subsection.  If we define $\Omega'_m$ by
\beq {1\over48}\Omega_m + {1\over36}\Omega_{X_m} = 
{1\over48}\Omega'_m + {1\over36}\Omega_{X}, \qquad
\Omega'_m = \Omega_m + {3\over4}\Omega_L + 2\Omega_{L X},\eeq
the mixed term $\Omega_{L X}$ drops out and we obtain
\bea \Omega'_m &=&
- {1\over4}\Omega_L - 8R\bR
+ 2G_{\alpha\dot\beta}G^{\alpha\dot\beta} + \D^2R + \bD^2\bR 
\ddd - {\partial\over\pp\ln\cM}
\[{1\over4}\(\D^2\ln\cM\Dd\ln\cM\Db\ln\cM\hc\)
- 2G_{\alpha\dot\beta}\Dc\ln\cM\Db\ln\cM
\right.\mmm\qquad\l + \(\ln\cM\lbr{1\over8}\bD^2\D^2\ln\cM 
+ \Dc(R\Da\ln\cM)\rbr
\hc\)\right.\mmm\qquad\l + \half\Dc\ln\cM\Da\ln\cM\Dd\ln\cM\Db\ln\cM
\right.\mmm\qquad\l - (\ln\cM)^2\({1\over4}\Dc\La + \ln\cM\Dc\Xa\)\].
\label{E10}\eea 
The first five terms on the right hand side of \myref{E10} are
invariant under modular and \ux\, transformations and have the correct
chiral and Weyl weights for some part of them be included in the
chiral projection of the modified linear superfield $L$; for example the 
second and third terms complete the PV contribution to the anomalous
coefficent of $\Omega_{G B}$.  The remainder may require 
additional counterterms for anomaly cancellation.  Defining $\Omega_r$ by
\beq \Omega'_m = - {1\over4}\Omega_L - 8R\bR
+ 2G_{\alpha\dot\beta}G^{\alpha\dot\beta} + \Omega_r
+ \D^2R + \bD^2\bR,\label{E12} \eeq
it is the coefficient of
\beq \del\ln\cM^2 = 2\del\ln\cM = 2\(\del H + \del\H\), 
\qquad \Dd H = \Da\H = 0,\eeq
under an infinitesimal modular and/or \ux\, transformation on
the corresponding part 
\beq \L_r  = - {1\over8\pi^2}{1\over48}\ell_r = 
- {1\over8\pi^2}{1\over48}
\superint E\Tr\eta\int2d\ln\cM\Omega_r(\ln\cM)\label{E14}\eeq
of the Lagrangian.  Under a finite transformation
\beq \Del\ln\cM = H +\H,\eeq
we have\footnote{The integrated anomaly for pure supergravity was
given in ref. \cite{bk}.}
\bea \Del\ell_r &=&  2\superint E\Tr\eta\lbr
\(\half\Dc\La + 2\Dc\Xa\)\[H\H + 2\(H + \H\)\ln\cM\]
\right.\mmm\qquad\l - {1\over4}\(\D^2H\Dd\H\Db\H
+ 2\D^2H\Dd\H\Db\ln\cM 
+ \D^2\ln\cM\Dd\ln\H\Db\H\hc\)\right.\mmm\qquad\l -
{1\over4}\(\D^2H\Dd\ln\cM\Db\ln\cM 
+ 2\D^2\ln\cM\Dd\H\Db\ln\cM\hc\)\right.\mmm\qquad\l
+ G^{\alpha\dot\beta}\(\Da H\Dd\H + \Da H\Dd\ln\cM + \Da\ln\cM\Dd\H\)
\right.\mmm\qquad\l
- \(\{H + \ln\cM\}\lbr{1\over8}D^2\bD^2\H + \Dc(R\Da H)\rbr
\hc\)\right.\mmm\qquad\l
- \(H\lbr{1\over8}\D^2\bD^2\ln\cM + \Dc(R\Da\ln\cM)\rbr
\hc\)\right.\mmm\qquad\l
- \half\Dc H\Da H\Dd\H\Db\H -
\Dc H\Da H\Dd\H\Db\ln\cM \right.\mmm\qquad\l
-\Dc H\Da\ln\cM\Dd\H\Db\H -
2\Dc H\Da\ln\cM\Dd\H\Db\ln\cM\right.\mmm\qquad\l -
\half\Dc H\Da H\Dd\ln\cM\Db\ln\cM -
\half\Dc\ln\cM\Da\ln\cM\Dd\H\Db\H\right.\mmm\qquad\l -
\Dc H\Da\ln\cM\Dd\ln\cM\Db\ln\cM -
\Dc\ln\cM\Da\ln\cM\Dd\H\Db\ln\cM\rbr.
\label{E11}\eea
Note that since $\Dc\La,\Dc\Xa$ are linear supermultiplets subject
to the condition \myref{E6}, it is obvious that the constant
$\ln\mu$ in the expression \myref{lnM} for $\ln\cM$ drops
out of \myref{E11}.

\subsubsection{Chern-Simons superfields}\label{csfields}
We wish to impose the modified linearity condition \beq
\chiproj(L + \Omega) = \bchiproj(L + \Omega) = 0, \eeq
where $\Omega$ is a superfield with \uk\, chiral weight
$w_\chi(\Omega) = 0$ and Weyl weight $w_W(\Omega) = 2$ such that
\beq \L(\Omega) = {\sqrt{g}\over32\pi^2}\superint E\Omega, \qquad \Omega =
e^{K/3}\Omega_0,\label{LO}\eeq
is invariant under superweyl transformations.  For example, the
standard case has $\Omega = \Omega_{YM}$.  In this case the
Lagrangian (\ref{LO}) reads in component form
\beq \L(\Omega_{YM}) = - {\sqrt{g}\over32\pi^2}F_a^{\mu\nu}F^a_{\mu\nu} +
\cdots = - {\sqrt{g_0}\over32\pi^2}F_a^{\mu\nu}F^a_{\mu\nu} + \cdots, \eeq
where we dropped fermions and auxiliary fields; these have
nontrivial transformations under the superweyl transformation~\cite{bggm}.  
%
%
%
%
%
%
%
Girardi and Grimm \cite{gg} constructed an explicit expression for the
Yang-Mills CS superfield in terms of prepotential superfields
$\varphi_\alpha,\phib,$ defined as the (ordinary) spinorial
derivatives of a hermetian superfield $U$ (called $\Upsilon$ in the
notation of \cite{gg}) such that the (covariant) chiral(anti-chiral)
projection of $\varphi(\bv)$ is the chiral(anti-chiral) Yang-Mills
superfield strength:
\beq \varphi_\alpha = - U^{-1}D_\alpha U, \quad \phib =
\Db U\bU, \quad \chiproj\phia = -8\Wa, \quad \bchiproj\phib =
8\Wb.\eeq
The CS superfield is constructed from these prepotentials and their
covariant spinorial derivatives:
\bea \Omega_{YM} &=& - {1\over8}\Tr\(\phic\Wa\) - {i\over32}\Db
\(\chid + \sigd\),  \nnn
\chid &=& {i\over2}\Tr\[\Yd\Dc\Ya + 2\Yc\Da\phid\],
\nnn
\sigd &=& \Tr\(\int_0^1dtY_t\[\Yc,Y_{\alpha\dot\beta}\]
+ {i\over2}\int_0^1dtY_t\[\Dc\Ya,\Yd\]\), \nnn
Y_A &=& \D_AU\bU, \quad Y_t = \pp_tU(t)\bU(t),\eea
and the interpolating superfield $U(t)$ is defined to satisfy: 
\beq U(0) = 1,\quad U(1) = U.\eeq 
The above construction is quite general.
In particular we can take $U$ to be a real function $U(Z,\Z,\vx)$ of the
chiral superfields and the \ux\, vector superfield. The above construction with 
\beq U\to U_K = e^{-K/8}\eeq 
is the CS form $\Omega_X$ for $\Xc\Xa$.  Similarly, 
\beq U\to U_m = \cM\eeq
gives the CS form $\Omega_L$ for $\Lc\La$. 
For these Abelian cases, the integration is trivial, giving, up to a 
linear superfield, the expression \myref{E7} for $\Omega_L$, and an
analogous expression with $\ln\cM\to - {1\over8}K$ for $\Omega_X$.
%

The CS forms $\Omega_X$ and $\Omega_L$ are explicitly dependent on the
dilaton through the presence of the K\"ahler potential $K$ and therefore
do not satisfy the condition \myref{opcond}.  In the remainder of this
section we show that the linear/chiral multiplet duality outlined in
\myref{genlin}--\myref{delgv2} of \mysec{strat} still holds.

\def\cS{{\cal S}}\def\cS{{\cal S}}

The action is
\beq \cS = - \int E\[3 - 2L s(L) + (S + \S)(L + \Omega)\], \qquad K
= k(L) + G(T + \T, \Phi),\label{actionApp}\eeq
with
\beq \Omega = e^{K/3}\Omega_0,\qquad S = \chiproj \Sigma,\eeq
and $L,\Sigma$ unconstrained.  To assure a canonical Einstein term in the chiral formulation we require
\beq 2L s(L) - L(S + \S) = 0.\label{einstApp}\eeq
First consider the standard case:
\beq {\pp\over\pp L}\Omega_0 =
{\pp\over\pp S}\Omega_0 =0.\eeq
The equation of motion for $\Sigma$ gives
\beq {1\over E}{\del\cS\over\del\Sigma} = -\chiproj(L + \Omega) = 0,\label{modlinApp}\eeq
which is just the modified linearity condition for $L$ and which implies
\beq \int E S(L+\Omega)\hc = 0,\eeq
so the action \myref{actionApp} reduces to
\beq \cS = - \int E\[3 - 2L s(L)\].\label{LactionApp}\eeq
On the other hand the equation of motion for $L$ is 
\bea {1\over E}{\del\cS\over \del L} &=& 2s(L)+ 2L s'(L) - S - \S +
     {\pp K\over\pp L}\ddd - {1\over3}{\pp K\over\pp L}\[2L s(L) - (S
     + \S)L\] \eee 2L s'(L) + K'(L) = 0,\label{einst2App}\eea 
\noindent
where the last line, which assures a canonical Einstein term in the
linear formulation, was obtained using \myref{einstApp}.  Integrating
this using \myref{einstApp} gives
\noindent
\beq S + \S = 2s(L) = - \int{d L\over L}{\pp K\over\pp
  L}.\label{dualityApp}\eeq
So for example if $K = \ln L$, $S + \S = 1/L$.  Using \myref{einstApp},
the action expressed in terms of $S$ reduces to
\beq \cS = - \int E\[3 + (S + \S)\Omega\].\label{SactionApp}\eeq

To cancel the anomaly we can add a constant of integration $V =
V(Z),\;Z\ne S,L$ to the right hand side of \myref{dualityApp}:
\noindent
\beq S + \S = 2s(L) = - \int{d L\over L}{\pp K\over\pp L} +
V.\label{duality2App}\eeq
Or, equivalently, we can add a term 
\beq - \int E L V \eeq
to the action \myref{actionApp}.

Now consider instead the case {$\Omega_0 = \Omega_0(\sigma)$}, $\sigma
= S + \S$.  Although the superfield $\sigma$ is no longer simply a
Lagrange multiplier, it is still a nonpropagating field that can be
removed by its equation of motion.  In this case $\Omega$ still drops
out of the equation of motion for L, leaving \myref{einstApp} and
\myref{einst2App}--\myref{SactionApp} unchanged.  However the equation
of motion for $\Sigma$ now reads
\beq {1\over E}{\del\cS\over\del\Sigma} = -\chiproj\[L + \Omega + (S + \S){\pp\over\pp S}\Omega\] = 0,\label{modlin2App}\eeq
\noindent
which gives a different modified linearity condition for $L$ and the
action \myref{actionApp} gets an extra term
\noindent
\beq \cS = - \int E\[3 - 2L s(L) + (S + \S)\(S{\pp\over\pp
  S}\Omega\hc\)\].\label{Laction2App}\eeq
\noindent
When we calculate the component Lagrangian, after inserting the
operator $\chiproj$ and integrating by parts and neglecting the
Einstein term $-3\int E$, we obtain
\noindent
\bea \cS &\ni& - {1\over8}\int{E\over R}\[s(L)\chiproj L - S(S
+\S)\chiproj{\pp\over\pp S}\Omega\]\hc\eee {1\over8}\int{E\over
  R}\[s(L)\chiproj\Omega + \{s(L) - S\}(S +
\S)\chiproj{\pp\over\pp\sigma}\Omega\]\hc.\eea
Using \myref{dualityApp} the last term cancels out and the first term is just the counterpart of the second term in \myref{SactionApp}
in the linear formulation.

Finally if we take $\Omega_0 = \Omega_0(L)$, the equation of motion
for $\Sigma$ just gives \myref{modlinApp}--\myref{LactionApp}.  but
using \myref{einstApp} the equation of motion for $L$ gives
\noindent
\beq {1\over E}{\del\cS\over \del L} = 2L s'(L) + k'(L) -
2s(L)e^{K/3}{\pp\Omega_0\over\pp L} = 0,\label{einst3App}\eeq
so now we get a different differential equation for $s(L)$ but we
still get the same action.

%% file: AppendixF.tex
\subsection{Notations and conventions}\label{notation}
\setcounter{equation}{0} 
In this Appendix we summarize our notation and conventions.
\subsubsection{Sign conventions}
Our Dirac matrices and space-time metric signature $(+---)$ are those
of Bjorken and Drell or Itzykson and Zuber.  We use upper case
notation $(R,\Gamma)$ for derivatives of the K\"aher metric $K_{i\m}$,
and lower case $(r,\gamma)$ for derivatives of the space-time metric
$g_{\mu\nu}$.  Our sign conventions for, respectively, the Riemann
tensor, Ricci tensor, and curvature scalar are
\bea r^\mu_{\nu\rho\sigma} &=& g^{\mu\lambda}r_{\lambda\nu\rho\sigma}
= \pp_\sigma\Gamma^\mu_{\nu\rho} - \pp_\rho\Gamma^\mu_{\nu\sigma}
+ \Gamma^\mu_{\sigma\tau}\Gamma^\tau_{\nu\rho}
- \Gamma^\mu_{\rho\tau}\Gamma^\tau_{\nu\sigma},
\quad\; r_{\mu\nu} = r^\rho_{\mu\rho\nu}, \quad\; 
r = g^{\mu\nu}r_{\mu\nu}.\eea
The general coordinate covariant derivative is defined by
\beq \nabla_\mu A_\nu = \pp_\mu A_\nu - \gamma^\rho_{\mu\nu}A_\rho,
\qquad {\rm etc.}\eeq
We use identical definitions for the K\"ahler curvature and
scalar field reparameterization derivatives with 
\beq g_{\mu\nu}\to K_{i\m} = K_{\m i}, \qquad \gamma\to\Gamma,\qquad
r\to R, \qquad \nabla_\mu\to D_i,D_{\m}.\eeq
We use the Yang-Mills sign conventions of \cite{crem} and \cite{fz}:
\bea \D_\mu\phi^b &=& \nabla_\mu\phi^b + iA^a_\mu(T_a)^b_c\phi^c,\\
\D_\mu\bph^{\bar b} &=& \nabla_\mu\bph^{\bar b} - 
iA^a_\mu(T_a)_{\bar c}^{\bar b}\bph^{\bar c},\qquad
(T_a)_{\bar c}^{\bar b} = (T^*_a)^b_c = (T_a)^c_b.\eea
Note that $\D_\mu$ is used for general coordinate and Yang-Mills
covariant derivatives, while $D_\mu$ is fully covariant.  Thus
for a fermion $\psi$, $D_\mu\psi$ includes the spin, Yang-Mills,
K\"ahler and reparameterization (as well as the affine connection
for the graviton $\psi_\mu$), and for a function of scalar fields
$z^i,\z^{\m}$,
\beq D_\mu f(z,\z) = \D_\mu z^i D_i f(z,\z) + \D_\mu z^{\m} D_{\m}
f(z,\z),\eeq
with repeated indices summed.

\subsubsection{Supergravity conventions}
We work in K\"ahler \uo\, superspace~\cite{bggm}. The tree level
Lagrangian we are working with is defined by the K\"ahler potential
$K$, superpotential $W$ and gauge kinetic function $f_{a b}$ given
in \myref{sdil}.  A generic chiral superfield is denoted by $Z^i$:
\beq Z^i = S,T,\Phi = (z^i,\chi^i,F^i),\eeq
with components 
\beq z^i = \l Z^i\r,\qquad \chi^i_\alpha = {1\over{\sqrt2}}\l\Da Z^i\r,
\qquad F^i = - {1\over4}\l\Dc\Da Z^i\r\equiv 
- {1\over4}\l\D^2 Z^i\r,\eeq
where $\l Z\r$ denotes the lowest component, and the antichiral
superfields are 
\beq \Z^{\m} = (Z^m)^\dag = (\z^{\m},\bar\chi^{\m},\bF^{\m}).\eeq
We also use the notation 
\beq Z^p = T^i,\Phi^a = (z^p,\chi^p,F^p),\qquad i = 1,2,3,\eeq
for the moduli 
\beq T^i = (t^i,\chi^{t^i},F^{t^i}) \eeq
and the gauge charged chiral multiplets 
\beq \Phi^a = (\phi^a,\chi^a,F^a).\eeq
The dilaton chiral supermultiplet decomposes as
\beq  S = (s,\chi^s,F^s), \qquad s = x + i y.\eeq
The Yang-Mills superfield strengths are
\beq \Wa^a = (\lambda^a_\alpha,F^a_{\mu\nu},D^a),\eeq
and the supergravity supermultiplet includes the vierbein $e^\alpha_\mu$,
the gravitino $\psi_\mu$ and the auxiliary fields 
$G_{\alpha\dot\beta},R,\bR$.
Solution of the tree level equations of motion determine the
auxiliary fields as
\bea \l F^i\r &=& - e^{-K/2}A^i = - e^{K/2}K^{i\m}\(\bar W_{\m} 
+ K_{\m}\bar W\), 
\qquad \l D^a\r = {1\over x}\D^a = {1\over x}K_i(T^a z)^i, \\
\l R\r &=& \half M = e^{- K/2}A = e^{K/2}W(z) = \(\l\bR\r\)^\dag, 
\qquad \l G_{\alpha\dot\beta}\r = 0,\eea
where here (and throughout the text) the notation $\l{\cal O}(\Psi)\r$
denotes the bosonic component of the functional ${\cal O}$ of the
superfields $\Psi$, that is, the $\theta = \bar\theta = 0$ component
with all fermions set to zero.  Our definition of $M$ is such that
the vacuum value $\langle M\rangle$ is the gravitino mass; it
differs by a factor $- 1/3$ from that of \cite{bggm}.  The
covariant derivatives 
\beq A_{i_1\cdots i_n} = D_{i_1}\cdots D_{i_n}A \eeq 
of the modular covariant operator
\beq A = e^K W,\qquad A \to e^{\bF}A,\qquad A_i \to e^{\bF}A_i,
\qquad {\rm etc.}\eeq
are defined in \myref{DA}.  Note that for the K\"ahler potential and
superpotential $W$ we use the conventional notation for ordinary
derivatives
\beq K_{i_1\cdots i_n} = \pp_{i_1}\cdots\pp_{i_n}K, \qquad
W_{i_1\cdots i_n} = \pp_{i_1}\cdots\pp_{i_n}W.\eeq

\subsubsection{PV superfields}
To regulate the superpotential couplings we introduce 
modular covariant PV fields 
\beq \dZ^P = \dot T^I,\dot\Phi^A\eeq 
that transform like $dZ^p$ under gauge and modular
transformations. To make the K\"ahler potential for these fields
modular invariant and their superpotential modular covariant, we need
to introduce three additional fields $\dZ^N,\; N = 1,2,3,$ and we
also use the notation
\beq \dZ^\rho = \dZ^N,\dZ^P.\eeq
These fields acquire mass through invariant couplings to the PV 
superfields 
\beq \dY_\rho = \dY_N,\dY_P,\eeq
where $\dY_P$ transforms like $K_{p\bar q}d\Z^{\bar q}$ under YM
transformations.  These have no other superpotential couplings; their
divergent contributions to field strength terms are canceled by the
PV fields, introduced in \myapp{orb1},
\beq \Psi^C = U^A,U_A,V^A,\Phi^N,\Phi^n,\qquad N,n = 1,2,3,\eeq
where $\Phi^{N,n}$ are gauge singlets, and $U^A,U_A,V^A$ form a real
(reducible) representation of the gauge group such that
\myref{cascond} is satisfied.  Their masses not invariant under \ux\,
and modular transformations; they reflect chiral matter contributions
to the anomaly.

To regulate all the nonrenormalizable terms in the superpotential and
YM couplings, we also need fields $\tZ^P$; they acquire invariant
masses through coupling to $\tY_P$. These transform, respectively,
like $dZ^p$ and $K_{p\bar q}d\Z^{\bar q}$ under YM and modular
transformations.

To regulate the renormalizable Yang-Mills couplings we introduce
chiral PV fields $\hY_P$ that transform like $K_{p\bar q}d\Z^{\bar q}$
under YM and modular transformations, and acquire invariant masses
through coupling to chiral superfields $\hZ^P$ that transform like
$\dZ^P$.  We also need chiral superfields $\varphi^a,\hph^a,\tph^a$ in
the adjoint representation of the YM gauge group. The fields
$\varphi^a,\hph^a,$ regulate gauge couplings to matter and to the
gravity sector, respectively, and couple to one another in an
invariant mass term; $\tph^a$ has no couplings to light matter and its
noninvariant mass term reflects the gaugino/gauge contribution to the
modular anomaly.

To regulate nonrenormalizable dilaton/YM couplings we introduce chiral
superfields $\phi^S$, that acquire invariant masses through
superpotential couplings to chiral fields $\phi_S$, and chiral
superfields $\theta_s$ and Abelian vector superfields $V_s$ that
acquire invariant masses through a superhiggs mechanism.  Finally, to
regulated additional nonrenormalizable gravity couplings we need
chiral superfields $\phi^C$, with noninvariant masses that reflect the
gravity sector contribution to the modular anomaly, and chiral
superfields $\theta_0$ and Abelian vector superfields $V_0$ that
acquire invariant masses through a superhiggs mechanism.

Unless otherwise specified, $\Phi^C$ denotes any chiral PV superfield
with PV signature
\beq \eta^C = \pm 1.\eeq

\subsubsection{The covariant derivative expansion}\label{cdexp}
Here we collect the operators~\cite{us2} that appear in the covariant
derivative expansion used in the evaluation of the variation of the
action.  If $F(x)$ is a scalar field operator, we define 
\beq \hF = \sum_{n=0}^\infty{(-i)^n\over n!}\(D\cdot{\pp\over\pp p}\)^n
F(x),\qquad D\cdot{\pp\over\pp p}F(x) \equiv [D_\mu,F(x)]{\pp\over\pp
p_\mu}.\label{hF}\eeq 
In addition we define 
\beq \G_\mu = \sum_{n=0}^\infty{n+1\over(n+2)!}  \(-iD\cdot{\pp\over\pp
p}\)^n\G_{\nu\mu}{\pp\over\pp p_\nu},\qquad \G_{\nu\mu} =
[D_\mu,D_\nu].\label{Gmu}\eeq 
The full formal expansions of the following space-time curvature-dependent operators 
\bea T^{\mu\nu} &=& g^{\mu\nu} - {1\over3}r^{\mu\rho\sigma\nu}
{\pp^2\over\pp p^\rho\pp p^\sigma} + {i\over6}\nabla^\tau
r^{\mu\rho\sigma\nu}{\pp^3\over\pp p^\rho\pp p^\sigma\pp p^\tau}
+ O\(\pp^4\over\pp p^4\),\label{Tmunu}\\
P^{\mu\nu}\gamma_\nu &=& P^\mu = \gamma^\mu -
{1\over6}r^{\mu\rho\sigma\nu}\gamma_\nu{\pp^2\over\pp p^\rho\pp p^\sigma}
+ O\(\pp^3\over\pp p^3\),\label{Pmunu}\\
\del_\mu &=& {i\over9}\(\nabla_\mu r_{\rho\nu} - \nabla_\nu r_{\rho\mu}\)
{\pp^2\over\pp p_\rho\pp p_\nu}
+ O\(\pp^3\over\pp p^3\),\label{delmu}\\
X &=& - {r\over3} + {i\over3}\nabla_\mu r{\pp\over\pp p_\mu}
+ O\(\pp^2\over\pp p^2\),\label{Xexp}\eea
are given\footnote{There is a sign
error in the third and second terms, respectively, of the expressions
for $T_{\mu\nu}$ and $X$ in Eq. (A.19) of~\cite{us2}.}
 in Appendix A of \cite{sigma}.